\documentclass[11pt,preprintnumbers,amsmath,onecolumn,nofootinbib]{revtex4}

\usepackage[latin1]{inputenc}

\usepackage[french,english]{babel}

\usepackage{epsfig}
\usepackage{graphicx}
\usepackage{bm}
\usepackage{amsmath}
\usepackage{amssymb}

\newcommand{\si}{\sigma}
\newcommand{\al}{\alpha}

\newcommand{\az}{\varphi}
\newcommand{\ro}{\rho}

\newcommand{\la}{\lambda}
\newcommand{\be}{\beta}

\newcommand{\tro}{\tilde{\rho}}

\newcommand{\Del}{\Delta}
\newcommand{\ka}{\kappa}

\newcommand{\oeq}{\begin{equation}}
\newcommand{\ceq}{\end{equation}}
\newcommand{\oeqn}{\begin{eqnarray}}
\newcommand{\ceqn}{\end{eqnarray}}

\renewcommand{\>}{\rangle}
\newcommand{\<}{\langle}
\renewcommand{\(}{\left(}
\renewcommand{\)}{\right)}
\renewcommand{\[}{\left[}
\renewcommand{\]}{\right]}
\renewcommand{\ll}{\left|}
\newcommand{\rl}{\right|}

\newcommand{\stf}{\,\,\,}
\newcommand{\sdf}{\,\,}
\newcommand{\stb}{\!\!\!}
\newcommand{\sdb}{\!\!}

\newcommand{\kfi}{|\phi \>}

\newcommand{\kpsi}{|\psi \>}

\newcommand{\kal}{|\al\>}

\newcommand{\kvac}{|-\>}
\newcommand{\knu}{|\nu\>}
\newcommand{\kmu}{|\mu\>}

\newcommand{\bfi}{\<\phi |}

\newcommand{\bvac}{\< - |}
\newcommand{\bmu}{\<\mu|}

\newcommand{\bpsi}{\<\psi|}

\newcommand{\oQ}{\hat{Q}}

\newcommand{\oP}{\hat{P}}

\newcommand{\oH}{\hat{H}}

\newcommand{\ox}{\hat{x}}
\newcommand{\oy}{\hat{y}}
\newcommand{\oz}{\hat{z}}

\newcommand{\oV}{\hat{V}}

\newcommand{\oD}{\hat{D}}
\newcommand{\oR}{\hat{R}}

\newcommand{\oro}{\hat{\rho}}
\newcommand{\oh}{\hat{h}}

\newcommand{\osi}{\hat{\sigma}}
\newcommand{\ovr}{\hat{\bf r}}
\newcommand{\ovp}{\hat{\bf p}}
\newcommand{\ovR}{\hat{\bf R}}

\newcommand{\op}{\hat{p}}

\newcommand{\of}{\hat{f}}

\newcommand{\oad}{\hat{a}^\dagger}
\newcommand{\oa}{\hat{a}}
\newcommand{\oA}{\hat{A}}
\newcommand{\oB}{\hat{B}}
\newcommand{\oF}{\hat{F}}
\newcommand{\oN}{\hat{N}}
\renewcommand{\ox}{\hat{x}}
\renewcommand{\oy}{\hat{y}}
\renewcommand{\oz}{\hat{z}}
\newcommand{\ob}{\hat{b}}
\newcommand{\obd}{\hat{b}^\dagger}

\newcommand{\ovsi}{\hat{\boldsymbol{\sigma}}}
\newcommand{\odel}{\hat{\delta}}
\newcommand{\ovk}{\hat{\bf k}}
\newcommand{\oO}{\hat{O}}
\newcommand{\ov}{\hat{v}}
\newcommand{\obe}{\hat{\beta}}
\newcommand{\obed}{\hat{\beta}^\dagger}

\newcommand{\del}{\delta\!}
\newcommand{\dt}{\frac{\partial}{\partial t}}

\renewcommand{\d}{{\mbox d}}

\newcommand{\hb}{\hbar}

\renewcommand{\vr}{{\bf r}}

\newcommand{\vv}{{\bf v}}

\newcommand{\vp}{{\bf p}}
\newcommand{\vP}{{\bf P}}
\newcommand{\ve}{{\bf e}}

\newcommand{\mH}{{\mathcal{H}}}

\newcommand{\mN}{{\mathcal{N}}}

\newcommand{\mA}{{\mathcal{A}}}
\newcommand{\mL}{{\mathcal{L}}}

\newcommand{\Tr}{\mbox{Tr}}

\newcommand{\hba}{\hat{\beta}}

\newcommand{\ket}[1]{\vert #1 \rangle}

\newcommand{\crea}{\hat{a}^\dagger}
\newcommand{\anni}{\hat{a}}

\begin{document}

\title{\Huge Microscopic approaches for nuclear Many-Body dynamics \\
\vspace{0.5cm}{\it \huge Applications to nuclear 
reactions\footnote{Lecture given at the ''Joliot Curie'' school, Maubuisson, september 17-22, 2007. A french version is available at the URL http://www.cenbg.in2p3.fr/heberge/EcoleJoliotCurie/coursannee/cours/CoursSimenel.pdf .}} \\
\vspace{1.5cm}}

\author{\Large Cédric Simenel and Benoît Avez}
\affiliation{DSM/DAPNIA/SPhN, CEA SACLAY, F-91191
Gif-sur-Yvette Cedex, France}
\author{\Large Denis Lacroix}
\affiliation{GANIL, B.P. 55027,
F-14076 CAEN Cedex 5, France}

\maketitle

\vspace{1.5cm}

\centerline{\large ABSTRACT}

\vspace{0.5cm}

These lecture notes are addressed to PhD student and/or researchers who want a general overview of 
microscopic approaches based on mean-field and applied 
to nuclear dynamics.
Our goal is to provide a good description of low energy heavy-ion collisions.
We present both formal aspects and practical applications of the time-dependent
Hartree-Fock (TDHF) theory. The TDHF approach gives a mean field dynamics
of the system under the assumption that particles evolve independently in their self-consistent average field. 
As an example, we study 
the fusion of both spherical and deformed nuclei with TDHF. We also focus on
nucleon transfer which may occur between nuclei below the barrier. These studies
allow us to specify the range of applications of TDHF in one hand, and, on the 
other hand, its intrinsic limitations: absence of tunneling below the Coulomb barrier, 
missing dissipative effects and/or quantum fluctuations. Time-dependent mean-field 
theories should be improved to properly account for these effects.
Several approaches, generically named "beyond TDHF" are presented which  
account for instance for pairing and/or direct nucleon-nucleon collisions.
Finally we discuss recent progresses in exact ab-initio methods based on the 
stochastic mean-field concept. 

\newpage

\def\tocname{Table of content}

\tableofcontents

\newpage

\section{Introduction}

\subsection{General considerations}

Heavy-Ion accelerators have given important information on nuclear reactions
with stable nuclei. More and more precise measurements shed light on the interplay 
between reaction mechanisms and the internal structure of the two reaction partners.
This interplay is perfectly illustrated by low energy reactions like fusion. For instance, 
fusion cross sections are influenced by vibrational and rotational modes
of the nuclei \cite{das98}. New radioactive low energy beams facilities such as SPIRAL2, open new 
opportunities for these studies.
Theoretically, microscopic models which incorporate both dynamical effects and nuclear 
structure in a common formalism should be developed. Dynamical theories based on the mean-field 
concept are of the best candidates. 

The present lecture notes give a summary of actual progress in mean-field transport models
dedicated to Heavy-Ion reactions at low energy. The starting point of different approaches, {\it i.e.} 
Time-Dependent Hartree-Fock with effective interaction, is described extensively in section \ref{sec:dyn}, 
while examples of applications are given in section   
\ref{sec:barriere}. TDHF should be extended to describe the richness of phenomena 
occurring in nuclear systems.
In section \ref{sec:audela}, we introduce transport theories that incorporate effects 
such as pairing correlations and/or direct nucleon-nucleon collisions. 

This lecture can be read at different levels. The reader interested in applications (section \ref{sec:barriere}) 
can skip most of the formal aspects described in section \ref{sec:dyn}. Section \ref{sec:audela} is relatively technical and 
requires a good understanding of the formal aspects described in section  \ref{sec:dyn}. Finally, minimal notion of second 
quantization which are used in these notes are summarized in appendix \ref{annexe:rappelsMQ}. 

\subsection{Microscopic Non-relativistic approaches}

Typical examples of quantum microscopic theories used to describe static properties of atomic nuclei 
are: various Shell Models, mean-field and beyond theories (Hartree-Fock-Bogoliubov, Random-Phase-Approximation,
Generator-Coordinate-Method...) \cite{rin80} or algebraic approaches like the Interactive-Boson-Model \cite{iac91}.

Why should we also develop quantum microscopic approaches to describe nuclear reactions? 
\begin{itemize}
\item[$\bullet$] By itself, the N-body dynamical problem is a challenging subject.
\item[$\bullet$] Most of the information one could get from nuclei are deduced from nuclear reactions. Therefore 
a good understanding of the reaction is mandatory.
\item[$\bullet$] As already mentioned, it is necessary to develop a common formalism for both 
static and dynamical properties.  
\item[$\bullet$] Microscopic theories have only few adjustable parameters (essentially the effective
interaction), and we do expect that the predicting power is accordingly increased compared to more
phenomenological approaches.
\item[$\bullet$] On opposite to most of macroscopic approaches which are specifically dedicated to a given 
mechanism, all dynamical effects should be "a priori" included. For instance, TDHF can be used either to study 
reactions between two nuclei or collective motion in a single nucleus.   
\item[$\bullet$] Using microscopic theories, we also expect to be able to deduce dynamical information on the behaviour of nucleons
in the nuclear medium like for instance in-medium nucleon-nucleon cross sections 
\end{itemize}
Last, microscopic approaches that will be described here are non relativistic.
Relativistic effects are not expected to affect atomic nuclei because
the typical velocities of nucleons are much less than the light speed $(v/c)^2\sim 1/10$.

\subsection{What means independent particles?}

It is relatively common to use words like "mean-field theories"
or "independent particles systems" as well as their complement 
"beyond mean-field" or "correlated states". Let us start with a 
discussion on this terminology. Then, we will be able to define 
the starting point of most of microscopic approaches dedicated to 
the description of nuclei.    

\subsubsection{Independent and correlated particles}

We first assume two identical free fermions, one in a state $\kmu$ and the 
other one in a state $\knu$. Except for the Pauli principle which forbids $\kmu = \knu$, 
the fact that one fermion occupies $\kmu$ is independent from the fact that the other occupies 
$\knu$. We say that the two particles are {\it "independent"}. 
The state vector associated to the two particles system accounts 
for anti-symmetrization and reads
$|\mu \nu\> = \(|1:\mu, 2: \nu\> - |1:\nu, 2: \mu\> \)/\sqrt{2}$.
Such a state is indifferently called {\it independent particle state} or {\it Slater determinant} (see appendix \ref{annexe:rappelsMQ}).
The concept of independence could be easily generalized to $N$ particles.
A system will be called here "independent particle system" if it could be written in a specific basis as an anti-symmetric product
of single-particle states. In second quantization form (appendix \ref{annexe:rappelsMQ}), such a state 
writes
\oeq
 | \phi _{\nu_1 \cdots \nu_N} \>  = \left( \prod_{i=1}^N \sdf \oad_{\nu_i} \right) \sdf  \kvac.
\label{eq:un_slat}
\ceq

On the opposite, we call "correlated" a state that cannot be written in terms of a single Slater determinant, i.e. $
 | \psi \>  = \sum_{\al}  \sdf c_\al \sdf  | \phi _\al \> 
$ where each $c_\al$ gives the probability of each configuration.
Let us for instance consider a state describing two particles that decomposes onto two Slater determinants, 
{\it i.e.} $|\psi> = c_1 |\mu_1 \nu_1> + c_2 |\mu_2 \nu_2>$.  Then the concept
of correlation becomes obvious because the occupation of one specific state
affects the occupation probability of the other state. In this particular example:
\begin{itemize}
\item[$\bullet$] If one particle is in the state $|\nu_1\>$, then the other is in $|\mu_1\>$.
\item[$\bullet$] If one particle is in the state $|\nu_2\>$, then the other is in $|\mu_2\>$.
\end{itemize}

\subsubsection{Mean-field approximation} 

{The aim of nuclear mean-field theories is to describe self-bound nuclei in their {\it intrinsic} frame 
where wave-functions are localized (on opposite to the {\it laboratory} frame).} 
A possible description of a self-bound localized system 
in terms of Slater determinants could be constructed 
from single-particle wave-functions of an harmonic oscillator or a Woods-Saxon potential.   
These potentials are then interpreted as effective average mean-fields that simulate the 
interaction between particles.
In other words "Each nucleon freely evolves in a mean-field generated by the surrounding nucleons". 
It is worth mentioning the necessity to consider intrinsic frame 
is very specific  to self-bound systems like nuclei. For instance, for electronic systems, electrons are always 
considered in the laboratory frame since they are automatically localized due to the presence of atoms 
and/or external fields. 

In this lecture, we consider more elaborated mean-field, 
generally "called" self-consistent mean-field, 
like those found in  
Hartree-Fock (HF) and Hartree-Fock-Bogoliubov (HFB) theories. 
In the first case, the mean-field is calculated from occupied
single-particles while the second case is more elaborated and requires the notion of quasi-particles (see section \ref{subsec:TDHFB}).
In section \ref{sec:TDHF}, TDHF equations are obtained 
by neglecting correlations.
{ Last, although it is not the subject of the present lecture, it is worth mentioning that the 
introduction of effective interactions for nuclear mean-field theories leads to a 
discussion on correlation much more complex than presented here (see \cite{sto07}
for a recent review).}

\subsubsection{Mean-free path and justification of mean-field approaches in nuclear physics}

Can independent particles approximation give a good description of nuclear systems including 
reactions between two nuclei? { The justification of mean-field theories is largely based on the 
empirical observation that many properties vary smoothly for nuclei along the nuclear charts: 
single-particle densities, energies...} Based on this consideration, macroscopic and mean-field models 
have been introduced and turns out to be very successful to describe nuclei. Therefore, by 
itself, the predicting power of independent particle approximation justify {\it a posteriori} 
its introduction.     

The large mean-free path of a nucleon in nuclear matter (larger than the size of the nucleus itself) 
gives another justification of the independent particle hypothesis \cite{boh69}. This implies 
that a nucleon rarely encounter direct nucleon-nucleon collision and can be, in a good approximation,
considered as free. This might appear surprising in view of the strong interaction between nucleons 
but could be understood as a medium effect due to Pauli principle.
Indeed, the phase-space accessible to nucleon after a direct nucleon-nucleon collision inside the nucleus 
is considerably reduced due to the presence of other surrounding nucleons (essentially all states below the 
Fermi momentum are occupied). { However, if the relative kinetic energies of two nucleons increases, which happens
when the beam energy increases, the Pauli principle become less effective to block such a collision and 
the independent particle approximation breaks down.}        

\subsubsection{Symmetries and correlations}

Small remarks against the intuition:  independent particle states used for nuclear systems contain correlations. 
This could be traced back to the fact that some symmetries of the original Hamiltonian are generally broken. 
For instance, nucleons described within the nuclear mean-field approach are {\it spatially correlated}
because they are localized in space. This is possible because mean-field is introduced in the 
{\it intrinsic frame} and translational invariance is explicitely broken.  
Indeed, if we do not break this symmetry, then the only mean-field solution would be a constant potential
and associated wave-functions would be plane waves. 
We then come back to the free particle problem which 
are not self-bound anymore. 

We illustrate here an important technique which consists in breaking explicitly symmetries to incorporate 
correlations which could hardly be grasped in an independent particle picture (see for instance \cite{ben03}). 
Among the most standard symmetries explicitly broken generally to describe nuclear structure, we can quote 
breaking of rotational invariance which authorizes to have deformed nuclei and help to recover some long range correlations\footnote{Let us consider an elongated nucleus ("cigare" shape), the fact that one nucleon is at one side 
of the cigare implies necessarily that other nucleons should be on the other side. This should be seen as long range 
correlations that affects nucleons as a whole.}.
Gauge invariance (associated to particle number conservation) is also explicitly broken in HFB theories in order 
to include short range correlations like pairing (see section \ref{subsec:TDHFB}).
The latter approach will still be called "mean-field". It however goes beyond independent 
particle approximations by considering more general Many-body states formed of product of independent quasi-particles
(a summary of terminology and approximations considered here is given in table \ref{tab:approches}).

Last, it is worth mentioning that any broken symmetry should normally be restored. This is generally done by using projection techniques
\cite{rin80,bla86,ben03}.

\subsubsection{Theories beyond mean-field}

In nuclear physics, mean-field is often considered as the "zero" order microscopic approximation.
Many extensions are possible (several "beyond mean-field" approximations will be presented in section 
\ref{sec:audela}). These extensions are generally useful to include correlations 
that are neglected at the mean-field level. 
As we will see, "beyond mean-field" approximation are absolutely necessary 
to describe the richness of effects in nuclear structure as well as in nuclear dynamics.    
In particular, not all correlations could be incorporated by only breaking symmetries and 
often one has to consider the state of the system as a superposition of 
many independent (quasi)particles states.

The previous discussion clearly points out that mean-field approaches might include correlations. 
Nevertheless, the terminology that is generally used (and that we continue to use here) 
is that {\it non-correlated state} will be reserved to Slater determinant states. A {\it 
correlated state} then refers to a superposition of Slater determinants.  
Table \ref{tab:approches} summarizes different approaches that will be discussed 
in this lecture and associated type of correlations included in each approach.

\begin{table}[th]
    \begin{center}
{ 
\begin{tabular}{|l|l|l|l|} \hline
Name & Approximation & Variational space & Associated observables \\
\hline
&&&\\
TDHF & mean-field & indep. part & one-body \\ 
&&&\\
TDHF-Bogoliubov & m.-f. + pairing  & indep. quasipart.  & generalized one-body \\
&&&\\
Extended-TDHF & m.-f. + collision (dissipation) & correlated states & one-body  \\ 
&&&\\
Stochastic-TDHF & m.-f. + collision  & correlated states & one-body   \\ 
&(dissipation+fluctuations)&&\\
&&&\\
Time Dependent Density Matrix & c.m. + two-body correlations & correlated states & one- and two-body\\ 
&&&\\
Stochastic Mean Field & Exact (within statistical errors) & correlated states & all   \\ 
(Functional integrals) &&&\\
\hline
\end{tabular}
}
\end{center}
\caption{Summary of microscopic approaches presented in these notes}
\label{tab:approches}
\end{table}

\subsection{Effective interaction and Energy Density Functional (EDF)}

We will use in the following a rather standard approach to the nuclear many-body problem.
Starting from a microscopic two-body Hamiltonian and using second quantization, 
we introduce the Hartree-Fock theory. It is however worth mentioning that the use of most recent realistic
two-body (and normally three-body) interactions will not lead to reasonable results (if any) at the Hartree-Fock level. In view of this difficulty, the introduction of effective interactions adjusted to nuclear properties was a major break-down. 
These interactions are not directly connected to the original bare interaction and are expected 
to include effects (in particular in-medium effects) that are much more involved than in the standard
Hartree-Fock theory. In that sense, mean-field theories in nuclear physics have many common aspects with 
Density Functional Theories (DFT) in condensed matter. However, in nuclear physics we often keep the concept 
of effective interactions.  

The most widely used interactions are contact interactions (Skyrme type) and finite range interactions (Gogny type).
The second type of interactions is still too demanding numerically to perform time-dependent calculations and only 
Skyrme forces are nowadays used for TDHF. 
To relax some of the constraints due to the use of 
effective interactions, the more general concept of Energy Density Functional (EDF) is introduced. 
In that case, the static and dynamical properties of the system are directly obtained by minimizing 
a functional of the one-body density, denoted by $E[\ro]$ \cite{dob07,ben07}.

\subsection{The N-body problem: basic formalism}

The evolution of a Many-Body state $\kpsi$ is given by the time-dependent 
Schr\"odinger equation   
\oeq
i\hb \sdf \dt \kpsi = \oH \sdf \kpsi .
\label{eq:schroed}
\ceq
This equation can equivalently be formulated in term of a variational principle.
In that case, the quantum action is minimized between 
two times $t_0$ and $t_1$ with respect to the wave-function variations  
\oeq
\delta \sdf \[\int_{t_0}^{t_1} \stb \d t \stf \bpsi \sdf \oH - i\hb \dt \sdf \kpsi\] = 0.
\label{eq:principe_variationnel}
\ceq

The Many-body Hamiltonian decomposes into a kinetic term and a two-body 
interaction term (for the sake of simplicity, we will not consider higher order interactions) 
\oeq
\oH = \sum_{i=1}^N \sdf \frac{\op{(i)}^2}{2m} + \sum_{i>j=1}^N \sdf \ov{(i,j)}.
\label{eq:oH1}
\ceq
In second quantization (appendix \ref{annexe:rappelsMQ}), the Hamiltonian writes
\oeq
\oH = \sum_{ij} \sdf t_{ij}\sdf \oad_i \oa_j + \frac{1}{4} \sum_{ijkl} \sdf  \bar{v}_{ijkl} \sdf \oad_i \oad_j \oa_l \oa_k 
\label{eq:oH2}
\ceq
where matrix elements associated to the kinetic energy and anti-symmetric two-body interaction are given respectively by 
\oeqn
t_{ij} &=& \frac{1}{2m} \sdf \<i | \op^2 | j \>  \label{eq:tij}\\
\bar{v}_{ijkl} &=& v_{ijkl} - v_{ijlk} \label{eq:vbar} \\
 v_{ijkl} &=& \<1:i, 2: j | \sdf \ov{(1,2)} \sdf | 1:k , 2: l \> . \label{eq:v12}
\ceqn
Note that here, the two-body state $|1:i,2:j\>$ is not anti-symmetric. This notation 
means that the particle "1" is in the state $|i\>$, while the particle "2" is in the state $|j\>$ 
(for further details on notations see appendix \ref{annexe:rappelsMQ}).

Let us again summarize the overall goal: Even if we were able to solve the exact Many-Body problem 
(Eq.~(\ref{eq:schroed})), which is 
not the case anyway in nuclear physics, we do not need to have all the information contained in the exact 
wave-function to understand 
physical processes related mainly to one-body degrees of freedom, like collective motion or fusion reactions.   
Construction of microscopic models are therefore guided by two principles: 
\begin{itemize}
\item[$\bullet$] we focus exclusively on what we really need,
\item[$\bullet$] we do relevant approximations for the considered problem.
\end{itemize}
A strategy of approximation consists in focusing on specific degrees of freedom. This 
is generally equivalent to minimize the variational principle (\ref{eq:principe_variationnel})
on a restricted subspace of the total Hilbert space of Many-body wave functions \cite{bla86}.
The selection of this subspace is crucial and is driven by the physical process considered.

Among the microscopic transport theories, the TDHF method, 
proposed by Dirac in 1930~\cite{dir30} as an extension of the 
Hartree-Fock static mean-field theory~\cite{har28,foc30} 
is a tool of choice. 
It corresponds to a mean-field theory where the only input
is the effective two-body interaction. In practice, TDHF Equations can be derived 
by restricting the variational space to Slater determinants.     

In the following, formal and practical aspects of TDHF are presented. Then examples 
of applications to reactions close to the fusion barrier are given.   
Successes and limitations of TDHF will illustrate physical effects missing in this 
approximation. "Beyond mean-field" theories that incorporate these effects are finally 
discussed. 

\section{Dynamical mean-field theories}
\label{sec:dyn}

Let us recall the two basic questions: 
\begin{itemize}
\item[$\bullet$] What is the relevant information for the description of nuclear reactions ?
\item[$\bullet$] What approximations can (and cannot) be done to treat this information ?
\end{itemize}
The knowledge of nuclei trajectories during collisions, their shapes or particle numbers 
already gives a good understanding of the reaction mechanisms. All these quantities are related 
to one-body degrees of freedom (see appendix \ref{annexe:1corps}). Therefore, if we are able 
to give a realistic description of the one-body density matrix, we will also correctly reproduce
these observables.    
As we will see, the reduction of the information to one-body degrees of freedom is 
intimately connected to the independent particle approximation discussed in the introduction.
This will answer the second question.    
Starting from the Schr\"odinger equation, the equation of motion 
for the one-body degrees of freedom (contained in the evolution of the one-body density)
is obtained. Then connections with the independent particle approximation 
are discussed. Finally, we describe practical aspects related to the resolution of TDHF. 

Readers not interested in formal details can jump to Eq.~(\ref{eq:TDDM}) which gives the most general 
form of the one-body density evolution for any Many-Body system.

\subsection{Expectation values of one-body observables}

\subsubsection{General expression}

Let us consider a system of $N$ particles $\kpsi$ (eventually correlated) and a one-body operator 
$\oF=\sum_{i=1}^N \of(i)$ (see Eq.~(\ref{eq:un_corps})). Starting from 
general expression of a Many-Body wave-function (Eq.~(\ref{eq:def_fo})) and using the closure 
relation  
(Eq.~(\ref{eq:fermeture})), we obtain  
\oeqn
\<\oF \>_\psi &=&
\bpsi \frac{1}{N!}  \int \stb \d \xi_1 ... \d \xi_N | \xi_1... \xi_N \> \<\xi_1... \xi_N|
 \sdf  \sum_{i=1}^{N} \of(i) \sdf  \frac{1}{N!}  \int \stb \d \xi'_1 ... \d \xi'_N | \xi'_1... \xi'_N \> \<\xi'_1... \xi'_N \kpsi
\nonumber \\
&=& \frac{1}{N!} \int \stb \d \xi_1 ... \d \xi_N 
 \d \xi'_1 ... \d \xi'_N \stf
\psi^*(\xi_1 ... \xi_N) \sdf  \psi(\xi'_1 ... \xi'_N)  \sdf  \sum_{i=1}^{N} \<\xi_1... \xi_N| \sdf \of(i)\sdf
| \xi'_1... \xi'_N \>
\label{eq:F}
\ceqn
where spin and isospin quantum numbers are included in the notation  
$\xi\equiv (\vr s \tau)$. Using the Wick's theorem (see appendix \ref{subannexe:wick}), leads to
\oeq
\<\xi_1... \xi_N| \sdf \of(i)\sdf| \xi'_1... \xi'_N \> = 
\left|
 \begin{array}{ccccccc}
\<\xi_1|\xi'_1\> & \cdots & \<\xi_1|\xi'_{i-1}\> & \<\xi_1|\of|\xi'_i\> & \<\xi_1|\xi'_{i+1}\> & \cdots &\<\xi_1|\xi'_N\> \\
\vdots & & \vdots & \vdots& \vdots& &\vdots \\
\<\xi_N|\xi'_1\> & \cdots & \<\xi_N|\xi'_{i-1}\> & \<\xi_N|\of|\xi'_i\> & \<\xi_N|\xi'_{i+1}\> & \cdots &\<\xi_N|\xi'_N\> \\
\end{array}
\right|.
\ceq
Using the anti-symmetry of each state of the basis, it is possible to reduce, 
in integral (\ref{eq:F}), 
the action of $\of$ to the first label "1" only.
Noting $f(\xi,\xi')=\<\xi| \of | \xi'\>$ the matrix elements of $\of$, we obtain
\oeq
\<\oF\>_\psi =  N \int \stb \d \xi  \d \xi'  \d \xi_2 ... \d \xi_N \stf
\psi^*(\xi\xi_2 ... \xi_N) \sdf  \psi(\xi' \xi_2 ... \xi_N)   \sdf f(\xi,\xi').
\label{eq:Ffifi}
\ceq

\subsubsection{One-body density matrix}
\label{subsubsec:mat_dens_1corps}

We introduce the one-body density matrix $\ro^{(1)}$ associated to the state $\kpsi$.
Its matrix elements in the basis $\{\xi\}$ are given by
\oeq
\ro^{(1)}(\xi,\xi') = N \sdf  \int \stb \d \xi_2 ... \d \xi_N \stf 
\psi^*(\xi'\xi_2 ... \xi_N) \sdf  \psi(\xi \xi_2 ... \xi_N) .
\label{eq:def_ro1}
\ceq
Accordingly, equation (\ref{eq:Ffifi}) can be simply expressed as 
\oeq
\<\oF\>_\psi =    \int \stb \d \xi  \d \xi'  \stf
 \ro^{(1)}(\xi',\xi)  \sdf f(\xi,\xi') 
= \Tr [\ro^{(1)} f].
\label{eq:Trrof}
\ceq
Therefore, all the information required to estimate a one-body observable is contained in 
the one-body density matrix. Such a matrix can always be associated to any system of $N$ particles\footnote{
It is worth mentioning that the discussion presented here can be generalized to $M$-body observables.
In that case, we can define the $M$-body density that contains all the information on $M$-body (and below) 
quantities (see appendix \ref{annexe:densite}).}.
In the following, we will essentially consider properties of the one-body density 
and omit exponent~$^{(1)}$.

Using second quantization, matrix elements of $\ro$ are defined as (see Eq.~(\ref{eq:defroM}))
\oeq
{\ro}_{ij}  = \bpsi \sdf \oad_{j} \oa_{i} \sdf \kpsi = \< \oad_j \oa_i  \>_\psi .
\label{eq:elt_dens}
\ceq
Eq.~(\ref{eq:def_ro1}) can indeed be recovered using  (\ref{eq:recouv}),
(\ref{eq:def_fo}) and (\ref{eq:fermeture}) 
\oeqn
\ro(\xi,\xi') &=& \< \oad(\xi')\oa(\xi)\>_\psi \nonumber \\
&=&  \frac{1}{N!} \int \stb \d \xi_1 ... \d \xi_N 
 \d \xi'_1 ... \d \xi'_N \stf
\psi^*(\xi_1 ... \xi_N) \sdf  \psi(\xi'_1 ... \xi'_N)  \sdf  \<\xi_1... \xi_N| \sdf  \oad(\xi')\oa(\xi) \sdf
| \xi'_1... \xi'_N \> \nonumber \\
&=& \frac{N^2}{N!}  \int \stb \d \xi_2 ... \d \xi_N 
 \d \xi'_2 ... \d \xi'_N \stf
\psi^*(\xi' \xi_2 ... \xi_N) \sdf  \psi(\xi \xi'_2 ... \xi'_N)  \sdf  \<\xi_2... \xi_N| \xi'_2... \xi'_N \> \nonumber \\
&=& N \sdf  \int \stb \d \xi_2 ... \d \xi_N \stf 
\psi^*(\xi'\xi_2 ... \xi_N) \sdf  \psi(\xi \xi_2 ... \xi_N) .
\ceqn

In the following, the one-body density components are introduced in a specific single-particle basis. 
For instance (see section \ref{subsec:aspects_pratiques}), TDHF calculations are often performed in coordinate 
representation. In that case, using equations (\ref{eq:ar}) and (\ref{eq:adr}), we have 
\oeqn
{\ro}(\vr s \tau , \vr' s' \tau') &=& \bpsi \sdf \oad(\vr' s' \tau') \sdf \oa(\vr s \tau) \sdf \kpsi 
 = \sum_{ij} \< \oad_i \oa_j \>_{\psi} 
\stf {\az_i^{s'\tau'}}^*\!(\vr' ) \sdf \az_j^{s\tau}(\vr)
\label{eq:rorst}
\ceqn
where single-particle wave-functions $\az^{s\tau}_i$ are defined in Eq.~(\ref{eq:fosp}).

Note that, we can always write the one-body density matrix as an operator acting 
in the Hilbert space of the single-particle 
wave-functions\footnote{This definition of operators $\oro$ should not be confused 
with the definition of one-body 
operators given in appendix~\ref{annexe:1corps} since it is not defined in the space of many-body 
wave function.}
\cite{rin80}
\oeq
\oro = \sum_{ij} {\ro}_{ij} \sdf |i\>\<j|.
\label{eq:oro}
\ceq
In particular, this operator depends on the system wave-function $\kpsi$ 
and then it is a funtion of the time during a collision.

Let us come back to  one-body observables.
The second quantization simplifies the calculation of their expectation values. 
Using Eq.~(\ref{eq:un_corps_2ndQ}), we  have directly
\oeq
\< \oF \>_\psi =  \bpsi\sdf\sum_{ij} \sdf f_{ij} \sdf  \oad_i\sdf  \oa_j \sdf \kpsi =  \sum_{ij} \sdf f_{ij} \sdf {\ro}_{ji} = \Tr\( \ro f\)
\label{eq:Fmoy}
\ceq
which is nothing but Eq. (\ref{eq:Trrof}). Therefore we see that the evolution of any one-body observable 
can be obtained from the evolution of the one-body density matrix.

\subsubsection{The independent particle case}
\label{sec:slater}

Let us consider the case of a system described by a Slater determinant
$|\phi\> \equiv | \phi _{\nu_1\cdots\nu_N} \>$. The associated one-body density is denoted by $\ro$.
In appendix \ref{annexe:videHF}, we show that any Slater can be considered as a vacuum (called HF vacuum) for specific operators
written as linear combination of the   
$\oad$ and $\oa$. We can thus apply the Wick's theorem. Methodology associated to the Wick's theorem as well as contractions 
are given in appendix \ref{annexe:correl_part_indep}. This technique (Eq.~(\ref{eq:etats_occ})) implies that only occupied states contribute 
to the summation in Eq.~(\ref{eq:rorst}). We finally end with 
\oeq
\ro(\vr s \tau, \vr' s' \tau') = \sum_{n=1}^N  \sdf {\az_{\nu_n}^{s'\tau'}}^*\!(\vr' ) \sdf \az_{\nu_n}^{s\tau}(\vr) .
\ceq
Therefore, for Slater determinants, the knowledge of occupied states is equivalent 
to the knowledge of one-body density matrix. The specificity  of independent particles 
systems is that {\it all} the information is contained in the one-body density. 
This is nicely illustrated by the fact that any $M$-body density matrix could be
expressed as an anti-symmetric product of one-body densities leading to vanishing 
$M$-body correlation matrices at all orders (see appendix~\ref{annexe:densite}). 

For dynamical problems where the Many-Body state is assumed to stay in an independent 
particle state at all times 
(this is the case of TDHF presented in section \ref{sec:TDHF}), it is sufficient 
to only  follow the one-body density in time. This is strictly equivalent to 
follow occupied states.

Note finally the following useful property: for an independent particle state, $\ro^2=\ro$. 
Indeed, the operator given by  Eq.~(\ref{eq:oro}) then writes 
\oeq
\oro = \sum_{n=1}^N \sdf |\nu_n\> \<\nu_n|.
\label{eq:oro_Slater}
\ceq
Therefore, $\oro$ is nothing but the projector on the subspace of occupied single-particle states of $|\phi_{\nu_1\cdots\nu_N}\>$.
Considering a set of orthonormal single-particle states, we have  
\oeq
 \oro^2= \sum_{m,n=1}^N |\nu_m\> \sdf \<\nu_m|\nu_n\> \sdf \<\nu_n| = \oro.
\label{eq:proj}
 \ceq
Assuming the above relation for the one-body density is equivalent to assume independent particle states.  

\subsubsection{Dynamical Evolution from the Erhenfest theorem}
\label{subsubsec:Ehrenfest}

As discussed above, the basic approximation is to focus on a specific class of observables, which are the one-body 
observables in our case. 
Considering any observable ~$\oF$, using (Eq.~(\ref{eq:schroed})) and its Hermitian conjugate
for any N-body state $\kpsi$, we have 
\oeq
\dt  \<\oF\>_\psi   = \(\frac{i}{\hb} \sdf\bpsi \sdf \oH\) \sdf \oF \sdf \kpsi + \bpsi \sdf \oF \sdf \( \frac{-i}{\hb} \sdf\oH \sdf \kpsi \) 
= \frac{i}{\hb} \< \, [\oH,\oF ] \, \>_\psi
\ceq
which is nothing but the standard Ehrenfest theorem for the operator $\oF$. 
The above equation is exact and valid for any (correlated or not) state and observable. 
Here, $\oH$ is the full microscopic Hamiltonian (see Eqs. (\ref{eq:oH1}) and (\ref{eq:oH2})). 
For a one-body operator, $\oF$ can be written as a linear combination of the $\oad_i\oa_j$. Therefore, it is sufficient to 
follow the expectation values of the $\< \oad_i\oa_j \>$ , which are nothing but the matrix elements of $\ro$ and are given 
by 
\oeq
\dt \sdf \<\oad_i\oa_j\>_\psi = \dt \sdf \ro_{ji} 
= \frac{i}{\hb} \< \, [\oH,\oad_i\, \oa_j ] \,\>_\psi.
\label{eq:erhenfest}
\ceq 

\subsection{Time-Dependent Hartree-Fock (TDHF)}
\label{sec:TDHF}

\subsubsection{Exact evolution of $\ro^{(1)}$}

Our goal is to provide the best description as possible of the one-body density matrix evolution.
The only approximation that will be made 
is that the system remains in an independent particle state at all time. Several derivations 
of TDHF exist so far. Here, we use the Ehrenfest theorem as a starting point.

Reporting the Hamiltonian expression  
(Eq.~(\ref{eq:oH2})) in the evolution of the $\ro_{ji}$  (Eq.~(\ref{eq:erhenfest})), we get
\oeq
i\, \hb \sdf \dt \sdf \< \oad_i \oa_j \>_\psi = \sum_{kl} \sdf t_{kl} \< [\oad_i\, \oa_j\sdf ,\sdf \oad_k \,\oa_l ]\,\>_\psi
+ \frac{1}{4} \sdf \sum_{klmn} \sdf \bar{v}_{klmn} \sdf \< \, [  \oad_i\, \oa_j\sdf ,\sdf   \oad_k \,\oad_l \,\oa_n \,\oa_m ]\,\>_\psi.
\label{eq:erhenfest_detail}
\ceq
Let us start with the first term associated to kinetic energy. Using 
Eqs. (\ref{eq:anticom1}) and (\ref{eq:anticom2}), we obtain
\oeqn
\< \, [ \, \oad_i \, \oa_j \, , \, \oad_k \, \oa_l \, ] \, \>_\psi
&=& \<  \oad_i \, \oa_j  \, \oad_k \, \oa_l  \>_\psi -  \<  \oad_k \, \oa_l  \, \oad_i \, \oa_j \>_\psi \nonumber \\
&=& \del_{jk} \,  \<  \oad_i \, \oa_l \>_\psi -  \<  \oad_i \, \oad_k \, \oa_j  \,  \oa_l \>_\psi  - \del_{il} \,  \<  \oad_k \, \oa_j \>_\psi +   \<  \oad_k \, \oad_i \, \oa_l  \,  \oa_j \>_\psi  \nonumber \\
&=& \del_{jk} \,  {\ro}_{li} - \del_{il} \,  {\ro}_{jk} .
\ceqn
The kinetic energy term reduces to 
\oeq
\sum_{kl} \sdf t_{kl} \< [\oad_i\, \oa_j\sdf ,\sdf \oad_k \,\oa_l ]\,\>_\psi
 = \sum_k \sdf \( t_{jk} \sdf {\ro}_{ki} - t_{ki} \sdf {\ro}_{jk} \).
\label{eq:terme_cinetique}
\ceq
For the two-body interaction, we should first express the expectation value of the commutator  
\oeqn
\< \, [ \, \oad_i \, \oa_j \, , \, \oad_k \, \oad_l  \, \oa_n \, \oa_m \, ] \, \>_\psi
&=& \<  \oad_i \, \oa_j  \, \oad_k \, \oad_l  \, \oa_n \, \oa_m  \>_\psi -  \<  \oad_k \, \oad_l  \, \oa_n \, \oa_m \, \oad_i \, \oa_j  \>_\psi \nonumber \\
&=& \<  \oad_i \, \oad_l  \, \oa_n \, \oa_m  \>_\psi \sdf \del_{jk} -  \<  \oad_i \, \oad_k  \, \oa_n \, \oa_m  \>_\psi \sdf\del_{jl} 
+ \<  \oad_i \, \oad_k \, \oad_l  \, \oa_j \, \oa_n \, \oa_m  \>_\psi  \nonumber\\
&& - \<  \oad_k \, \oad_l  \, \oa_n \, \oa_j  \>_\psi \sdf \del_{mi} +  \<  \oad_k \, \oad_l  \, \oa_m \, \oa_j  \>_\psi \sdf \del_{ni} 
- \<  \oad_k \, \oad_l \, \oad_i  \, \oa_n \, \oa_m \, \oa_j  \>_\psi. \nonumber \\
\label{eq:groscom}
\ceqn
The two terms with 6 annihilation/creation operators cancel out. Other terms are 
nothing but components of the two-body density matrix (defined in appendix \ref{annexe:densite}).
The two-body density can be decomposed into a sum of an anti-symmetric product of two one-body density (the 
uncorrelated part) plus the so-called {\it two-body correlation} matrix, denoted by $C^{(2)}$ (see appendix \ref{annexe:correl_2corps}).
Using this decomposition (Eq.~(\ref{eq:ropsi})), Eq.~(\ref{eq:groscom}) writes
\oeqn
\left< \, \[ \, \oad_i \, \oa_j \, , \, \oad_k \, \oad_l  \, \oa_n \, \oa_m \, \] \, \right>_{\psi}
&=& \( \ro_{{mi}}  \ro_{{nl}} -  \ro_{{ml}}  \ro_{{ni}} + C_{mnil}\) \del_{jk} + \( \ro_{{mk}}  \ro_{{ni}} -  \ro_{{mi}}  \ro_{{nk}} + C_{nmik} \) \del_{jl} \nonumber \\
&&+ \( \ro_{{jl}}  \ro_{{nk}} -  \ro_{{jk}}  \ro_{{nl}} + C_{njkl}\) \del_{mi} + \( \ro_{{jk}}  \ro_{{ml}} -  \ro_{{jl}}  \ro_{{mk}} + C_{mjlk} \) \del_{ni} \nonumber \\
\ceqn
where exponents $^{(1)}$ and $^{(2)}$ in equation (\ref{eq:ropsi}) have been omitted for simplicity. 
Altogether, the two-body interaction contribution to the one-body evolution reduces to 
\oeqn
 \frac{1}{4} \sdf \sum_{klmn} \sdf \bar{v}_{klmn} \sdf \left< \, \[  \oad_i\, \oa_j\sdf ,\sdf   \oad_k \,\oad_l \,\oa_n \,\oa_m \]\,\right>_\psi &=& 
\frac{1}{2} \sdf \sum_{klm} \sdf \[ \bar{v}_{jklm} \sdf\( \ro_{{li}} \ro_{{mk}} - \ro_{{lk}}  \ro_{{mi}} +C_{lmik} \) \right. \nonumber \\
&&\left.+    \bar{v}_{klim} \sdf\( \ro_{{jl}} \ro_{{mk}} - \ro_{{jk}}  \ro_{{ml}} + C_{mjkl} \) \]\nonumber \\
&=& \sum_{klm} \[ \bar{v}_{jklm} \sdf \(\ro_{{li}} \ro_{{mk}} +\frac{1}{2} C_{lmik}\)\right.\nonumber \\
&&\stf \stf \stf\left.- \bar{v}_{klim} \sdf\( \ro_{{jk}} \ro_{{ml}}+ \frac{1}{2} C_{jmkl} \) \]
\label{eq:inter_corel}
\ceqn
where we have used $\bar{v}_{klmn} = - \bar{v}_{klnm} = - \bar{v}_{lkmn}$ and the anti-symmetry of $C^{(2)}$ \footnote{The fact that 
$C^{(2)}$
is anti-symmetric is a consequence of the anti-symmetry of $\ro^{(2)}$ and can be deduced from anti-commutation rules for fermions (\ref{eq:anticom1})
entering in the definition of $\ro^{(2)}$ (Eq.~(\ref{eq:defroM})).}. The two-body contribution can finally be written as
\oeq
\sum_{k} \sdf \( U[\ro]_{jk} \sdf \ro_{{ki}} - U[\ro]_{ki} \sdf \ro_{{jk}} \)
+ \frac{1}{2} \sum_{klm} \(  \bar{v}_{jlkm} \sdf C_{kmil} -   \bar{v}_{klim} \sdf C_{jmkl} \) 
\label{eq:potentiel}
\ceq
where $U[\ro]$ is the Hartree-Fock self-consistent mean-field. The latter can be written with the use of partial 
traces as  
\oeq
U[\ro]_{ij} = \sum_{kl} \sdf\bar{v}_{ikjl} \sdf \ro_{{lk}}
=\<i|\, \Tr_2\{\bar{v}(1,2) \sdf \ro(2)\}\, |j\>
=\Tr_2\{\bar{v}(1,2) \sdf \ro(2)\}_{ij}.
\label{eq:Uro}
\ceq
The trace is only made on the second particle "2" while the labels $i$ and $j$ correspond to the particle labeled by "1". 
The matrix $U[\ro]$ therefore corresponds to a one-body mean-field operator. 
In the following, we will often use the notation 
$\bar{v}_{12}\equiv \bar{v}(1,2)$ and $\ro_2\equiv \ro(2)$. 
Then, Eq. (\ref{eq:Uro}) simply reads $U[\ro]_1= \Tr_2\{\bar{v}_{12} \sdf \ro_2\}$.

The second term in equation (\ref{eq:potentiel}) reflects the effect of correlations on the evolution of one-body degrees of freedom. 
It can also be written with a partial trace on the particle~"2"
\oeq
\sum_{klm}  \bar{v}_{ilkm} \sdf C_{kmjl}  
= \Tr_2\{ \bar{v}_{12} \sdf C_{12} \} _{ij}.
\label{eq:vc}
\ceq
Then, using Eqs. (\ref{eq:elt_dens}), (\ref{eq:erhenfest_detail}), (\ref{eq:terme_cinetique}), (\ref{eq:potentiel}) and  (\ref{eq:vc}) 
we finally deduce that the most general expression of the one-body density for any correlated system with a two-body interaction 
can be written as 
\oeqn
i\, \hb \sdf \dt \sdf {\ro}_{ji} &=& \sum_{k} \sdf \[ \( t_{jk} + U_{jk}[\ro] \) \sdf \ro_{{ki}} - \( t_{ki} + U_{ki}[\ro] \) \sdf \ro_{{jk}}\] \nonumber \\
&&+ \frac{1}{2} \[ \sdf \Tr_2 \{ \bar{v}_{12} \sdf C_{12} \}_{ji} - \Tr_2 \{ C_{12} \sdf \bar{v}_{12} \}_{ji} \].
\ceqn
The anti-symmetry of $\bar{v}_{12}$ and $C_{12}$ implies that 
 $ \Tr_2 \{ \bar{v}_{12} \sdf C_{12} \} =2 \sdf  \Tr_2 \{{v}_{12} \sdf C_{12} \}$
and finally leads to the more compact form
 \oeq
i\, \hb \sdf \dt \sdf {\ro}_{ji} =
\[\sdf h[\ro]\, ,\, \ro\sdf \]_{ji} + 
 \stf \Tr_2 \[ \sdf {v}_{12} , \sdf C_{12}\sdf  \]_{ji},
 \label{eq:TDDM}
\ceq
where 
\oeq
h[\ro]=t+U[\ro]
\label{eq:h}
\ceq
is the matrix associated to the one-body Hartree-Fock Hamiltonian. 
Note that, up to here, since no approximation has been made, the dynamical evolution of the one-body density is exact.
However, the equation above requires {\it a priori} to also follow the two-body correlations in time which may be to 
complicated. 

\subsubsection{Time-dependent mean-field approximation}

In  TDHF, two-body 
correlations are neglected at all time, {\it i.e.} 
$C^{(2)}=0$.
This is equivalent to assume that the system remains in an independent 
particle state at all time (see appendix \ref{annexe:correl_part_indep}).
Starting from Eq. (\ref{eq:TDDM}) and neglecting the correlation term, 
we finally get the TDHF equation for the one-body density matrix $\ro$ 
\oeq
i\,\hb \sdf \dt \ro = \[\sdf h[\ro]\, ,\, \ro \] .
\label{eq:TDHF}
\ceq
In this section, we will now concentrate on the latter equation which could also be written 
in terms of operators as 
\oeq
i\,\hb \sdf \dt \oro = \[\sdf \oh[\ro]\, ,\, \oro \] ,
\label{eq:opTDHF}
\ceq
where  $\oh$ and $\oro$ act both on the Hilbert space of single-particle states.
In a complete basis of this space $\{|i\>\}$ with the closure relation $\sum_i |i\>\<i| = \hat{1}$, we have 
\oeqn
\oro |i\> &=& \sum_{j} \sdf \ro_{ji} \sdf|j\>,~~~~ 
\oh |i\> = \sum_{j} \sdf h[\ro]_{ji} \sdf|j\>.
\ceqn
Let  us now give some properties of 
equation (\ref{eq:opTDHF}). First $\oro^2=\oro$ is preserved.
Therefore at all time, the density could be decomposed on a set of single-particle states $\oro = \sum_{n=1}^N |\nu_n\>\<\nu_n|$ 
(Eq.~(\ref{eq:oro_Slater})). The TDHF equation can equivalently be written in terms of $N$ coupled 
self-consistent equations on the single-particle states
\oeq
i\, \hb \sdf \dt |\nu_n(t)\> = \oh[\ro(t)] \sdf | \nu_n(t) \>, \stf \stf 1\le n \le N.
\label{eq:TDHFfo}
\ceq
Indeed, starting from the expression of $\oro$ and using the above Schr\"odinger-like equation for single-particle states, we recover the TDHF equation (Eq. (\ref{eq:opTDHF}))
that
$$
i\hb \frac{\partial}{\partial t} \oro = i\hb \sum_{n=1}^N \[\(\dt |\nu_n\>\) \<\nu_n|+|\nu_n\> \( \dt \<\nu_n| \) \] 
= [\oh,\oro] .
$$
This shows the equivalence of the single-particle representation and density formulation.

Though Eqs (\ref{eq:TDHFfo}) take the form of Schr\"odinger equations, they are
non llinear because 
the Hamiltonian depends on the one-body density. As a consequence it depends explicitly 
on time.
We clearly see here some difficulties of mean-field theories. Indeed, we want to describe 
the system evolution between the initial and final time with a Hamiltonian which is itself depending on the evolution. 

\subsection{The Skyrme effective interaction}

Mean-field equations have been derived for any general two-body Hamiltonian.
However, in practice, the interaction is chosen to simplify numerical aspects. 
Most (if not all) applications of TDHF in the nuclear context have been performed 
using the Skyrme like \cite{sky56} interaction. 
The most widely used Skyrme force writes
\oeqn
\ov (1,2) &=& t_0 \sdf \( 1+x_0\, \oP_\si \) \sdf \odel \nonumber \\
&+& \frac{1}{2} \sdf t_1 \sdf \( 1+x_1\, \oP_\si \) 
\sdf \(\ovk^2 \sdf  \odel - \odel \sdf \ovk^2 \) \nonumber \\
&+& t_2 \sdf \( 1+x_2\, \oP_\si \) 
\sdf \(\ovk \cdot  \odel \sdf  \ovk \) \nonumber \\
 &+& \frac{1}{6} \sdf t_3 \sdf \( 1+x_3\, \oP_\si \) 
 \sdf \ro^\al\! (\ovR) \sdf \odel \nonumber \\
 &+& i\, W_0 \sdf \ovsi \cdot \(\ovk \times \odel \, \ovk \) 
\label{eq:skyrme}
\ceqn
where $\odel = \del\(\ovr(1)-\ovr(2)\)$, $\ovk = \(\ovp(1)-\ovp(2)\)/\hb$ 
(relative impulsion), $\ovsi = \ovsi(1)+ \ovsi(2)$, $\ovR = \(\ovr(1)+\ovr(2)\)/2$,
$\ovsi(i) = \osi_x\!(i) \, \ve_x+ \osi_y\!(i) \, \ve_y + \osi_z\!(i) \, \ve_z$, $\osi_{x/y/z}(i)$ 
are operators acting on the spin of particle $i$ and are given in terms of  
Pauli matrices acting on the spin space.
 $\oP_\si = \(1+  \ovsi(1) \cdot \ovsi(2) \)/2$ corresponds to the exchange 
of the spin. $\ro(\vr) \equiv \sum_{s\tau} \ro^{(1)}(\vr s \tau,\vr s \tau )$ 
is the particle density at $\vr$.
"$t_1$" and "$t_2$" terms are non-local in space and simulate the short range part of the interaction.
Finally "$W_0$" is the spin-orbit term.

The very interesting aspect of this interaction is its zero range nature, which greatly 
simplifies the mean-field expression in coordinate space.
Parameters ($t_{0-3}$, $x_{0-3}$, $W_0$ and $\al$) are generally adjusted to 
reproduce nuclear properties like saturation, incompressibility ... of nuclear matter 
and selected properties of finite nuclei 
(see for instance \cite{cha97,cha98,mey00}).

An important aspect of the fitting procedure which has direct 
implication on nuclear reactions calculations is the following: for nuclear structure 
calculation, center of mass contribution are removed 
(see section 3.2. of ref. \cite{cha98}) to obtain a better description 
of nuclei in their intrinsic frame. For collisions, only the intrinsic frame 
of the total system is considered and the same correction used for a single nucleus
could not be used anymore. 
This is the reason why specific forces which explicitly 
do not account for center of mass correction have been developed like SLy4$d$ \cite{kim97} 
where parameters of the force have been adjusted to reproduce the same properties as
Sly4 \cite{cha98} except that the center of mass correction is neglected.

Note finally that the "$t_3$" term in Eq. (\ref{eq:skyrme}) explicitly  depends 
on the system density.  For specific integer values of $\alpha$, the density dependent 
two-body interaction could be interpreted as a higher order interaction. In practice, non-integer 
values of $\alpha$ turns out to be more effective in reproducing nuclear properties. This however 
has important consequences. In particular, since the interaction depends on the system 
on which it is applied, strictly speaking, we cannot really use the terminology "interaction". 
One often use the very notion of Energy Density Functional to avoid confusion.     
In addition, translational invariance (but not Galilean invariance  \cite{tho62})
is explicitly broken. 

\subsection{Numerical implementation of TDHF: practical aspects}
\label{subsec:aspects_pratiques}

Several applications of TDHF have been performed in the last decade for 
nuclear collective motion studies~\cite{sim03,nak05,alm05,uma05,rei06,rei07} 
and nuclear reactions~\cite{kim97,sim01,sim04,uma06a,mar06,uma06b,uma06c,uma06d,guo07,uma07,
sim07a,sim07b,guo08,uma08} with various numerical methods to solve the TDHF equation.
We present here a method used  to implement TDHF for nuclear collisions.
Different steps of a calculation are presented to better illustrate numerical constraints.
To apply TDHF, we should 
\begin{itemize}
\item[$\bullet$] Construct the HF ground states of each of the collision partners. 
This requires to first solve their HF equations. 
\item[$\bullet$] Starting from two Slaters at an initial distance $D_0$, one should construct a single Slater associated to the composite system. 
\item[$\bullet$] The nuclei should be initially positioned and boosted to properly account for the reaction properties (Beam Energy, impact parameter...)
\item[$\bullet$] The dynamical evolution should be performed iteratively  to solve the self-consistent TDHF equations. This is 
generally done by solving the Time Dependent Schr\"odinger equations on occupied states.
\item[$\bullet$] Finally, we should compute a set of observables to get informations on the reaction itself.  
\end{itemize}
Since all the information is contained in the one-body density, only one-body wave-function need to be considered.

Sections \ref{subsubsec:etats_HF} to \ref{subsubsec:boost_gallileen} describe respectively the 
construction of initial Slater determinants and how initial conditions for reactions according to Rutherford trajectories are 
imposed.
Section \ref{subsubsec:evolution_dynamique} presents the numerical implementation of mean-field transport equations.

\subsubsection{Hartree-Fock initial state}
\label{subsubsec:etats_HF}

We assume that the collision partners are initially in their ground states. 
Consistently with the TDHF approach, we should consider that the ground state 
of both nuclei are solution of the self-consistent Hartree-Fock (HF) states. 
The one-body density associated to each state is solution to the stationary version 
of the TDHF equation (Eq.~(\ref{eq:opTDHF})): 
\oeq
 \[\sdf \oh[\ro]\, ,\, \oro \] =0.
\label{eq:HF}
\ceq
This equation is valid in any basis. Then, a specific basis should be chosen 
to explicitly solve the equation. Since $\oh[\ro]$ and $\oro$ do commute, we can 
choose common eigenstates. We denote this basis  by $\{\kal\}$ with 
$\oh \sdf \kal = e_\al \sdf \kal$ and $\oro \sdf \kal = n_\al \sdf \kal $.
Here occupation numbers verifies $n_\al=0$ ou 1. 
The  $N$-body Slater determinant state is constructed from the occupied states (with $n_\al=1$).

Eigenvalues $e_\alpha$ of $\oh$ can eventually be interpreted as single-particle energies \cite{vau72}.
In the HF approximation, the ground state is obtained by filling the $N$ lowest energy
single-particle states.  Therefore, we only need to find the $N$ lowest eigenstates of $\oh$.

\subsubsection{Imaginary-time method}

The imaginary-time method \cite{dav80} is a widely used method to find 
the lowest eigenvalues of an operator (whose eigenstates is bound from below).
Let us illustrate this method for a particle in an external field.
The method consists in starting from an initial wave-function $|\nu_e\>$ which is not 
{\it a priori} an eigenstate of the one-body Hamiltonian, denoted by $\oH$. This state can be decomposed onto 
the true eigenstates ($\oH|\mu_n\> = E_n |\mu_n\>$)
\oeq
|\nu_e\> = \sum_n \sdf c_n |\mu_n\>.
\ceq
We apply the operator $e^{-\be \oH}$ on the initial state 
\oeq
e^{-\be \oH} |\nu_e\> = \sum_n c_n e^{-\be E_n} |\mu_n\> 
= e^{-\be E_0} \sum_n c_n e^{-\be ( E_n-E_0)} |\mu_n\>.
\ceq
The lowest energy  eigenstate $|\mu_0\>$ associated to $E_0$ can then be obtained from
\oeq
|\mu_0 \> = \lim_{\be \rightarrow \infty} \frac{e^{-\be \oH} \sdf|\nu_e\>}
{\<\nu_e|\sdf e^{-2\be \oH}\sdf | \nu_e\>^{1/2}}.
\label{eq:tps_imag}
\ceq
Indeed, we have $E_n\ge E_0$ and then $e^{-\be (E_{n}-E_0)} \rightarrow 0 $ for $\be\rightarrow \infty$ except for
$n=0$, $i.e$  only the ground state component does not vanish. 
The denominator 
is required because $e^{-\be \oH}$ is not unitary ($\be \in \mathbb{R}$).
The terminology "Imaginary-time" comes obvisouly from 
the fact that $e^{-\be \oH}$ looks like the propagator in time $e^{-i \oH t/ \hb}$
where the time is a complex quantity $t=-i\hb\be$. It is finally worth to mention 
that the initial state should be guessed in order to contain at least a small fraction of 
the lowest eigenstate. 

The imaginary-time method should be further improved to obtained eigenstates 
of the single-particle HF energies. Indeed, two additional  difficulties exist: 
\begin{itemize}
\item[$\bullet$] Since the system is composed of $N$ particles, 
we need the $N$ lowest eigenstates of $\oh$ and therefore we need to guess $N$ initial starting points 
for the single-particle wave-functions.  
\item[$\bullet$] The Hamiltonian $\oh$ is  non-linear and 
depends explicitly on the system density $\ro$.
\end{itemize}
To solve the first difficulty, we apply the imaginary time method 
imposing 
 that the states remain orthonormal (through a Graham-Schmidt orthogonalization for instance).  
We then expect to converge towards the $N$ lowest eigenstates if the states are not initially orthogonal to the "true" 
eigenstates. A common choice for the initial states are those of Harmonic  or Nilsson potentials.

To solve the second difficulty, we procede 
iteratively with small imaginary-time steps $\Delta \be$. At each step, the 
density is calculated from the states and the mean-field Hamiltonian 
is modified accordingly.

A schematic representation of the different steps of the imaginary-time procedure 
is 
\oeq
 \begin{array}{ccc}
\{ |\nu_1^{(n)}\> \cdots |\nu_N^{(n)} \>\} \stf\stf \stf \Rightarrow  \stf\stf \stf\ro^{(n)} 
& \Rightarrow &\oh^{(n+1)} = \la \oh[\ro^{(n)}]  + (1-\la) \oh^{(n)} \\
\Uparrow \stf\stf\stf\stf\stf\stf& & \Downarrow \\
|\nu_i^{(n+1)}\> = \frac{1}{\mN_i} \(|\nu_i'\> - \sum_{j=0}^{i-1} \sdf 
\< \nu_j^{(n+1)}| \nu_i'\> \sdf |\nu_j^{(n+1)}\> \)
& \Leftarrow & |\nu_i'\> = (1-\Del \be \, \oh^{(n+1)}) \sdf |\nu_i^{(n)}\>  \\
\end{array}.
\label{eq:schema_tps_im}
\ceq
Here $\mN_i$ denotes the normalization of the state "i". Note that, "$\la$" 
is an extra parameter generally used to slow down the convergence and avoid numerical 
instability \cite{bon05}. In practice, it has the effect to mix the mean-field Hamiltonian 
 at a step "n" with the one at the step "n-1". 
 The small time-step increment are generally performed using simple 
 development of the exponential, {\it i.e.}
$e^{-\Delta\be \, \oh^{(n)}}\simeq (1-\Del \be \, \oh^{(n)})$. Again, since the latter is not unitary
a Graham-Schmidt method is used at each iteration.

In practice, a specific Hilbert space basis should be first chosen to numerically 
express the different steps depicted in (\ref{eq:schema_tps_im}). The most common 
choice is either the space (resp. momentum) coordinate representation $|\vr s \tau\>$ 
( resp. $|\vp s \tau\>$) or harmonic oscillator basis $|nljm\>$. In both cases, since 
an infinite number of states could not be considered, the basis has to be truncated. In coordinate 
case, this is achieved by restricting the space to a discrete finite box while in the Harmonic oscillator basis 
a limited number of major shells is considered. Once the basis 
is selected, all ingredients of the theory are expressed in this basis.

\begin{figure}
\begin{center}
\epsfig{figure=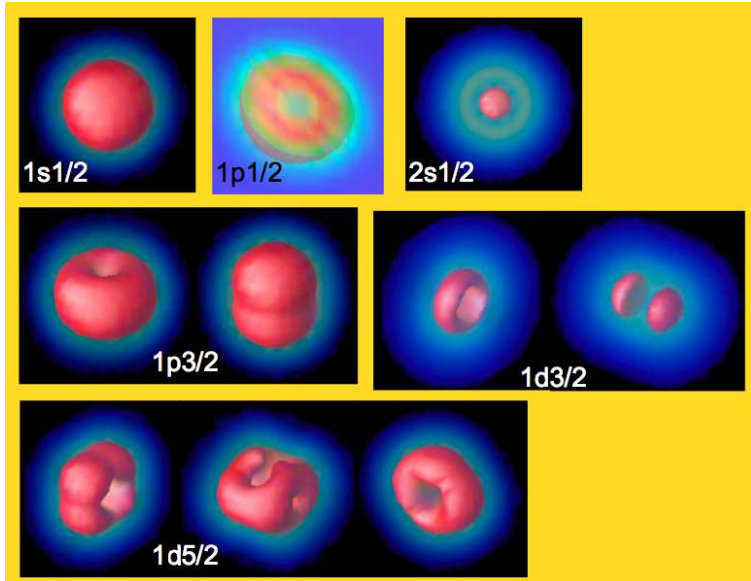,width=10cm} 
\caption{Spatial densities of single-particle neutron states in $^{16}$O. The state 
 1$p$1/2 is cut at the middle to show the hole in its center.}
\label{fig:fo}
\end{center}
\end{figure}

At the end of the iterative procedure, a set of $N$ single-particle states is obtained that completely defines
the HF ground state. Unoccupied levels could also be obtained in a similar way 
although they do neither affect the density nor the mean-field. 
In Figure \ref{fig:fo}, individual densities $\sum_s |\az_\nu(\vr)^{s\tau}|^2$ associated 
to neutron states in $^{16}$O and obtained with the imaginary-time method are 
presented. In that specific case, the discretized coordinate space was used. 
Technically, the box should be taken sufficient large to avoid the effect of the 
boundary. Here {\it Hard Boundary} conditions are retained imposing that 
wave-functions cancel out outside the box. Note finally, that in Fig. \ref{fig:fo}, 
only the $1s1/2$,
$1p3/2$ and $1p1/2$  single-particle states are occupied \footnote{Each state presented here
is doubly degenerated due to the explicit assumption of time-reversal invariance in the calculation.}.

\subsubsection{The two nuclei case: initial state construction}
\label{sec:CIstat}

Since each nucleus has been constructed in its HF ground state, 
we start from two independent particles systems to construct the initial 
TDHF condition. However, we have shown previously that  
Eq.~(\ref{eq:TDHF}) describes {\it a priori} the evolution of a {\it single} 
Slater determinant.

\begin{figure}
\begin{center}
\epsfig{figure=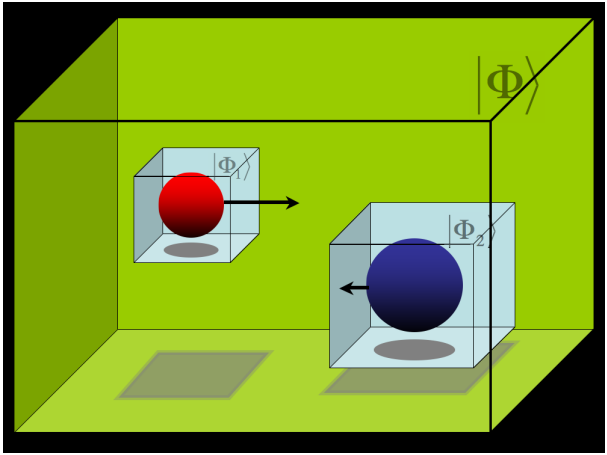,width=8cm} 
\caption{Schematic representation of the 2 initial HF boxes positioned in a bigger box 
for TDHF calculations.}
\label{fig:reseaux}
\end{center}
\end{figure}

It is however always possible to construct a Slater 
determinant 
$\kfi = | \phi_{\nu_1\cdots \nu_A}\>$
associated to $A=A_1+A_2$ independent particles from two Slater determinants
 $|\phi_1\>=| \phi_{\al_1\cdots \al_{A_1}}\>$ and 
$|\phi_2\>=| \phi_{\be_1\cdots \be_{A_2}}\>$ if the two systems are initially well separated. 
Let us consider the two one-body densities  
$\oro_1 = \sum_{n=1}^{A_1} |\al_n\>\<\al_n|$ and 
$\oro_2 = \sum_{n=1}^{A_2} |\be_n\>\<\be_n|$.
For each densities we have 
$\oro_i^2=\oro_i$ (see section \ref{sec:slater}).
If we now construct the square of the total density $\oro = \oro_1+\oro_2$
\oeq
\oro^2 = \oro_1^2 + \oro_2^2 + \oro_1\sdf \oro_2 +\oro_2\sdf  \oro_1 
= \oro + \sum_{m=1}^{A_1} \sum_{n=1}^{A_2} \( |\al_m\> \< \al_m|\be_n\> \< \be_n|
+ \mbox{H.c.}\)
\ceq
where H.c. stands for {\it Hermitian conjugated}. 
Therefore, to have 
$\oro^2=\oro$, single-particle states should not overlap. 
In practice, this is possible due to "Hard boundary" conditions in the HF calculations which impose 
that the single-particle wave-functions 
vanish at some distance. Therefore, we should just be careful to consider 
a box for TDHF that is large enough to avoid overlap between the two boxes used 
in the HF case (see figure \ref{fig:reseaux}).

\subsubsection{Dynamical evolution of nuclei}
\label{subsubsec:boost_gallileen}

The TDHF theory is a quantum theory since it treats explicitly the $N$ particles with
 wave-function. However, the restriction to independent particles states does not 
allow, in general, for a  {\it probabilistic} interpretation of reactions channels. Accordingly, 
some quantum aspects are missing, we will clearly see this pathology in fusion reactions 
(where the fusion probability will either be 0 or 1). Essentially,  TDHF gives {\it classical
trajectories} for the evolution of centers of mass.  

We consider the system in its total center of mass frame. The impact parameter and beam energy (which fixes the initial 
velocity $v$ of the projectile at infinite distance) allow us to determine the initial TDHF 
condition. At this initial time ($t=0$), the two nuclei have a relative distance $D_0$. In general, a Rutherford
trajectory is assumed to account for Coulomb trajectory from infinity to $D_0$. 
This hypothesis 
is consistent with the assumption that the two nuclei are in their ground state at $t=0$,
{\it i.e.} we assume that no energy has been transfered from the relative motion to 
internal degrees of freedom up to $D_0$.

Using notations of section \ref{sec:CIstat}, a velocity $\vv_i$ is applied to each nucleus ($i=1$ or 2) 
imposing the impulsion $\vP_i=A_im\vv_i$. This is performed, by applying a 
translation of each $\ro_i$ in momentum space \cite{tho62}
\oeq
\oro_i(t=0) = e^{i\,m\,\vv_i\cdot \ovr/\hb} \sdf \oro_i \sdf e^{-i\,m\,\vv_i\cdot \ovr/\hb}
\label{eq:CIro}
\ceq
where the position operator $\ovr = \ox \, \ve_x + \oy \, \ve_y + \oz \, \ve_z$ 
acts in single-particle space. Note that, here $\oro_i$ is reserved to the static HF while $\oro_i(t=0)$ with time in parenthesis 
corresponds to boosted HF.

In practice, since one usually follows directly single-particle wave-functions, the translation 
in momentum space is directly applied to the waves 
\oeqn
\az^{st}_{\al_n} (\vr,t=0)&  = e^{i\,m\,\vv_1 \cdot \hat \vr } \sdf \az^{st}_{\al_n} (\vr) &
\stf \stf 1\le  n \le A_1 \nonumber \\
\az^{st}_{\be_n} (\vr,t=0)&  = e^{i\,m\,\vv_2 \cdot \hat \vr } \sdf \az^{st}_{\be_n} (\vr) &
\stf \stf 1\le  n \le A_2.
\ceqn
Once the two nuclei are positioned on the network and properly boosted, there is 
no more reason to distinguish single-particle states from one or the other collision partner.

\subsubsection{Numerical methods for dynamics}
\label{subsubsec:evolution_dynamique}

To solve the system of equations of motion, we should solve TDHF equations 
for occupied states (Eq.~(\ref{eq:TDHFfo})). The main difficulty 
is the fact that the Hamiltonian itself depends on time.
{ As a consequence, as in the imaginary-time case, specific procedure 
should be implemented to account for self-consistency in propagators.} 
We consider a small time step increment $\Del t$ and perform 
the time evolution iteratively. Over small time intervals $\Del t$, the 
Hamiltonian is almost constant. However, to conserve the  total energy, 
the numerical algorithm should be symmetric with respect to time-reversal operation.
This implies to consider the Hamiltonian value at time $t+\frac{\Del t}{2}$ 
for the evolution of wave-functions from $t$ to $t+\Del t$ \cite{flo78} \footnote{This algorithm 
is similar to a Runge-Kutta method.} 
\oeq
|\nu(t+\Del t)\> \approx e^{-i \frac{\Del t}{\hb} \oh\(t+\frac{\Del t}{2}\)} \sdf |\nu(t)\>.
\ceq
A schematic illustration of the real time propagation could be written as: 
\oeqn
&& \begin{array}{ccccc}
\{ |\nu_1^{(n)}\> \cdots |\nu_N^{(n)} \>\} & \Rightarrow &\ro^{(n)}& \Rightarrow 
& \oh^{(n)}\equiv\oh[\ro^{(n)}] \\
\Uparrow  & & & & \Downarrow \\
|\nu_i^{(n+1)}\> =  e^{-i\frac{\Delta t}{\hb} \, \oh^{(n+\frac{1}{2})}} \sdf |\nu_i^{(n)}\>
& & & &  |\tilde{\nu}_i^{(n+1)}\> = e^{-i\frac{\Delta t}{\hb} \, \oh^{(n)}} \sdf |\nu_i^{(n)}\> \\ 
\Uparrow & & & &   \Downarrow \\
 \oh^{\(n+\frac{1}{2}\)}\equiv\oh\[\ro^{\(n+\frac{1}{2}\)}\] & \Leftarrow &
\ro^{\(n+\frac{1}{2}\)}= \frac{\ro^{(n)} + \tro^{(n+1)}}{2} & \Leftarrow & \tro^{(n+1)} \\
\end{array} \nonumber\\
&&
\label{eq:algo}
\ceqn
where $|\az^{(n)}\>$ corresponds to an approximation of $|\az(t_n=n\Del t)\>$.
In this algorithm, starting from the density at time $t$, a first estimate 
of the density at time $t+\Del t$, denoted by  $\tro^{(n+1)}$ is obtained.
The Hamiltonian used in the propagator is then computed using the average 
density obtained from  $\ro^{(n)}$ and $\tro^{(n+1)}$. Then, the real new density 
at time $t+\Del t$ is obtained using this Hamiltonian. As in the imaginary-time 
case, an approximate form of the exponential is generally used which in some cases, 
breaks the unitarity (even in the real time evolution) and orthonormalization 
of the single particle states must be controlled.

\section{Application of TDHF to reactions around the fusion barrier}
\label{sec:barriere}

\begin{figure}
\begin{center}
\epsfig{figure=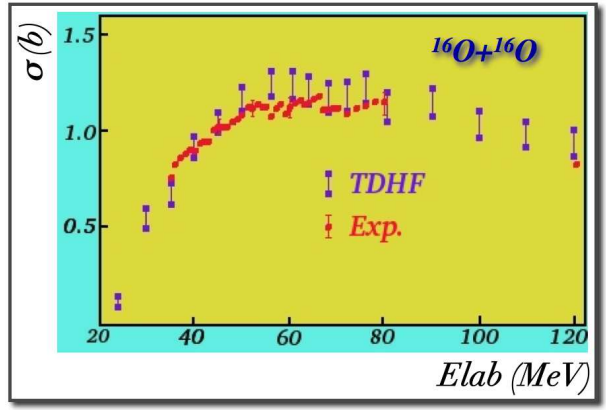,width=8cm} 
\caption{The  $^{16}$O+$^{16}$O  fusion was one of the first application 
of TDHF calculation (adapted  from \cite{bon78}) 
.}
\label{fig:16O_16O}
\end{center}
\end{figure}

TDHF has been applied to nuclear physics more than thirty years ago. First applications were 
essentially dedicated to fusion reactions \cite{bon76,bon78,flo78,neg82}.
Figure \ref{fig:16O_16O}, adapted from  P. Bonche {\it et al.} \cite{bon78}, 
illustrates the predicting power of TDHF for fusion cross sections.
The main difference with actual calculations is that, at that time,
 several symmetries 
 were used 
to make the calculation tractable. The major advantage of 
imposing symmetry was to reduce the dimensionality (and therefore the numerical effort). 
The major drawback was the reduction of applicability. The second difference with nowadays 
calculations comes from the fact that simplified forces were used, missing the 
richness of effective interactions used in up to date nuclear structure studies.   
Recently, all symmetry assumptions have been relaxed and forces using 
all terms of the energy functional have been implemented \cite{uma06a}.
We present here several applications in 3D coordinate space with the full SLy4$d$ \cite{kim97} Skyrme force.

First, general aspects related to fusion reactions are presented, like fusion barrier properties, cross sections...
Then, we illustrate how a microscopic dynamical model 
can be used to get informations from the underlying physical process. 
Besides fusion probability, transfer of nucleon will be discussed. 
Finally, limitations of standard mean-field models are discussed.  

\subsection{Selected aspects of fusion}

\subsubsection{Definition of fusion}

Nuclear fusion is a physical process where two initially well separated 
nuclei collide and form a compound nucleus which has essentially 
lost the memory on entrance channel (Bohr hypothesis). This hypothesis is rather 
simple while the experimental measurement is rarely trivial. Indeed, energy and 
angular momentum conservations imply compound nucleus generally formed at rather 
high internal excitation and angular momentum. As a consequence, the fused 
system cools down by gamma, particle emissions and/or eventually fission. 
Overall, several decay channels are competing leading to a broad range of final phase-space 
which eventually overlaps with direct reactions (and more generally pre-equilibrium) processes.  
This is for instance the case of quasi-fission (where the systems keep partial 
memory of the entrance channel) which leads to mass and charge distribution which can eventually 
be similar to fission. Similarly complete and incomplete fusion are sometimes 
difficult to disentangle due to the presence of direct break-up channels (which are enhanced in 
the weakly bound nucleus case). Therefore,  fusion events are  sometimes experimentally difficult 
to distinguish from other processes. 

\subsubsection{One dimensional approximation}

\begin{figure}
\begin{center}
\epsfig{figure=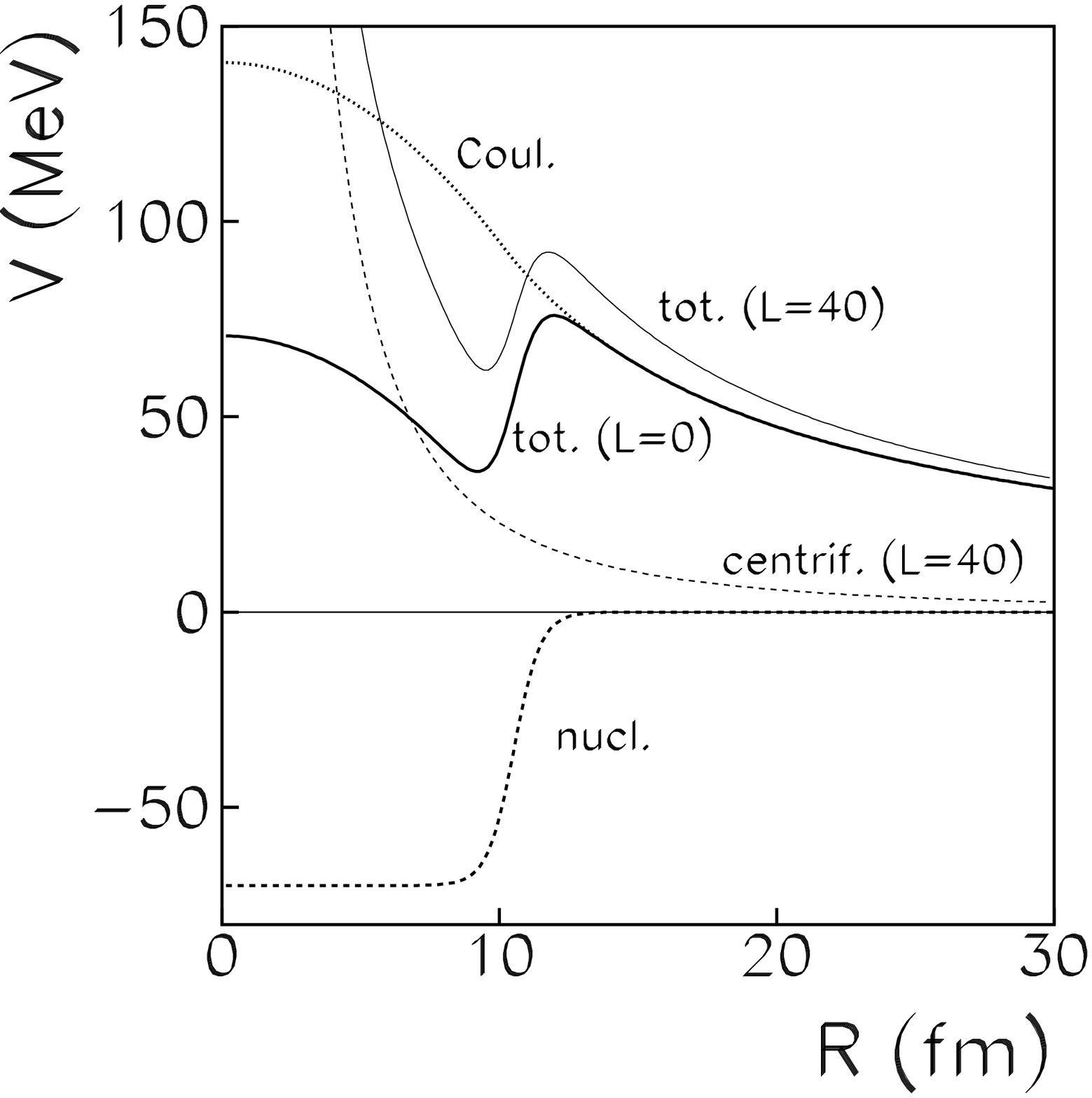,width=8cm} 
\hfill
\epsfig{figure=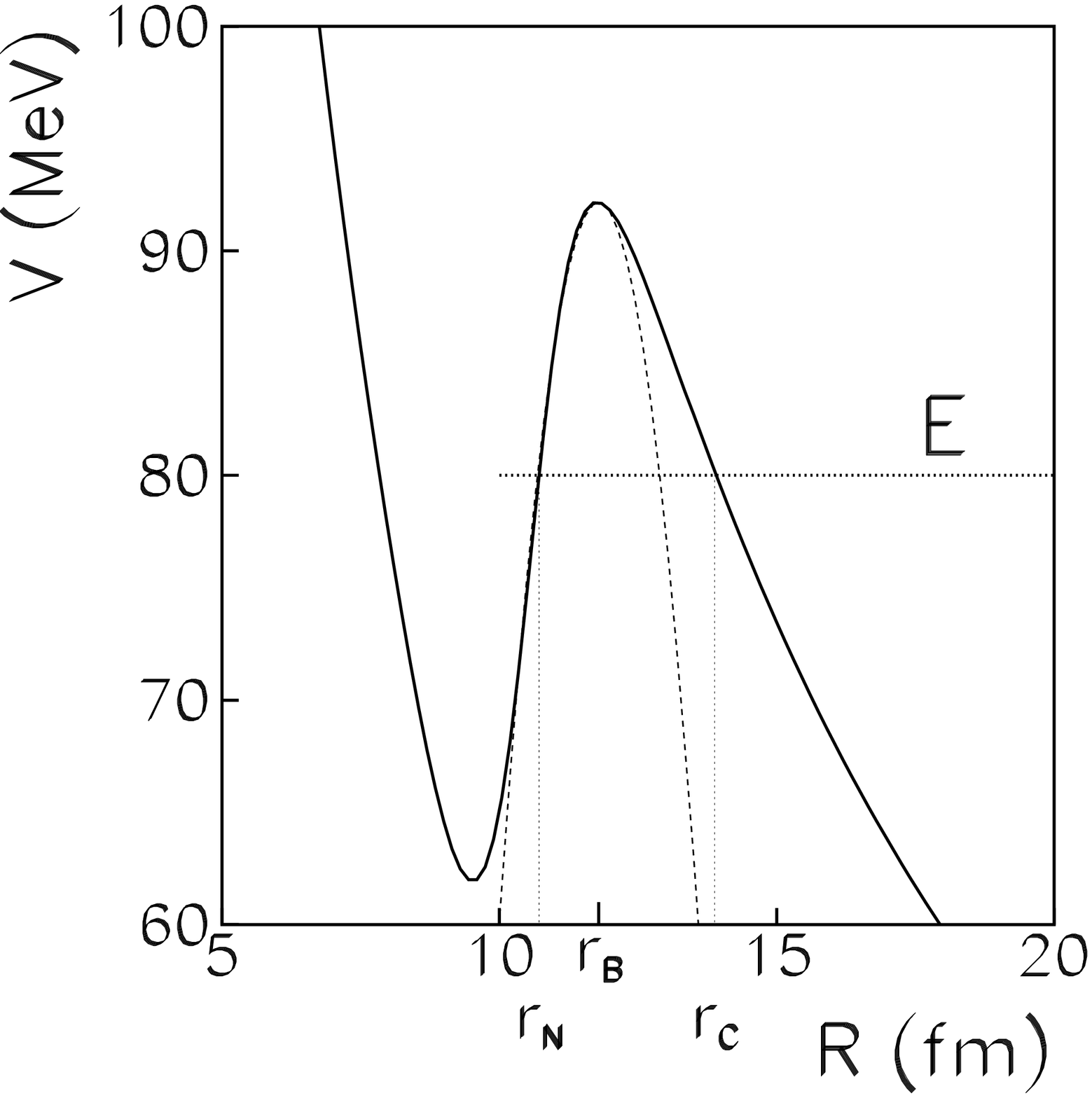,width=8cm} 
\caption{Left: example of nuclear, Coulomb, centrifugal (with $\ell =40$), 
and total ($\ell =0$ and $\ell =40$) potentials for $^{16}$O+$^{208}$Pb.
The Wong parameterization is used for the nuclear part~\cite{won73}.
Right: zoom around the barrier for $\ell =40$. For a given 
energy $E$, the inner and outer turning points at  $r_N$ and 
 $r_C$ respectively, as well as the parabolic approximation (dashed line) are presented.}
\label{fig:nanni1}      
\end{center}
\end{figure}

The simplest approach to fusion reaction is to consider the relative distance $r$ between 
the centers of mass of the nuclei as the most relevant degree of freedom. Fusion reactions then reduce to the dynamical evolution in
a one-dimensional 
potential where the potential is deduced from the long-range Coulomb repulsive interaction of the two nuclei and 
from their short range mutual nuclear attraction (see figure \ref{fig:nanni1}).
We assume that fusion takes place when the system reaches the inner part of the fusion barrier ($r<r_N$ 
on figure \ref{fig:nanni1}).
Within this approximation, fusion cross sections can be expressed as a 
function of the transmission probability $T_\ell(E)$ for each energy E and angular momentum $\ell$
\begin{equation}
\si_{fus}(E)~=~\sum_\ell {\pi \hbar^2 \over 2\mu E}~ (2\ell+1)T_\ell(E),
\label{sigma}
\end{equation}
where $\mu$ denotes the reduced mass. Below the fusion barrier, {\it i.e.} $(E<B)$, 
fusion is possible only by quantum tunneling.
Using the WKB (Wentzel-Kramers-Brillouin) approximation leads to the following
transmission coefficients  
\begin{equation}
T_\ell(E)=\left[1+\exp\left({2 \over \hbar}~\int_{r_C}^{r_N} dr \sqrt{2\mu (U(\ell,r)-E)}\right)\right]^{-1}
\label{WKB}
\end{equation}
where $U(\ell,r)$ is the total (nuclear + Coulomb + centrifugal) potentials while  
$r_C$ and $r_N$ correspond to "turning points" at energies $E$  (see Fig. \ref{fig:nanni1}).
Another simplification can eventually be made using a parabolic approximation.
In that case, the potential is approximated by an inverted parabola with curvature $\hbar \omega_B$.
This approximation is justified close to the fusion barrier only and leads to analytical expression 
for the transmission coefficients  
\begin{equation}
T_\ell(E)~=~{ 1 \over 1+ \exp\left[2\pi(B-E)/\hbar\omega_B\right]}~~.
\label{parabola}
\end{equation}
Finally, summation on different $\ell$ leads to the Wong formula \cite{won73}
\begin{equation}
\si_{fus}(E)~=~{R_C^2 \hbar\omega_B \over 2 E}\ln \left\{1+\exp \left[{2\pi \over \hbar \omega_B}(E-B)\right]\right\}~~.
\label{wong}
\end{equation}
\begin{figure}
\begin{center}
\epsfig{figure=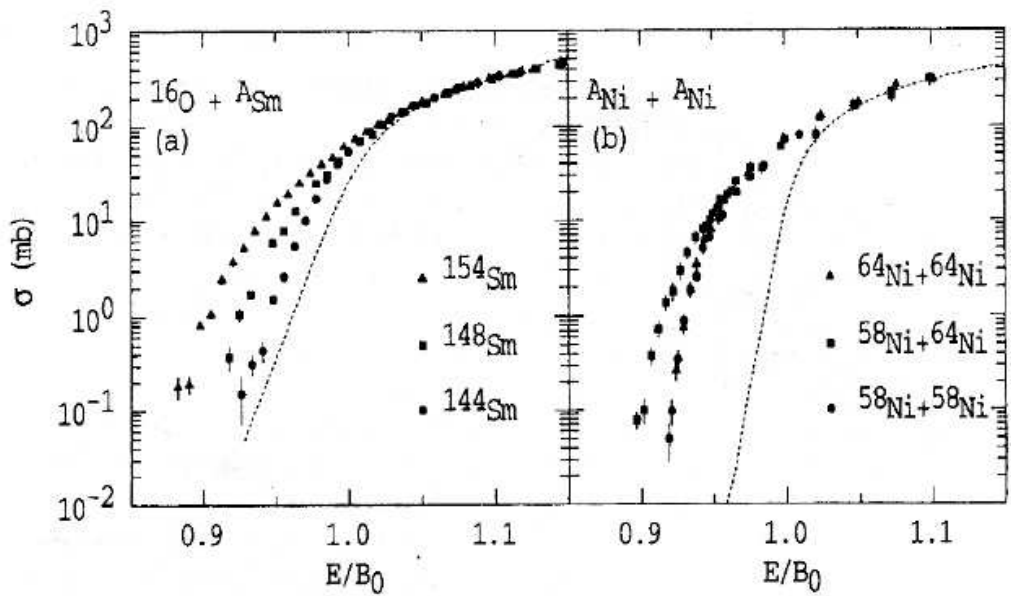,width=14cm} 
\caption{Fusion cross sections for different systems as a function of center  of 
mass energy (divided by their respective fusion barrier energy). Left: 
 $^{16}$O projectile on different Samarium isotopes. Right:   Nickel on  Nickel collisions. 
Curves correspond to calculations deduced from barrier penetration model with standard parameterization
of nucleus-nucleus potentials (extracted from \cite{das98}).}
\label{fig:2}      
\end{center}
\end{figure}
Predictions of this one-dimensional approximation are compared to experimental observations for O+Sm and Ni+Ni systems
in Fig. \ref{fig:2}. The comparison is relatively satisfactory above the barrier but fails to reproduce 
sub-barrier cross-sections which are underestimated by several orders of magnitude
by the Wong formula.

In addition, experimental data clearly show large differences from one isotope to the 
other which underlines the inherent nuclear structure effects and could not be simply 
explained by the change of the radii. Note finally that improvements
where the parabola and/or the WKB approximations are not made
do not improve significantly the comparison. 

\subsubsection{Coupling between relative motion and internal degrees of freedom}

The discrepancy between experiments and simple estimates can directly be traced back to the fact that 
we assumed nuclei as rigid objects without internal structure. In fact, fusion is affected 
by the reorganization of internal degrees freedom as the two nuclei approach. This induces a 
coupling between the internal degrees of freedom and the relative motion.    
The only way to include this effect is to add additional degrees of freedom 
in the description of fusion. { At the macroscopic level, this is generally achieved 
by introducing, for instance, deformation, orientation, neck parameters, mass and charge asymmetries leading
to more complex "multi-dimensional" potentials (see \cite{lac02} and references therein).} This will be illustrated 
with TDHF calculations below, in particular to study the effects of deformation. 

Sub-barrier fusion is a perfect illustration of effects induced 
by couplings to internal degrees of freedom. These couplings can lead to 
an ensemble of fusion barriers called {\it barriers distribution} and can 
give enhancement of cross sections by several order of magnitude 
in the sub-barrier fusion regime\footnote{Note that, as a counterpart, fusion above 
the barrier are usually reduced compared to the one-dimensional case.}. 
A proper description of these effects could only be achieved if inelastic excitations, 
collective modes, transfer and all relevant processes are properly accounted for (see \cite{baha,das98}).

Note finally that additional important effects could appear when weakly bound 
nuclei are involved. In particular, break-up channels and new collective modes 
may become important. From an experimental point of view, influence 
of these new effects on reduction/increase of fusion cross sections is still 
under debate. 
With future low energy radioactive beams, we do expect to get additional 
informations on the reaction mechanisms with weakly bound nuclei around the barrier. 

\subsubsection{Barriers distribution}

The experimental barriers distribution $D_B^{exp}(E)$ is obtained from the 
excitation function $\si_{fus}(E)$ using the relation \cite{row91}
\oeq
D_B^{exp}(E) = \frac{\d ^2 \(\si_{fus} E\)}{\d E^2}.
\label{eq:D_B}
\ceq
This function could be interpreted as the probability that the system has 
its barrier at energy $E$.

Let us illustrate this with a simple cases. Consider a classical model with a 
single barrier at energy $B$. In that case, $D_B$ identifies with the Dirac function $\delta (E-B)$.
The latter formula is consistent with the fusion probability found in classical systems (i.e. $\hb\rightarrow 0$)
deduced from the Wong formula (\ref{wong}). The fusion probability is zero below the barrier 
and equal to $\pi R_B^2(1-\frac{B}{E})$ for $E\ge B$. We therefore deduce that the distribution barrier reads 
$D_B(E) = \pi R_B^2 \delta (E-B)$ which is the expected behavior. For quantum systems, $D_B(E)$ spreads over a wider range 
of energies.    

\subsection{Fusion barriers and excitation functions with TDHF}

In this section, TDHF results essentially extracted from Refs. \cite{sim07b,uma06b} 
are presented

\subsubsection{Fragments trajectories}

\begin{figure}
\begin{center}
\epsfig{figure=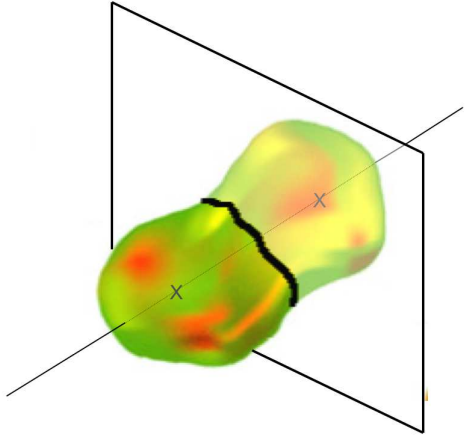,width=8cm} 
\caption{Schematic illustration of separation plane between two fragments after 
the contact.}
\label{fig:fragments}
\end{center}
\end{figure}

A way to characterize TDHF trajectories consists in matching 
the microscopic theory with the one-dimensional model described previously. 
{The main difficulty is then to define the relative distance between the two nuclei.
When fragments are well separated, such a distance could be easily defined. On the opposite, after 
the touching, this becomes more complicated and could only be achieved for short time after the touching.
Fortunately, around the barrier, nuclei are generally still well separated.} 
In practice, the main axis of the reaction could be extracted (it generally corresponds to 
the main axis of deformation)~\cite{was08}. Then, along this axis, the local minimum of the density profile 
defines the separation plane (see figure \ref{fig:fragments}). Note that this distance does not cancel out 
even for one sphere.

\subsubsection{Barriers distribution for two initially spherical nuclei}

\begin{figure}
\begin{center}
\epsfig{figure=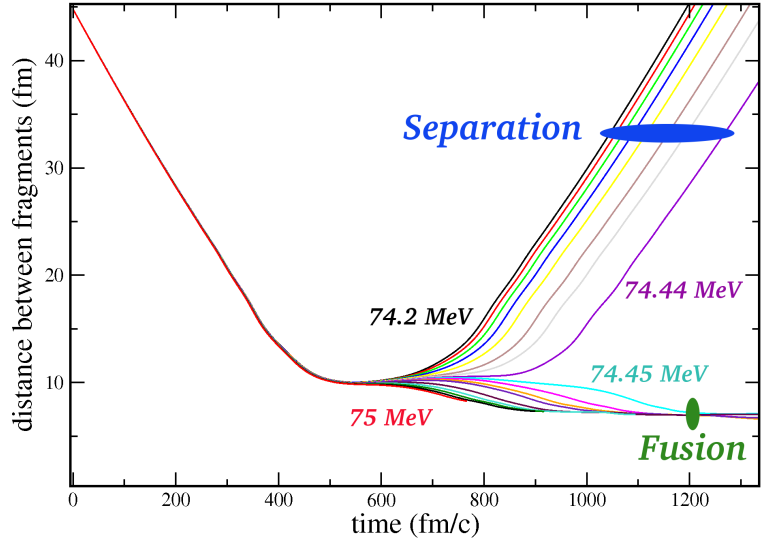,width=10cm} 
\caption{Relative distance between fragments as a function of time for head-on 
$^{16}$O+$^{208}$Pb reactions.}
\label{fig:distances}
\end{center}
\end{figure}

\begin{figure}
\begin{center}
\epsfig{figure=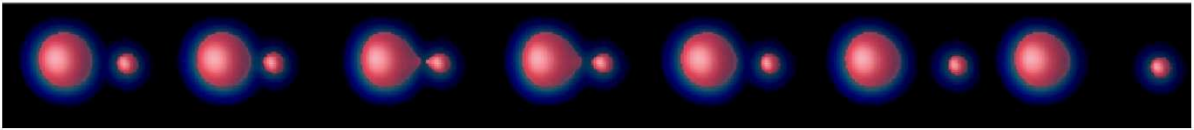,width=16cm} 
\caption{Density evolution for the reaction $^{16}$O+$^{208}$Pb corresponding to head-on collision at center of mass 
energy 74.44 MeV (just below fusion barrier). The red surfaces correspond to 
an iso-density  half 
of the saturation density. Each figure is separated by a time step of 135 fm/c.}
\label{fig:dens_74_44}
\end{center}
\end{figure}

\begin{figure}
\begin{center}
\epsfig{figure=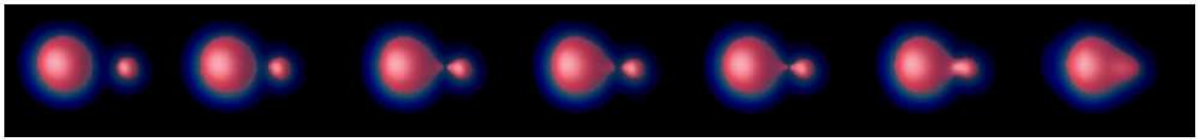,width=16cm} 
\caption{Same as figure \ref{fig:dens_74_44} for center of mass energy 74.45 MeV (just above fusion barrier).}
\label{fig:dens_74_45}
\end{center}
\end{figure}

We first consider $^{16}$O+$^{208}$Pb reaction. The two nuclei are initially in 
their ground state and are at a relative distance $D_0 = 44.8 $ fm.
Evolutions of relative distances for head-on collisions  ($b=0$~fm) obtained from 
different center of mass energies (from 74.2 et 75 MeV) are displayed in Fig.~\ref{fig:distances}~\footnote{Each trajectory presented in figure  \ref{fig:distances} have been 
computed using 4 hour CPU time on a NEC/SX-8 processor.}. 
Due to the narrow range of initial center of mass energies, relative distance evolutions 
before touching are all similar. On opposite, after $t=600$~fm/c, we clearly see 
two groups of trajectories:    
\begin{itemize}
\item[$\bullet$] $E\le 74.44$ MeV, the two fragments re-separate.
\item[$\bullet$] $E\ge74.45$ MeV, the relative distances remain small  ($r ~\le$ 10 fm).
\end{itemize}
The latter case corresponds to the formation of a compound nucleus after the collision. A precise analysis shows 
that the predicted fusion barrier lies between 74.44 and 74.45 MeV. Experimentally, it is found around 74 MeV, 
with a width of 4 MeV. It is rather interesting again to mention that TDHF seems to precisely describe fusion barrier 
while no parameters of the effective interaction has been adjusted on reactions.    
In order to better understand phenomena occuring in TDHF around the fusion barrier, 
the local part of the one-body density ($\ro(\vr)=\sum_{s \tau}\ro(\vr s\tau,\vr s \tau)$) evolutions 
are displayed in Figs. \ref{fig:dens_74_44}  and  \ref{fig:dens_74_45} respectively 
for energies just below and above the estimated fusion barrier. 
In the former case, the composite system hesitates to fusion. It forms a "di-nuclear" system for relatively 
long time ($\sim 500$ fm/c) before re-separating. During this time, nucleons are exchanged between the two partners. 
In the second case, the system 
passes the fusion barrier (it is just 10 keV above the barrier). More generally, the two figures illustrates the richness 
of physical phenomena contained in TDHF: surface diffusivity, neck formation at early stage of fusion process, quadrupole/octupole shapes 
of compound nucleus...       

\begin{figure}
\begin{center}
\epsfig{figure=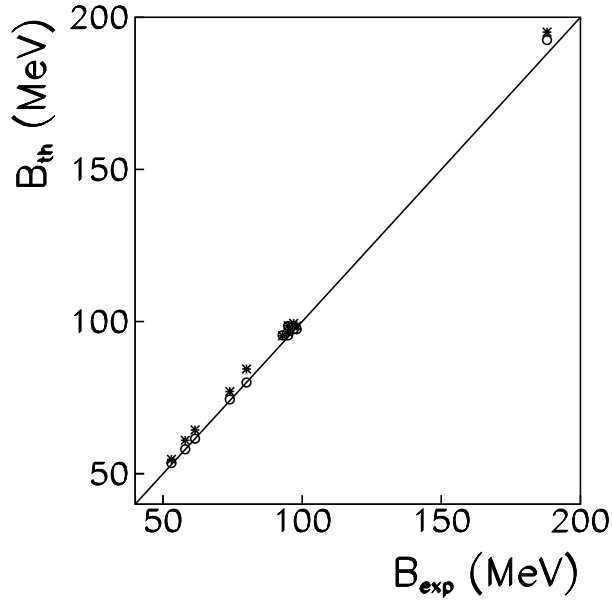,width=8cm} 
\caption{Macroscopic fusion barriers from the Bass parametrization (stars) 
compared to barriers extracted from TDHF (circles) 
as a function of experimental barriers 
(centroids of fusion barrier distributions except for the last point 
corresponding  to  $^{48}$Ti+$^{208}$Pb
where the barrier distribution is obtained from quasi-elastic scattering~\cite{mit07}). 
Figure extracted from \cite{sim07b}.}
\label{fig:th_exp}
\end{center}
\end{figure}

A similar agreement between experimental and calculated fusion barriers
 is found in other systems 
as shown in Fig. \ref{fig:th_exp}. Several projectiles starting from light 
$^{16}$O to medium mass $^{58}$Ni nuclei and targets from $^{40}$Ca
to $^{238}$U have been considered. The lowest energy barrier corresponds to  
$^{40}$Ca+$^{40}$Ca while the highest is obtained for $^{48}$Ti+$^{208}$Pb.
We clearly see in this figure that barriers extracted from TDHF, where the only inputs are the 
effective forces parameters~\cite{kim97}, give a better agreement with data 
than Bass empirical barriers~\cite{bas77,bas80}. 

\subsubsection{ Barrier distribution from collisions between a spherical and a deformed nucleus}

\begin{figure}
\begin{center}
\epsfig{figure=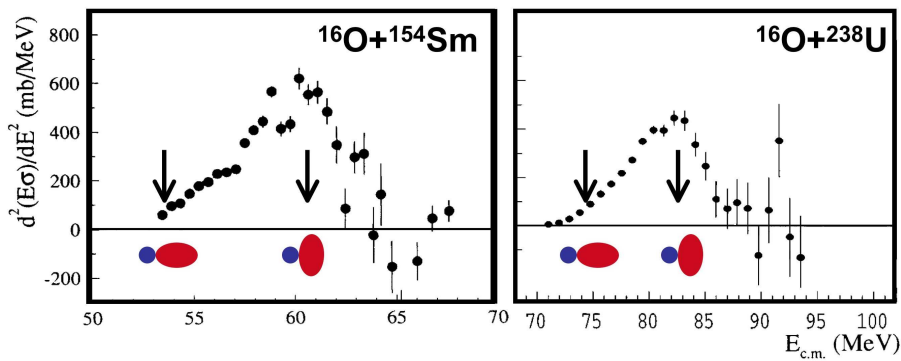,width=12cm} 
\caption{Typical experimental barrier distributions obtained for a light spherical nucleus
on a heavy deformed one. Arrows indicate barriers deduced from TDHF assuming different
orientations of the deformed nucleus at the touching point.}
\label{fig:def}
\end{center}
\end{figure}

Fusion involving at least one deformed nucleus is helpful to illustrate 
the appearance of several fusion barriers which could be easily 
interpreted in terms of a classical variable, the "relative orientation" of the two nuclei.
Figure \ref{fig:def} gives examples of experimental barriers distribution (Eq.~(\ref{eq:D_B})) 
for reactions involving one spherical light nucleus and a heavy deformed one. 
These distributions have a typical width of $\sim 10$ MeV. 
Tunneling effect can generally account for $\sim 2-3$ MeV widths~\cite{row91}.
An alternative 
interpretation should then be invoked to understand such a spreading. 

One can understand this spreading as an effect of 
different orientations
 of the deformed nucleus at the touching point.  
Indeed, for an elongated nucleus, if the deformation axis matches the reaction 
axis (at $b=0$ fm for instance), { then the barrier is reduced due to the 
plug in of nuclear effects at a larger distance}. On opposite, when the deformation axis is 
perpendicular to the collision axis, the barrier is increased. This two extreme cases correspond 
to an upper and lower limits (indicated by arrows on Fig. \ref{fig:def}) for the barrier distribution.   
\begin{figure}
\begin{center}
\epsfig{figure=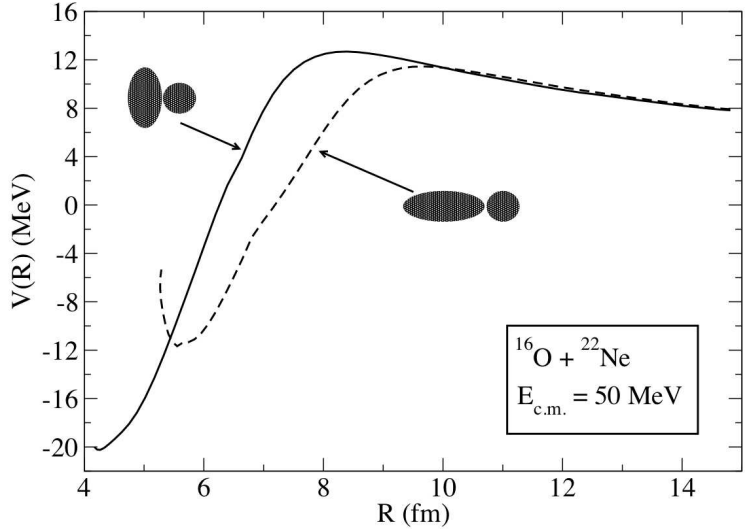,width=10cm} 
\caption{Example of nucleus-nucleus potential for reactions involving one spherical and one deformed 
nucleus for two relative orientations (taken from ref. \cite{uma06b}).}
\label{fig:umar}
\end{center}
\end{figure}

This effect is illustrated in figure \ref{fig:umar} where potentials extracted
from TDHF calculations are displayed as a function of relative distance for two 
different initial  orientations of a prolate nucleus. 
In this case, when the orientation and collision axis are parallel, 
the contact between nuclei takes
place at larger relative distance as compared to the perpendicular case
which leads to a more compact configuration. The fusion 
barrier is then reduced for the parallel case because the Coulomb repulsion
is smaller when the nuclear attraction starts to be significant. 
We also note in Fig.~\ref{fig:def} that barriers distributions are peaked for compact
configurations. This
means that the probability that the system goes through a high energy barrier is higher than the one associated 
to low energy barriers. This is due to the fact that the 
deformed nucleus is prolate and not oblate.
Let us consider for simplicity that the collision axis is the $z$ axis
 and that the deformation axis could only take three 
orientations along $x$, $y$ or $z$ with equal probabilities. 
For a prolate nucleus (the elongation occurs along the deformation axis), 
the compact configuration (associated to higher barriers)
is reached if the deformation axis is $x$ or $y$,
then with a probability~$\frac{2}{3}$.
Having a lower barrier is then less probable, corresponding
to a deformation along $z$ with a probability~$\frac{1}{3}$.
The opposite occurs with an oblate nucleus 
(the elongation is perpendicular to the deformation axis). In this case,
 only an orientation along the $z$ axis leads to a compact configuration,
then with a probability $\frac{1}{3}$.
This simplified discussion gives a qualitative understanding of figure~\ref{fig:def}. 

In the simple discussion above, we have assumed that the 
distribution of relative orientations is isotropic. 
This hypothesis generally breaks down 
due to the long range Coulomb interaction which tends to polarize the nuclei 
during the approaching phase. In the case of a prolate nucleus,
the Coulomb repulsion being stronger on the closest tip of the deformed nucleus,
the net effect is to favor orientations 
where the deformation axis is perpendicular to the collision axis.
This polarization effect modifies the simple isotropic picture, 
in particular when the deformed nucleus is light 
and its collision partner heavy~\cite{sim04}.

In summary, we have seen that barrier position and height
 can be affected by the structure of the collision partners. 
It is hazardous to conclude that fusion could be a tool to infer 
 nuclear structure properties.
 However, with precise fusion measurement, one
 can clearly get informations on deformation properties which might be hardly 
reached with other classical techniques of spectroscopy.

\subsubsection{Excitation functions}

The calculations presented in the previous section were performed using TDHF 
at zero impact parameter. To compute excitation functions (Fig. \ref{fig:16O_16O}), 
calculations should include all impact parameters up to grazing. 
The fusion cross section is given by Eq.~(\ref{sigma}) where $T_\ell(E)$ 
is nothing but the fusion probability for a given center of mass energy $E$ 
and angular momentum $\sqrt{l(l+1)}\hbar$. 
The independent particle hypothesis implies that  
$T_\ell(E)=1$ for $l\le l_{max}(E)$ and $0$ for $l>l_{max}(E)$. 
This finally leads to the so-called 
"quantum sharp cut-off formula" \cite{bla54}
\oeq
\si_{fus} (E) = \frac{\pi \hbar^2}{2\mu E} \sdf (l_{max}(E)+1)^2.
\ceq
To avoid discontinuities due to the cut-off and integer values of 
 $l_{max}(E)$,
$(l_{max}(E)+1)\hbar$ is generally approximated by its semi-classical equivalent $\mL_c=\sqrt{2\mu E}\, b_c$.
The latter corresponds to the classical angular momentum threshold for fusion
and $b_c$ denotes the maximum impact parameter
below which fusion takes place \cite{bas80}.
This replacement is justified by the fact that $(l_{max}+1)^2$ and
$\mL_c^2/\hbar^2$ are both greater than $l_{max}(l_{max}+1)$ and lower than $(l_{max}+1)(l_{max}+2)$. 
Accordingly, we finally obtain the standard classical expression for fusion cross sections
 $\si_{fus}(E) \simeq \pi \mL_c^2/2\mu E = \pi b_c^2$.

\begin{figure}
\begin{center}
\epsfig{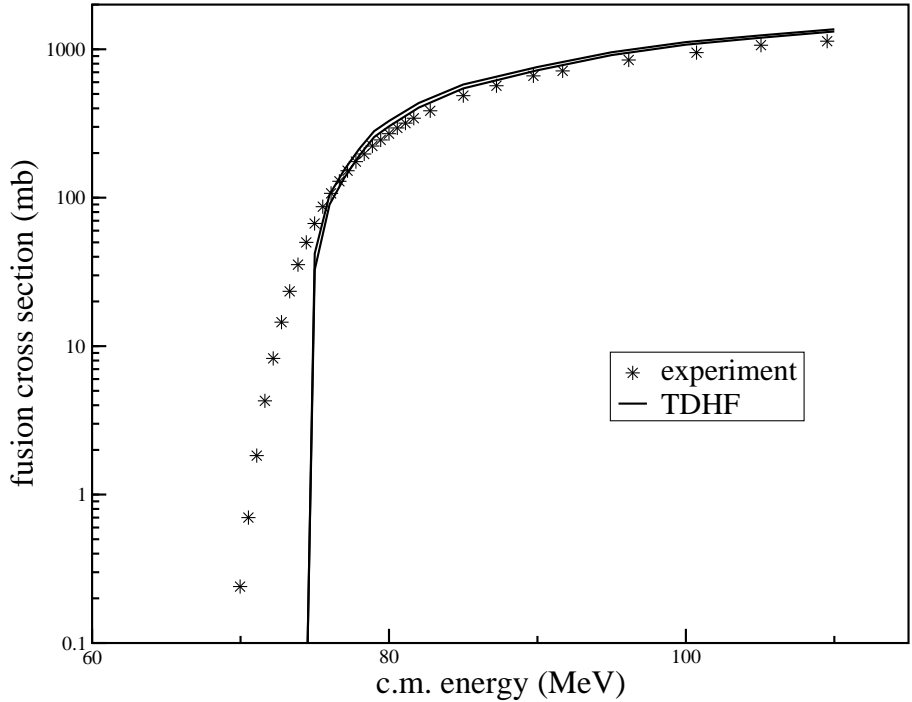} 
\caption{Experimental fusion cross section
(stars) compared to cross section deduced from TDHF (lines) for $^{16}$O+$^{208}$Pb.
The two lines correspond respectively to lower and upper limits of theoretical 
cross sections.}
\label{fig:fus}
\end{center}
\end{figure}

Figure \ref{fig:fus} presents a comparison between calculated and experimental cross sections 
for the  $^{16}$O+$^{208}$Pb system. 
Discretizing the impact parameters
gives an upper and lower bound for fusion cross sections.
We essentially see that above the fusion barrier, TDHF calculations reproduce rather well 
the experimental observations 
(the cross section is however slightly overestimated by 16\%) 
while at energies below the Coulomb barrier, the calculation misses the quantum tunneling
contribution.

\subsection{Nucleon transfer around the fusion barrier}

In this section, we study nucleon transfer below the fusion barrier with TDHF.
The basic observable associated to transfer is simply the mass (or nucleon number) 
of the two fragments after re-separation. If the latter differs 
from the entrance channel, this is an obvious signature of transfer~\cite{uma08}. 
Another signature would be the variance of nucleon number in the fragments.

\subsubsection{Transfer identification}

As illustrated in figure \ref{fig:dens_74_44}, two nuclei can form a di-nuclear system with a 
neck and then re-separate. There is {\it a priori} no reason that these two fragments conserve the
same neutron and proton numbers as in the entrance channel (except for symmetric reactions).
Indeed, between the touching and re-separation, nucleons can be exchanged.
In TDHF calculation, this exchange is treated through the time-dependent distortion of single-particle
wave-function which can eventually be partially transfered from one partner to the other.

The following operator written in r-space defines the number of particles in the right side of the separation plane   
(defined arbitrarily as $x>0$):
\oeq
\oN_D = \sum_{s\tau} \sdf \int \stb \d \vr \stf \oad(\vr s \tau) \sdf \oa(\vr s \tau) 
\sdf \mH(x)
\label{eq:NG}
\ceq
where $\mH(x)$ is the Heavyside function equal to 1 if $x>0$ and 0 elsewhere.

Denoting by $\< i | j \>_D = \sum_{s\tau} \int \sdb \d \vr \stf {\az_i^{s\tau}}^*(\vr) \sdf {\az_j^{s\tau}}(\vr)\sdf \mH(x)$
the overlap (limited to the right side) between two single-particle states and using Eq.~(\ref{eq:etats_occ}) 
for an independent particles state $\kfi$, we obtain (in the specific basis 
which diagonalizes the one-body density associated to $\kfi$)
\oeq
\<\oN_D\>_\phi = \sum_{ij} \sdf \< i | j \>_D \sdf\< \oad_i \, \oa_j\>_\phi 
= \sum_i \sdf \<i|i\>_D \sdf n_i.
\label{eq:N_D}
\ceq

\begin{figure}
\begin{center}
\epsfig{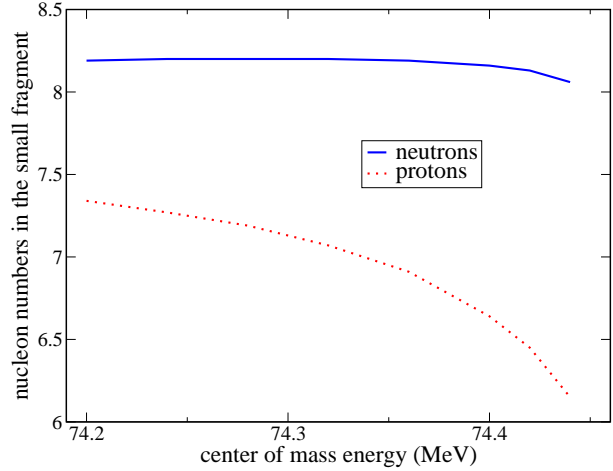} 
\caption{Average neutron (solid line) and proton (dotted line) numbers 
of the lightest fragment in exit channel of head-on 
$^{16}$O+$^{208}$Pb collisions below the Coulomb barrier as a function of the center of mass energy. }
\label{fig:transfert}
\end{center}
\end{figure}

Figure \ref{fig:transfert} gives the average final neutron and proton numbers
of the smallest fragment in exit channels of $^{16}$O+$^{208}$Pb reaction as a function of center 
of mass energy. We see that the more the energy increases, the more  the $^{16}$O looses protons.
At an energy just below the barrier, 
it has transfered around $2$ protons to the $^{208}$Pb. 
A possible explanation is the fast $N/Z$ equilibration which is expected 
to take place at contact. 
Indeed, considering the case where exactly two protons have been transfered leads to 
the chemical equation 
$$
 ^{16}\mbox{O}(1)+^{208}\!\mbox{Pb}(1.54) \rightarrow ^{14}\!\mbox{C}(1.33)+^{210}\!\mbox{Po}(1.5)
$$
where quantities in parenthesis correspond to the $N/Z$ values.
We see that the $N/Z$ initial asymmetry is almost equilibrated in the exit channel. 
This equilibration process also takes place at the first instant of fusion, 
and is increased when $N/Z$  differences between partners increase
\cite{bon81,cho93,sim01,sim07a}.

Note finally that these results on transfer agree qualitatively 
with what is observed experimentally~\cite{vul86}.
Indeed, it is 
observed that the one proton transfer channel from $^{16}$O to $^{208}$Pb
dominates below the Coulomb barrier, while at higher energies
the two protons transfer becomes the main transfer channel.

\subsubsection{Many-Body states associated to each fragment}

Well below the fusion barrier, transfer is prohibited and 
exiting fragments essentially reflect mass and charge 
properties of the entrance channel. In this case, the $^{16}$O remains
an eigenstate of the proton and neutron number operators\footnote{We may 
introduce the terminology that a Many-Body state has a "good" neutron (or proton) number
when the state is an eigenstate of the particle number $\oN = \sum_i |i\> \<i|$.}.
For these state, the variance of particle number operator (in the right side) is strictly 
zero:
$\si_D = \sqrt{\<\oN_D^2\>-\<\oN_D\>^2} = 0$.

This property is lost at higher energies where transfer occurs.
Then each fragment in the exit channel does not have a "good" particle number 
and can therefore not be associated anymore to a single Slater determinant.
It corresponds to a (eventually complicated) correlated
state even if the wave-function associated to the total system is a Slater determinant. 

Let us calculate the variance of $\oN_D$ after the reaction. We follow 
here the technique of Dasso {\it et al.}~\cite{das79}.
Using anti-commutation relations (\ref{eq:anticom2}), closure relation 
 $\sum_i |i\>\<i|=\hat{1}$, and Eq.~(\ref{eq:ro2}), we get in the same way 
 as for Eq.~(\ref{eq:N_D})
\oeqn
\<\oN_D^2\>_\phi &=& \sum_{ijkl} \sdf \< i | j \>_D \sdf\< k | l \>_D \sdf\< \oad_i \, \oa_j \,  \oad_k \, \oa_l \>_\phi \\
 &=& \sum_{ijkl} \sdf \< i | j \>_D \sdf\< k | l \>_D \sdf \(\< \oad_i \, \oa_l \>_\phi \sdf \del_{jk}
 +  \< \oad_i \,  \oa_j \>_\phi \sdf \< \oad_k \,  \oa_l \>_\phi 
 -  \< \oad_i \,  \oa_l \>_\phi \sdf \< \oad_k \,  \oa_j \>_\phi \)\nonumber \\
&=&  \sum_i \sdf  n_i \sdf \<i|i\>_D +  \sum_{i,j} \sdf n_i \sdf n_j \sdf \(\<i|i\>_D \sdf \<j|j\>_D
- \ll \<i|j\>_D \rl^2\).
\ceqn
The variance must then follow 
\oeq
\si_{D}^2 = \< \oN_D^2 \>_\phi -  \< \oN_D \>_\phi^2 = \sum_{i=1}^N \sdf \< i|i\>_D 
- \sum_{i,j=1}^N \sdf \ll\< i|j\>_D \rl^2.
\label{eq:siGN}
\ceq
Applying this formula to the small fragment in the exit channel
 at a center of mass energy 74.44 MeV, 
we get $\si_{D_p} \simeq 0.5$ for protons and  $\si_{D_n} \simeq 0.3$ for neutrons. 
This deviation from zero clearly indicates that the Many-Body state on each side of the 
separation plane is not a pure state with a good particle number anymore but a complicated 
mixing of states with various particle numbers. In addition, even if  the neutron 
number is almost constant in average, 
this non-zero variance is a signature of neutron transfer.   

Note finally, that the variance 
is maximum when the overlap (restricted to the right side) between 
two wave-functions is zero, {\it i.e.} if $\<i|j\>_D\propto \del_{ij}$.
Then Eq.~(\ref{eq:siGN}) identifies with a binomial distribution 
\oeq
\si_{D}^2 = \sum_{i=1}^N \sdf \< i|i\>_D \sdf \( 1- \< i|i\>_D \).
\ceq
This gives an upper limit for the variance 
\oeq
\si_{D} \sdf \le \sdf \sqrt{\<\oN_D\>_\phi\sdf  \(1-\frac{\<\oN_D\>_\phi}{\<\oN\>_\phi}\)}
\sdf  \le \sdf \sqrt{N/4}.
\label{eq:borne}
\ceq
This upper limit is an intrinsic limitation of independent particle systems 
which will not be able to reproduce distribution with fluctuations greater than this limit. 
This is a clear limitation of TDHF, which in general underestimates widths of fragments mass and charge
distribution observed in deep inelastic collisions \cite{das79,goe82}. {The account for correlation 
beyond the single-particle approximation is the only way to escape this difficulty.}

\subsection{Summary: success and  limitations of TDHF}

In this section, we have compared TDHF calculation with experiments on the different following physical effects and quantities

\begin{itemize}
\item[$\bullet$] Barriers positions and height
\item[$\bullet$] Shape of barrier distribution for deformed nuclei 
\item[$\bullet$] Excitation function 
\item[$\bullet$] Populations of fragments mass and charge below the fusion barrier.
\end{itemize}
Overall,  TDHF gives a reasonable qualitative (and sometimes quantitative)  
agreement with experiments.
We have also pointed out some of the limitations of TDHF:
\begin{itemize}
\item[$\bullet$] Essentially tunneling below the Coulomb barrier is not accounted for
\item[$\bullet$] Fluctuations of one-body observables like particle numbers are underestimated. 
\end{itemize}

Beside nuclear reactions, differences between TDHF calculations and experiments could also 
be observed in giant collective vibrations. For instance, damping of collective motion 
could not be properly accounted for at the mean-field level \cite{lac04}. This could again 
be traced back to missing two-body correlations effects. 
The rest of this lecture is devoted to transport theories going beyond mean-field.

%
%

\section{Dynamical theories beyond mean-field}
\label{sec:audela}


In the previous sections, we have illustrated some successes of mean-field theories in the description of nuclear dynamics.
The application of TDHF to nuclear excitations or reactions was a major advance. At the same time, 
the mean-field theory neglects physical effects which play a major role 
in nuclei. For instance short and long range correlations in static nuclei could
 only be accounted for by a proper treatment 
of pairing effects and configuration mixing \cite{ben03}. Conjointly, as the collision energy between two
nuclei increases,  
the Pauli principle becomes less effective to block direct nucleon-nucleon collisions. Then, two-body correlations 
should be 
explicitly accounted for. The description of such two-body effects is crucial since, for instance, 
this would be the only way
to understand how a nucleus could thermalize. During the past decades, several approaches have been developed to 
introduce correlations beyond 
mean-field. We summarize here some of these extensions.       

\subsection{Time-dependent mean-field with pairing correlations}
\label{subsec:TDHFB}

Pairing correlations are sometimes needed to describe nuclear systems. 
However, they are neglected in the independent particles 
approximation~\cite{rin80,bri05}. Mean-field approaches could naturally incorporate 
this effect by considering more general Many-Body states as products of 
independent quasi-particles. 
This leads to the so called Time-Dependent Hartree-Fock-Bogoliubov (TDHFB) theory.\\

\subsubsection{Quasi-particle vacuum }       

Wick's theorem (see appendix \ref{annexe:rappelsMQ}) simplifies 
the calculation of observables expectation values on vacuas. 
In appendix \ref{annexe:videHF}, we show, on one hand, 
that an independent particles state $\kfi$ is a vacuum (called HF vacuum)
for a specific set of operators,
and, on the other hand, is a simple example of "quasi-particle" vacuum. 
The canonical and unitary transformation connecting quasi-particle creation and annihilation operators (denoted by $\obed_\alpha$ and $\obe_\alpha$ respectively) to particle ones is given by Eq.~(\ref{transfoBogo}).
It is called {\it Bogolyubov transformation}~\cite{bog58}.
The vacuum associated to these quasi-particle operators is given by 
\begin{equation}
\ket{\psi_{bogo}}\sim \prod_{k}\hba_k \ket{-},
\label{vacuum}
\end{equation}
which obviously insures $\obe_i \ket{\psi_{bogo}} = 0$. This state 
is then called a {\it quasiparticle vacuum}.

Let us illustrate how pairing could be included in such a Many-Body state.
We consider the specific case where the transformation is simply written as
\begin{eqnarray}
  \obe_p=u_p \oa_p - v_p \oad_{\bar{p}},\ \ \ \ 
  \obe_{\bar{p}}=u_p \oa_{\bar{p}} + v_p \oad_p. 
  \label{specialHFB}
\end{eqnarray}
In this case, $v_p^2$ corresponds to the occupation probability of the state $p$ while $u_p^2 = 1-v_p^2$.
Eq.~(\ref{specialHFB}) is nothing but the standard BCS transformation~\cite{bar57} where 
single-particle states $p$ 
and $\bar p$ form a Cooper pair. We use the standard  convention $\bar{p} = -p < 0$. 
Starting from Eq.(\ref{vacuum}), grouping quasi-particles operators by pairs and using anti-commutation rules
(\ref{eq:anticom1}-\ref{eq:anticom2}), we can rewrite the Many-Body state as     
\begin{eqnarray}
\ket{\psi_{bogo}} \sim \prod_{p>0}\left(u_p + v_p~\oad_p \oad_{\bar p} \right)\kvac.
\end{eqnarray}
The above expression corresponds to the standard BCS state expressed in terms of pairs 
of correlated nucleons $\{p,\bar{p}\}$. In fact, any general Bogolyubov transformation can be written 
in a BCS form using the Bloch-Messiah-Zumino decomposition 
theorem~\cite{blo62,zum62,rin80}.
We therefore see that to account for more general Many-Body states is a way 
to incorporate correlations beyond mean-field. 


\subsubsection{Expectation values of observables on quasi-particle vacua}

Similarly to HF vacua, the Wick's theorem applies to Quasi-particle vacua.
In this case, however, contractions associated to the one-body 
density     
$\rho_{\alpha\beta} = \langle \oad_\beta \oa_\alpha\rangle$ 
are not the only one which differ from zero. The anomalous density, 
with matrix elements
$\kappa_{\alpha\beta} = \langle \oa_\beta \oa_\alpha \rangle$ 
(which also implies $\kappa^*_{\alpha\beta} = \langle \oad_\alpha \oad_\beta \rangle$)
should also be  taken into account. 
The latter contractions cancel out for independent particle systems.
The Bogolyubov transformation (Eq.~(\ref{transfoBogo})) can be inverted to express the $\oad$ and $\oa$ 
in terms of quasi-particles operators $(\obed,\obe)$ :
\begin{eqnarray}
\left\{
\begin{array} {cc}
\oa_\alpha   = & \sum_{i} U_{\alpha i} \obe_i + V^*_{\alpha i} \obed_i \\
\oad_\alpha   = & \sum_{i} V_{\alpha i} \obe_i + U^*_{\alpha i} \obed_i .
\end{array}
\right.
\end{eqnarray}
Using these expressions in $\rho$ et $\kappa$ and the fact that only $\langle \obe_i \obed_i \rangle$ differ
from zero, we deduce
\begin{equation}
\ro_{\alpha \beta} = \sum_i V_{\beta i}V^*_{\alpha i} = 
\left(V^* V^T\right)_{\alpha \beta},~~\kappa_{\alpha \beta}= \left(V^* U^T\right)_{\alpha \beta} ~.
\end{equation}
These contractions are used to write the generalized density matrix defined as 
\begin{eqnarray}
{\cal R}  = \left( 
\begin{array} {cc}
\left( \langle \crea_j \anni_i \rangle \right) & \left(\langle \anni_j^{ } \anni_i \rangle \right)\\
&\\
\left(\langle \crea_j \crea_i \rangle \right) &  \left(\langle \anni_j \crea_i \rangle   \right)
\end{array} 
\right)
 = \left( 
\begin{array} {cc}
\rho & \kappa \\
- \kappa^* & 1-\rho^*  
\end{array} 
\right).
\label{matriceR}
\end{eqnarray}
The anomalous density enables us to treat correlations that were neglected
at the independent particles level.
Indeed, elements of the associated two-body correlation now read
\begin{eqnarray}
\ro^{(2)}_{ijkl}=\langle ij | \oro_{12} | kl \rangle =
\langle \oad_k \oad_l \oa_j \oa_i \rangle &=& \overline{ \oad_k \oa_i }~\overline{ \oad_l \oa_j } 
- \overline{\oad_k \oa_j}~\overline{ \oad_l \oa_i} + \overline{ \oad_k \oad_l} ~\overline{ \oa_j \oa_i}\nonumber \\
&=& \rho_{ik} \rho_{jl} - \rho_{il} \rho_{jk} + \kappa_{ij} \kappa^*_{kl} .
\label{Wick2corps}
\end{eqnarray}
Countrary to Slater determinant, 
 $C_{12}$ does not {\it a priori} cancels out. 
 We further see that the HFB theory leads to a
{\it separable} form of the two body correlation
\begin{eqnarray}
C_{ijkl} &=& \kappa_{ij} \kappa^*_{kl}.
\label{eq:ckk}
\end{eqnarray}
In turns, HFB is more complex than the HF case. For instance, 
the state is not anymore an eigenstate of the particle number operator. 
The symmetry associated to particle number conservation is explicitly broken.
Fluctuations associated to the particle number $\oN = \sum_\alpha \oad_\alpha \oa_\alpha$ now write
\begin{eqnarray}
\langle \oN^2 \rangle - \langle \oN \rangle^2 = 2 \, \Tr (\kappa \kappa^\dagger) = 2 \, \Tr(\rho-\rho^2).
\end{eqnarray} 
In general these quantity is non-zero for a quasi-particle vacuum.
The fact that the particle number is not conserved implies 
that it has to be constrained in average 
in nuclear structure studies (this is generally done by adding a specific Lagrange multiplier to the variational principle).  
It is worth mentioning that in a TDHFB evolution using a (possibly effective) two-body interaction (Eq.~(\ref{eq:oH2})), 
the expectation value of $\oN$ is a constant of motion. 
Therefore, no constraint on particle number is necessary in the dynamical case.

\subsubsection{TDHFB Equations}

Different techniques could be used to derive the equation of motion in the HFB approximation~\cite{rin80,ben03,bla86}. 
Here, the same strategy as section \ref{sec:TDHF} is followed. 
The evolution of the Many-Body HFB state is given by the evolution of its normal 
and anomalous  densities $\rho_{ij}=\langle \crea_j \anni_i \rangle$ 
and $\kappa_{ij}=\langle \anni_j \anni_i \rangle$ which are given by the 
Ehrenfest theorem 
\begin{eqnarray}
i \hbar \frac{d}{dt} \rho_{ji}   &=& i \hbar \frac{d}{dt}\langle \crea_{i}\anni_{j} \rangle 
= \langle \left[\crea_{i}\anni_{j},\hat{H}\right]\rangle, \label{OBDM}\\
i \hbar \frac{d}{dt} \kappa_{ji} &=& i \hbar \frac{d}{dt}\langle \anni_{i}\anni_{j} \rangle 
= \langle \left[\anni_{i}\anni_{j},\hat{H}\right]\rangle.\label{OBPT}
\end{eqnarray}

{ }~{ } \\

\paragraph{One-body density evolution \\}

Eq.~(\ref{OBDM}) can be identified to Eq.~(\ref{eq:TDDM})
which has been obtained for any correlated state 
and  any two-body Hamiltonian. The only difference 
comes from the simplified correlations  
for quasi-particle vacua (Eq.~(\ref{eq:ckk}))
\begin{eqnarray}
i \hbar \frac{d}{d t} \rho_{ji} &=& 
\[\,h[\ro],\ro\,\]_{ji}+\frac{1}{2}\sum_{klm} \left( \bar{v}_{jklm} \kappa_{ik}^* \kappa_{lm} - \bar{v}_{klim} \kappa_{kl}^* \kappa_{jm} \right) \nonumber \\
&=& \sum_{k} \left( h_{jk} \rho_{ki} - h_{ik} \rho_{jk} - \Delta_{jk} \kappa_{ki}^* + \kappa_{jk} \Delta_{ki}^* \right) \label{eq:ro_TDHFB}
\end{eqnarray}
where we have introduced the pairing field 
\oeq
\Delta_{ij}=\frac{1}{2}\sum_{kl} \bar{v}_{ijkl} \kappa_{kl}. 
\label{eq:Delta}
\ceq

{ }~{ }  \\

\paragraph{Evolution of $\kappa$ \\}

Similarly,  the equation of motion for the anomalous density 
$\kappa$ reads 
\begin{eqnarray}
i \hbar \frac{d}{d t} \kappa_{ji} &=& \sum_{kl} t_{kl} \langle \left[ \anni_{i} \anni_{j} , \crea_{k} 
\anni_{l} \right] \rangle + \frac{1}{4} \sum_{klmn} \bar{v}_{klmn} \langle \left[ \anni_{i} \anni_{j} , \crea_{k} \crea_{l} \anni_{n} \anni_{m} \right] \rangle \nonumber \\
i \hbar \frac{d}{d t} \kappa_{ji} &=& \sum_{k} (t_{jk} \kappa_{ki} - t_{ik} \kappa_{kj})  +  \sum_{klm} (\bar{v}_{kjlm} \rho_{lk} \kappa_{mi} - \bar{v}_{kilm} \rho_{lk} \kappa_{mj} ) \nonumber \\ 
&\ & + \frac{1}{2}\sum_{klm} (\bar{v}_{kjlm} \rho_{ik} \kappa_{lm} - \bar{v}_{kilm} \rho_{jk} \kappa_{lm})+\frac{1}{2}\sum_{mn} \bar{v}_{jimn} \kappa_{mn} \nonumber \\
&=& \sum_{k} \left( h_{jk} \kappa_{ki} + \kappa_{jk} h_{ki}^* - \Delta_{jk} \rho_{ki}^* - \rho_{jk} \Delta_{ki} \right) 
+ \Delta_{ji}. \label{OBPT2}
\end{eqnarray}
Eqs. (\ref{eq:ro_TDHFB}) and (\ref{OBPT2}) give the evolution of the matricies $\ro$ 
and~$\kappa$
\begin{eqnarray}
i \hbar \frac{d}{d t}\rho &=& \left[h,\rho \right] + \kappa \Delta^* - \Delta \kappa^*, \label{eq:TDHFBrho} \\
i \hbar \frac{d}{d t}\kappa &=& h \kappa + \kappa h^* - \rho \Delta - \Delta \rho^* + \Delta. \label{eq:TDHFBkappa}
\end{eqnarray}
Finally, using the generalized density matrix $\mathcal{R}$ and the generalized HFB Hamiltonian $\mathcal{H}$ defined as
\begin{eqnarray}
{\cal H}  \equiv \left( 
\begin{array} {cc}
h & \Delta \\
- \Delta^* & - h^*
\end{array} 
\right),
\end{eqnarray}
Eqs.~(\ref{eq:TDHFBrho}) and~(\ref{eq:TDHFBkappa}) 
can be written in a more compact form
\begin{eqnarray}
i \hbar \frac{d{\cal R}}{dt} = \left[{\cal H} , {\cal R} \right] .
\label{eq:tdhfbR}
\end{eqnarray}
The above TDHFB equation generalizes the TDHF case
(Eq.~(\ref{eq:TDHF})) by accounting for pairing effects on dynamics. 


\subsubsection{Application of TDHFB theory}

In practice, numerical implementation of TDHFB is much more complex than the TDHF case.  
This is certainly the reason why, although first applications of TDHF started more than 30 years 
ago, very few attempts to apply TDHFB exist so far. This could be traced back to
conceptual and practical difficulties  inherent to this theory. 
The first difficulty comes from the fact that TDHFB equations should {\it a priori}
be solved in a complete single-particle basis while at the HF level only occupied states 
are necessary. In practice, specific methods should be used to truncate the basis. 
A second additional problem is the effective force that should be used in the pairing channel.
Though zero range forces are clearly very useful at the HF level, they 
lead to ultra-violet divergences in the pairing channel.
In practice, either a finite-range interaction has to be used or 
a specific regularization or renormalization scheme 
should be applied~\cite{dob84,bul02a,bul02b,bul03}. Although these methods 
provide reasonable solutions for the static case,
their application to nuclear dynamics is not straightforward.

Nowadays, the TDHFB theory has essentially been applied in the small amplitude limit 
(leading to the so-called QRPA theory~\cite{rin80,kha02})
or semi-classical limit (hydrodynamical models~\cite{tor87,abr06}). 
Only recently, the dynamical problem has been solved using the nuclear Gogny 
interaction~\cite{has07}. 



\subsection{When is the independent particle approximation valid ?}
\label{sec:depart_tdhf}

In previous sections, we have considered dynamical evolutions where the many-body state is forced
to remain in a specific class of trial state (Slater determinants or more general quasi-particle vacua).
This assumption could only give an approximation of the exact dynamics and we do expect in 
general that the system will deviate from this simple state hypothesis. We give here some 
motivations of the introduction of theories beyond mean-field.

Let us recall the original goal which is to describe as best as possible the dynamics of a self-bound complex 
quantum system. We first assume that a system is initially 
properly described by a Slater determinant\footnote{The discussion below can
 easily be generalized to an initial quasi-particle vacuum}, i.e. 
$\left| \psi(t=0) \right\rangle = \kfi$ with 
$\kfi = \Pi_{\alpha=1}^N \oad_\alpha \left| - \right\rangle$ where the label $\alpha$ 
refers to initially occupied states (hole states). The exact dynamics of the system is given by the time-dependent 
Schroedinger equation (Eq.~(\ref{eq:schroed})). Unless the Hamiltonian 
contains one-body operators only, the mean-field theory can only 
approximate the exact evolution of the system.

\subsubsection{Decomposition of the Hamiltonian on particle-hole (p-h) basis}
To 
precise the missing part, we complete the occupied states by a set (possibly infinite) of unoccupied single-particle
states (also called particle states) labeled by $\bar \alpha$ (associated to 
the creation/annihilation $\oad_{\bar \alpha}$ and  $\oa_{\bar \alpha}$). 
The completed basis verifies
\begin{eqnarray}
\sum_\alpha \left| \alpha \right\rangle \left\langle \alpha \right| + 
\sum_{\bar \alpha} \left| \bar \alpha \right\rangle \left\langle \bar \alpha \right| \equiv \oro + (1-\oro) = \hat{1}.
\end{eqnarray} 
From the above closure relation, any creation operator associated to a single-particle state 
$\left| i \right\rangle$ decomposes 
as 
\begin{eqnarray}
\oad_i &=& \sum_\alpha \oad_\alpha \left\langle  \alpha \left.  \right| i \right\rangle + 
\sum_{\bar \alpha} \oad_{\bar \alpha} \left\langle  \bar \alpha \left.  \right| i \right\rangle.
\label{eq:oadiph}
\end{eqnarray}
The particle-hole basis is particularly suited to express any operator applied to the state $\kfi $ due 
to the properties
\begin{eqnarray}
\hat a^\dagger_\alpha \kfi = \hat a_{\bar \alpha} \kfi =0.
\label{eq:ph}
\end{eqnarray} 
For instance, 
restarting from the general expression of $\oH$ (Eq.~(\ref{eq:oH2})) the different single-particle states $(i,j,k,l)$ can be expressed in the 
particle-hole basis (Eq.~(\ref{eq:oadiph})).
Then, using anti-commutation relations (\ref{eq:anticom1}) et (\ref{eq:anticom2}), $\bar{v}_{ijkl}=-\bar{v}_{jikl}=-\bar{v}_{ijlk}=\bar{v}_{jilk}$
and $\Tr_{12}\(\bar{v}_{12}\ro_1\ro_2(1-P_{12})\)=2\Tr_{12}(\bar{v}_{12}\ro_1\ro_2)$ 
as well as (\ref{eq:oro_Slater}),  (\ref{eq:Uro}), (\ref{eq:h})  et (\ref{eq:ph}),
we finally end with
\begin{eqnarray}
\begin{array} {lllcr}
\oH \kfi &=& \Big\{ 
E[\ro]+ \sum_{\bar \alpha \alpha} \sdf h[\rho]_{\bar{\al}\al}\sdf
\hat a_{\bar{\alpha}}^{\dagger} \hat a_{\alpha} \hspace*{1.cm} &  \Longleftrightarrow &\hat H_{MF}[\rho]
 \\
 \\
&&+ \frac{1}{4} \sum_{\bar \alpha  \bar \beta \alpha \beta }  
\sdf {\bar v}_{\bar{\al}\bar{\be}\be \al}  \sdf
a^\dagger_{\bar \alpha} a^\dagger_{\bar \beta} a_{\beta} a_{\alpha} & \Longleftrightarrow & \oV_{res}[\rho] \\
&&\Big\}\kfi 
\end{array}
\label{eq:hphi}
\end{eqnarray}
where $E[\ro] = \bfi \oH_{MF} \kfi = \Tr \[\rho \( t+\frac{1}{2}U[\rho] \)\]$
corresponds to the 
Hartree-Fock energy.

The above 
expression is helpful 
to understand the approximation made at the mean-field level.
In previous section, 
we have shown that mean-field provides the best approximation for one-body degrees of freedom using the Ehrenfest theorem. In appendix \ref{annexe:Thouless}, we show that 
the mean-field evolution can be deduced using the Thouless theorem \cite{tho61} and 
an effective Hamiltonian where $\oV_{res}$ is neglected.

\subsubsection{Limitation of the mean-field theory}

In expression (\ref{eq:hphi}), a clear separation is made between 
what is properly treated at the mean-field level ($E_0[\rho]$ and 
$\hat H_{MF}[\rho]$) and what is neglected, i.e. $\hat V_{res}[\rho]$. The latter is called 
'residual interaction'. At this point several comments are in order:
\begin{itemize}
\item[$\bullet$] The validity of the mean-field approximation depends on the intensity of the residual interaction
which itself depends on the state $\left| \Phi \right\rangle$ and therefore will significantly depend 
on the physical situation. Starting from simple arguments \cite{lic76}, 
the time $\tau_{SD}$ over which the Slater determinant picture breaks down could be expressed as:
\begin{eqnarray}
\tau_{SD} &=& \frac{\hbar}{2}  \Big(\frac{1}{N}
\sum_{\bar \alpha \bar \beta \alpha \beta} 
|\left\langle \bar \alpha \bar \beta \left| \bar v \right| \alpha \beta \right\rangle|^2 \Big)^{-1/2}.
\end{eqnarray} 
In the nuclear physics context, typical values of the residual interaction leads to $\tau_{SD} \simeq 100-200$ fm/c.
Therefore, even if the starting point is given by an independent particle wave-packet, the exact 
evolution will deviate rather quickly from the mean-field dynamics. This gives strong arguments in favor
of theories 
beyond TDHF in nuclear physics. 
\item[$\bullet$] An alternative expression of the residual interaction which is valid in any basis, is
\begin{eqnarray}
V_{res}[\rho]_{12} &=& \frac{1}{4} ( 1-\rho_1 )( 1-\rho_2 ) \bar v_{12} \rho_1 
\rho_2 .
\label{eq:vres}
\end{eqnarray}  
This expression illustrates that the residual interaction associated to a Slater determinants could be seen as 
a "dressed" interaction which properly accounts for the Pauli principle. Physically, the residual interaction corresponds
to direct nucleon-nucleon collisions between occupied states (2 holes) which could only scatter toward unoccupied
states (2 particles) due to Pauli blocking. 
We say sometimes that the residual interaction has a 2 particles-2 holes (2p-2h) nature.      
\end{itemize}    
Due to the residual interaction, the exact many-body state will decompose in a more and more complex 
superposition of Slater determinants during the time evolution.
 As stressed in the introduction of this lecture, due to the complexity 
of the nuclear many-body problem, the exact dynamics is rarely accessible. In the following section, methods 
to include correlations beyond mean-field, like direct nucleon-nucleon collisions or pairing, are discussed.

\subsection{General correlated dynamics: the BBGKY hierarchy}

Using the Ehrenfest theorem (section \ref{subsubsec:Ehrenfest}), we have shown that the mean-field theory is particularly 
suited to describe one-body degrees of freedom. A natural extension of mean-field consists in following 
explicitly two-body degrees of freedom. 
Considering now the Ehrenfest theorem for the one and two-body degrees of freedom 
leads to two coupled
equations for the one and two-body density matrix components 
$\rho^{(1)}_{ij} = 
\langle \oa^\dagger_j \oa_i \rangle$ and $\rho^{(2)}_{ij kl} = 
\langle \oa^\dagger_l \oa^\dagger_k \oa_i \oa_j \rangle$ 
\begin{equation}
\left\{ 
\begin{array}{cl}
i\hbar \frac{\partial }{\partial t}\rho _{1}
=&\left[ t_1,\rho _{1}\right] + \frac{1}{2}{\rm \Tr}_{2}\left[ \bar v_{12},\rho_{12}\right]  \\ 
&  \\ 
i \hbar \frac{\partial }{\partial t}\rho_{12} =& [t_1 + t_2 + \frac{1}{2} \bar v_{12}, \rho_{12}] 
+ \frac{1}{2} \Tr_3 \left[ \bar v_{13} + \bar v_{23} , \rho_{123} \right]. \\ 
\end{array}
\right.  \label{eq:BBGKY}
\end{equation}
Above equations are the two first equation of a hierarchy of equations, known 
as the Bogolyubov-Born-Green-Kirkwood-Yvon (BBGKY) hierarchy~\cite{bog46,bor46,kir46}
where the three-body density evolution is also coupled to the four body density and so on and so 
forth. Here, we will restrict to the equations on $\rho^{(1)}$ and $\rho^{(2)}$ which have often served as
the starting point to develop transport theories beyond mean-field 
\cite{cas90,rei94,abe96,lac04}.

\subsection{The Time-Dependent Density-Matrix Theory}
Previously, we have shown that the mean-field dynamics neglect two-body and higher correlations. 
Then the equations on $\rho^{(1)}$ reduce to TDHF. A natural extension which includes two-body effects 
is to treat explicitly two-body correlations and neglect only three-body ($C_{123} = 0$) 
and higher correlations\footnote{Introducing 
the permutation operator $P_{12}$ between two particles, defined as $P_{12} \left| ij \right\rangle = 
\left| ji \right\rangle$. The two-body correlation matrix is given by:
\begin{equation}
C_{12}=\rho _{12}-\rho _{1}\rho _{2}(1-P_{12})  \label{eq:c2}
\end{equation}
while the three-body correlations $C_{123}$ reads 
\begin{equation}
\begin{array}{ll}
C_{123}= & \rho _{123}-\rho _{1}C_{23}\left( 1-P_{12}-P_{13}\right) -\rho
_{2}C_{13}\left( 1-P_{21}-P_{23}\right) \\ 
& -\rho _{3}C_{12}\left( 1-P_{31}-P_{32}\right) -\rho _{1}\rho _{2}\rho
_{3}\left( 1-P_{13}\right) \left( 1-P_{12}-P_{23}\right).
\end{array}
\label{eq:c3}
\end{equation}}.  
The resulting theory where the one-body density~$\rho_1$ and the two-body correlation~$C_{12}$ are followed in time is generally 
called Time-Dependent Density-Matrix (TDDM) theory (see for instance \cite{cas90}) 
\begin{equation}
\left\{ 
\begin{array}{cll}
i\hbar \frac{\partial }{\partial t}\rho _{1}
=&\left[ h_{1}[\rho],\rho _{1}\right] + \frac{1}{2}{\rm Tr}_{2}\left[ \bar v_{12},C_{12}\right]  \\ 
&  \\ 
i\hbar \frac{\partial }{\partial t}C_{12}
=&\left[ h_{1}[\rho]+h_{2}[\rho],C_{12}\right] \\
&~+\frac{1}{2} \Big\{( 1-\rho_{1})(1-\rho _{2}) \bar v_{12} \rho _{1}\rho _{2}-%
\rho _{1}\rho _{2} \bar v_{12} ( 1-\rho _{1})(1-\rho _{2}) \Big\} & \Longleftrightarrow { B}_{12} \\
&~+ \frac{1}{2} \Big\{ \left( 1-\rho _{1}-\rho _{2}\right) \bar{v}_{12}C_{12}-C_{12}\bar{v}_{12}
\left( 1-\rho_{1}-\rho _{2}\right) \Big\} 
& \Longleftrightarrow {P}_{12}  \\
& \begin{array} {l}
+ {\rm Tr}_{3}\left[ \bar{v}_{13},\left( 1-P_{13}\right) \rho _{1}C_{23}\left(
1-P_{12}\right) \right]  \\
+ {\rm Tr}_{3}\left[ \bar{v}_{23},\left( 1-P_{23}\right) \rho _{1}C_{23}\left(
1-P_{12}\right) \right] .
\end{array} & \Longleftrightarrow {H}_{12}
\end{array} \right. 
\label{eq:tddm} 
\end{equation}
where we have dissociated explicitly three terms which 
will be responsible for the build up of correlations in time. 
The Born term, $B_{12}$, contains the physics of direct in-medium nucleon-nucleon 
collisions. Comparing $B_{12}$ and Eq.~($\ref{eq:vres}$), we see that it is directly proportional 
to the residual interaction. Indeed, starting from a Slater determinant 
($C_{12}(t_0) = 0$), this is the only term that do not cancel out in the evolution of $C_{12}$
over a short time scale. In particular, it will be responsible for the departure from an independent 
particle picture.
The physical interpretation of the term ${P}_{12}$ and ${H}_{12}$ is less straightforward. 
For instance, it has been 
shown that ${P}_{12}$ could be connected to pairing correlations  \cite{toh04} (see discussion below) while 
$H_{12}$ contains higher order $p-p$ and $h-h$ correlations. It is finally worth mentioning, that the last term could eventually be 
modified to better account for conservation laws (see discussion in \cite{pet94}).

Application of the TDDM theory faces two major difficulties. First, since 
we are considering explicitly two-body degrees of freedom, we have to deal 
numerically with huge matrices and appropriate truncation schemes should be performed.
Second, numerical applications are mainly possible with contact interaction (Skyrme like).
These interactions, which are zero range in $r$-space are thus of infinite range in momentum
space. This unphysical behavior of the interaction is critical in practice, since during
nucleon-nucleon collisions, particles may scatter towards too high momentum. No clear solution
to this problem exists so far in the TDDM theory \cite{lac04}.    
Due to these difficulties, only  few applications have been carried out so far for
collective vibrations \cite{deb92,luo99,toh01,toh02b}, and very recently for nuclear
collisions \cite{toh02}.

\subsection{Link between TDDM and TDHFB}

The connection between the TDDM and TDHFB has been clarified in Ref. \cite{toh04}. Assuming a
separable correlation in the p-p and h-h channels given by equation (\ref{eq:ckk}) gives 
 \begin{eqnarray}
\frac{1}{2}\left< \lambda \left| \Tr_2 \left[\bar v_{12}, C_{12} \right] \right| \lambda' \right>
&=& \frac{1}{2} \sum_{kmn} \left< \lambda k \left| \bar v_{12} \right| mn \right>
\left< m n \left| C_{12} \right| \lambda' k \right>
- \frac{1}{2} \left< \lambda k \left| C_{12}  \right| mn \right>\left< m n \left|  v_{12}\right| \lambda' k \right>
\nonumber \\
&=& \Delta_{\lambda k} \kappa^*_{\lambda' k} - \kappa_{\lambda k} \Delta^*_{\lambda' k} = (\kappa \Delta^* - \Delta \kappa^*)_{
\lambda \lambda'}
\end{eqnarray}
where $\Delta$ is nothing but the pairing field introduced above. Then, the 
one-body density evolution reduces to
\begin{eqnarray}
i \hbar \frac{d}{dt}\rho =  \left[h[\rho], \rho \right] + \kappa\Delta^* - \Delta\kappa^* 
\end{eqnarray}
In Ref. \cite{toh04}, it has been shown that neglecting  $B$ and $H$ in the second equation 
of (\ref{eq:tddm}) leads to a TDHF like equation. Keeping only $P$ and assuming (\ref{eq:ckk}) leads to
\begin{eqnarray}
i\hbar \frac{d}{dt} C_{ijkl} &=& 
i\hbar \Big\{ \frac{d \kappa_{ij}}{dt} \kappa^*_{kl} 
+ \kappa_{ij} \frac{d \kappa^*_{kl}}{dt} \Big\} \nonumber \\
&=& \[\sum_m (h_{im} \ka_{mj}+h_{jm} \ka_{im})
+ \frac{1}{2} \sum_{mnpq}(\delta_{im}\delta_{jn}
-\delta_{im}\ro_{jn}-\delta_{jn}\ro_{im}) \bar{v}_{mnpq} \ka_{pq}\]\ka^*_{kl} \nonumber \\
&&- \ka_{ij}\[\sum_m ( \ka^*_{ml}h_{mk}+ \ka^*_{km}h_{ml})
+ \frac{1}{2} \sum_{mnpq} \ka^*_{mn}\bar{v}_{mnpq} (\delta_{kp}\delta_{lq}
-\delta_{kp}\ro_{ql}-\delta_{lq}\ro_{pk})\]
\end{eqnarray}
where we have identified the  terms proportional to $\ka_{ij}$ et $\ka^*_{kl}$. 
We then get
\oeq
i\hb \sdf \partial_t \ka_{ij}= \sum_m (h_{im}\ka_{mj}+\ka_{im} h_{mj}^*)
+ \frac{1}{2} \sum_{mn} (\bar{v}_{ijmn}-\Sigma_p\sdf \ro_{ip}\bar{v}_{pjmn}-\Sigma_p\sdf\ro_{jp}\bar{v}_{ipmn})\ka_{mn}.
\ceq
Using the expression of the pairing field (Eq.~(\ref{eq:Delta})), we finally 
recover the TDHFB equation (\ref{eq:TDHFBkappa}).
The above equation does not insure that the correlation matrix remains separable during 
the time-evolution. However, assuming that $C_{ijkl}(t) \simeq \kappa_{ij}(t) \kappa^*_{kl}(t)$
is valid for all time, the equations of motion identify with the TDHFB equation. 
It is worth mentioning that the above technique gives an alternative derivation of the TDHFB equation starting from TDDM which in addition illustrates the physical content of $P$.    

\subsection{Extended and Stochastic Time-Dependent Hartree-Fock}
\label{sec:extended}

The pairing correlations become less important when the internal excitation 
of the system increases. In this case, direct nucleon-nucleon collisions are expected 
to dominate correlations beyond mean-field.
A possible way to treat the latter and    
to escape part of the complexity of TDDM is to focus on
the one-body density while the effect of correlations on this quantity is only
treated approximatively. 
This could be done by neglecting the terms $P$ and $H$ in
the evolution of $C_{12}$~\cite{won78,won79,dan84,bot90,ayi80}. Then, the equation for the
two-body correlations takes a simple form
\begin{equation}
i\hbar \frac{\partial }{\partial t}C_{12}-\left[ h_{1}[\rho] + h_{2}[\rho] ,C_{12}\right]
=B_{12}  \label{eq:c12f12}.
\end{equation}
Solving this equation
formally, we can develop the correlations over a time interval
from an initial time $t_{0}$ to a time $t$ as 
\begin{equation}
C_{12}(t)=-\frac{i}{\hbar }\int_{t_{0}}^{t} \stb \d s\stf  U_{12}\left( t,s\right)
B_{12}\left( s\right) U_{12}^{\dagger} \left( t,s\right) + \delta C_{12}(t)
\label{eq:c12t}
\end{equation}
where $U_{12}=U_{1}\otimes U_{2}$ represents the independent 
particle propagation of two
particles  with $\displaystyle U(t,s)=\exp
\left( -\frac{i}{\hbar }\int_{s}^{t}h[\rho (t^{\prime })]dt^{\prime }\right)$. 
In expression (\ref{eq:c12t}), the first term in the right hand side represents correlations
developed by the residual interaction during the time interval. The second
term describes propagation of the initial correlations $C_{12}(t_{0})$ from $%
t_{0}$ to $t$, i.e. $\delta C_{12}(t)=U_{12}(t,t_{0})C_{12}(t_{0})U_{12}^{\dag }(t,t_{0})$.
Reporting this expression in the evolution of $\rho_1$, we end up with a generalization of the TDHF 
theory (where we omit the label "1" in $\rho_1$)
\begin{equation}
i\hbar \frac{\partial }{\partial t} \rho = [h [\rho],\rho] + K[\rho] + \delta K(t)
\label{eq:ESTDHF}
\end{equation}
where $K[\rho]$, called {\it collision term}, reads 
\begin{eqnarray}
K[\rho] &=& -\frac{i}{\hbar }\int_{t_{0}}^{t}\stb \d s\stf {\rm Tr}_{2}[v_{12},U_{12}(t,s)B_{12}(s)U_{12}^{\dagger }(t,s)] 
\label{eq:k}
\end{eqnarray}
while $\delta K(t)$ is given by
\begin{eqnarray}
\delta K(t) =& \Tr_{2}[v_{12},\delta C_{12}(t)] .
\label{eq:dk} 
\end{eqnarray}
The term $\delta K(t)$, which accounts for the initial correlation $C_{12}(t_0)$, 
dominates around $t_0$. This term, in principle, contains all order correlations
that are accumulated up to time $t_0$, and it is a priori a very complicated quantity.
Its treatment is clearly out of the scope of a one-body transport theory and 
statistical assumption is generally made on the initial 
correlations. 
It is assumed that the exact two-body correlations
accumulated until $t_0$ exhibits random fluctuations. As a result, the
average value of the initial correlations vanishes. This
assumption is known as the "molecular chaos assumption" in classical
transport theory and it corresponds to factorization of two-particle
phase-space density before each binary collision \cite{kad62,hua62}. 
Eq. (\ref{eq:ESTDHF})  is then replaced by an ensemble of one-body 
evolutions
\begin{equation}
i\hbar \frac{\partial }{\partial t} \rho^n = [h [\rho^n],\rho^n] + K[\rho^n] + \delta K^n(t)
\label{eq:ESTDHFsto}
\end{equation}
where $"n"$ now refers to the particular stochastic path and  
where $\delta K^n(t)$ is a fluctuating operator which vanishes in average. 
The complete statistical description of $\delta K^n(t)$ can be found in 
Refs.~\cite{ayi01,lac04}. Eq. (\ref{eq:ESTDHFsto}) is the starting point of 
most of the microscopic transport theories that are applied nowadays in Heavy-Ion 
reactions at intermediate energies. However, due to its complexity, mainly 
the semi-classical version of Eq. (\ref{eq:ESTDHFsto}), known as Boltzmann Langevin theory, 
has been applied to realistic situations (for a recent review, see \cite{cho04}). Such 
a semi-classical approximation is not expected to apply in low energy nuclear reactions
and we will concentrate here on its quantum version.

\subsubsection{Average evolution: irreversible process in Extended TDHF}

We first concentrate on the average evolution and illustrate advantages 
of the introduction of a collision term on top of the mean-field dynamics.
Averaging Eq. (\ref{eq:ESTDHF}) over different trajectories leads to an equation 
where $\rho^n$ is replaced by the average one-body density, denoted by $\overline \rho$
and where only the collision term $K[\rho]$ remains. The resulting theory 
is called Extended TDHF with a non-Markovian collision term (or with "memory effects"). 
The terminology "non-Markovian" (in opposition to "Markovian") comes from the fact that the system 
at time $t$ depends not only on the density at time $t$ but also on its full history 
due to the presence of a time integral in Eq. (\ref{eq:k}).

\begin{figure}[tbhp]
\begin{center}
\epsfig{file=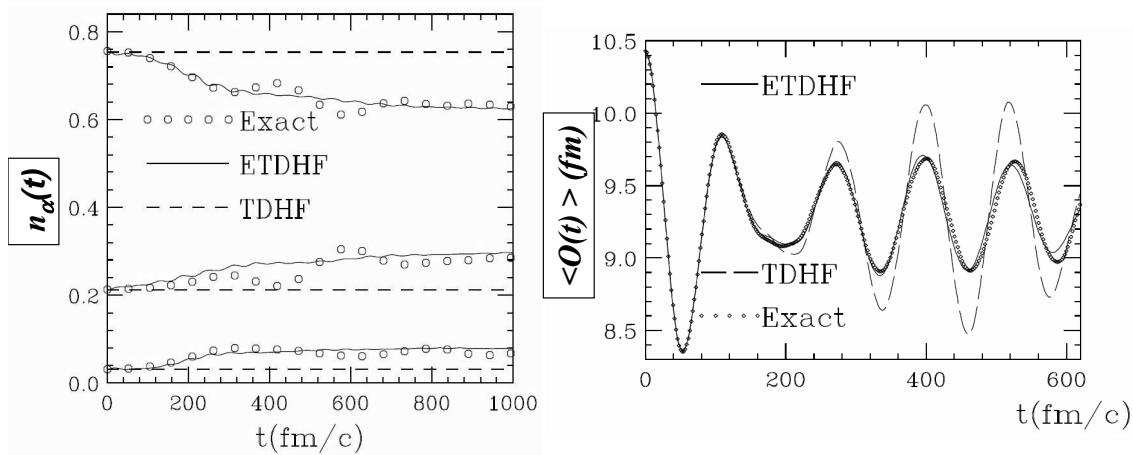,height=15.cm,angle=-90.}
\end{center}
\caption{Exact (circles), TDHF (dashed line) and Extended TDHF (solid line) 
evolutions of a two nucleon correlated system. Left: Occupation numbers, right: one-body centroid (adapted from 
\cite{lac99}.}
\label{fig:etdhf}
\end{figure}

At any time, eigenstates of 
$\overline \rho(t)$, denoted by $\left| \alpha (t) \right\rangle$ could be found. 
In this basis, called hereafter natural basis or canonical basis, 
$\overline \rho(t)$ reads
\begin{eqnarray}
\overline \rho(t) &=& \sum_\alpha \left| \alpha(t) \right\rangle n_\alpha(t) \left\langle \alpha(t) 
\right|.
\label{eq:averrho}
\end{eqnarray} 
Using the weak coupling approximation in combination with the first order
perturbation theory, the ETDHF equation can be transformed into a generalized
master equation for occupation numbers which accounts for the Pauli principle  
\begin{equation}
\frac{d}{dt}n_{\alpha }(t)=\int_{t_{0}}^{t} \stb \d s \stf\left\{ \left( 1-n_{\alpha
}\left( s\right) \right) {\mathcal{W}_{\alpha }^{+}}\left( t,s\right)
-n_{\alpha }\left( s\right) {\mathcal{W}_{\alpha }^{-}}\left( t,s\right)
\right\}. 
\label{eq:master}
\end{equation}
where the explicit form of the gain $\mathcal{W}_{\lambda }^{+}$ and loss $\mathcal{W}_{\lambda
}^{-}$ kernels can be found in Ref.~\cite{lac99}. Therefore, in contrast to TDHF where occupation
numbers are constant during the time evolution, in ETDHF the $n_\alpha$ evolve and could
eventually relax toward equilibrium. Such a relaxation is the only way to properly account 
for thermalization process in nuclei. The inclusion of correlation effect with 
Extended TDHF has been tested in the simplified case of two interacting nucleons in one dimension~\cite{lac99}.
In this case, the exact dynamics could be solved numerically. In Fig.~\ref{fig:etdhf},
starting from an initially uncorrelated state, the exact evolution of single-particle 
occupation numbers is compared to the Extended TDHF prediction. 
Fig. \ref{fig:etdhf}, shows that the Extended TDHF is able to reproduce fairly well the evolution 
of one-body occupation numbers and one-body observables over long time. This result is     
very promising and indicates that Extended TDHF seems to be an appropriate theory for the description
of dissipation when the residual interaction is weak. The above application has also demonstrated the 
importance of memory effects to properly describe quantum interacting systems although it significantly 
increases numerical efforts.

\subsubsection{Discussion on Stochastic mean-field dynamics}

Due to the underlying hypothesis of Extended TDHF, such an average theory is expected 
to apply only over short time scale. The long time dynamics requires to include 
both dissipation and fluctuations and therefore to solve explicitly the stochastic version 
of the transport theory (Eq. (\ref{eq:ESTDHFsto})). Up to now, stochastic mean-field has been 
essentially applied in the semi-classical limit neglecting explicitly non-Markovian effects.
Only recently, the theory proposed in Ref.~\cite{ayi01} was used to study small amplitude
collective vibrations in quantum system \cite{lac01,lac04} including full memory effects.
In that case, the description of damping of giant resonances is greatly improved by 
including both dissipative and fluctuating kernels. However, due to numerical as well 
as conceptual difficulties, the application of the above stochastic quantum mechanics to large amplitude 
collective motion, like nuclear collisions, remains an open problem. 

At this point, it is worth mentioning that extensive
works have been devoted to the formal derivation of dissipative quantum mechanics \cite{kad62}
and/or related stochastic equations for fermions,
including Markovian and non-Markovian effects \cite{ayi80,bal81,ayi88,ayi01,rei92a,rei92b,ohn95}.
These approaches have in common that the
residual part of the interaction introduces disorder on top of the mean field. Such theories end with 
rather complex transport equations which are hardly applicable in realistic situations like Heavy-Ion collisions. 
Connections have been made recently between the dissipative equations of the one-body
density and quantum jumps between Slater determinants using either perturbation theory or the connection 
between ETDHF \cite{rei92a,rei92b} 
and Lindblad equations \cite{lac06a} generally found in the theory of open 
quantum systems \cite{bre02}.  These theories, in which the many-body density are explicitly replaced 
by an average over Slater determinant densities, i.e. 
$D = \overline{\left| \phi^{n} \right\rangle \left\langle \phi^{n} \right|}$ where each Slater determinant evolves according to a Stochastic TDHF equation, open new perspectives.

\subsection{Functional integrals and exact treatment of Many-Body correlated systems with Stochastic 
mean-field theories}

\subsubsection{General Discussion}

Starting from a simple Slater determinant state 
$\left| \psi(t = 0) \right\rangle  = \left| \phi(t_0) \right\rangle$, 
 correlations will develop
in time and we do expect that the exact Many-Body state writes
\begin{eqnarray}
\left| \psi(t) \right\rangle  = \sum_k c_k (t) \left| \phi_k(t) \right\rangle
\end{eqnarray}   
where $\left| \phi_k \right\rangle$ denotes a complete (eventually time-dependent) 
basis of Slater-Determinant states. Accordingly, the many-body density writes 
(see appendix \ref{annexe:densite})
\begin{eqnarray}
\oD(t) &=& \sum_{k,k'} c_k (t)c^*_{k'} (t) \left| \phi_k(t) \right\rangle \left\langle \phi_{k'}(t) \right|.
\label{eq:exact}
\end{eqnarray} 
The extended and stochastic version of TDHF presented in sections   
\ref{sec:extended}, implicitly assume that the many-body density can be 
properly approximated by its diagonal components \cite{rei92a,lac06a} 
\begin{eqnarray}
\oD(t) &\simeq& \sum_{k} P_k  \left| \phi_k(t) \right\rangle \left\langle \phi_k(t) \right|
\end{eqnarray} 
where $P_k = |c_k (t)|^2$. Then a probability $P_k$ obeys master equations which eventually 
could be simulated by quantum jumps. The resulting density is obtained through the average
over different stochastic paths, {\it i.e.}
\begin{eqnarray}
\oD(t) & \simeq & \overline{\left| \phi_k(t) \right\rangle \left\langle \phi_k(t) \right|}.
\end{eqnarray}      
Physically, this could be understood as follows. The irrelevant degrees
of freedom (complex internal degrees of freedom) interact with the relevant degrees of freedom
(single-particle degrees of freedom) and induce a fast decay towards zero of the 
off-diagonal matrix elements. 
This phenomenon is known as a decoherence process \cite{kue73,kie03}.
It is clear, that such an approximation is associated with a loss of some quantum effects such as interferences
between different channels. Therefore we do expect that most of the extensions of TDHF presented above, as TDHF, 
will miserably fail to account for most of the true Many-Body quantum effects.

The goal of the present section is to demonstrate that one could always treat exactly the Many-Body density given by 
Eq. (\ref{eq:exact}) using an appropriate stochastic process between Slater determinants. Since there is no free 
lunch, the exact density is then obtained by an average 
\begin{eqnarray}
\oD(t) & \simeq & \overline{\left| \phi_k(t) \right\rangle \left\langle \phi_k'(t) \right|}
\end{eqnarray}  
where states in the left differ from states on the right.

\subsubsection{Functional Integral technique in a model case}

Functional integrals techniques have often been used to replace the exact Many-Body problem 
by an average over different "effective" one-body problem \cite{lev80a,lev80b,neg88}.
In Ref.~\cite{koo97}, the general strategy to obtain ground state properties
of a many-body system using Monte-Carlo methods, the so called Shell-Model Monte-Carlo, is described. 
Recently, this technique has been combined with mean-field theory to obtain Stochastic TDHF equations 
which in average leads to the exact evolution \cite{car01,jui02}. We give here a brief description of 
the method.

\subsubsection{Simple Introduction to functional integrals}

We again consider that at a given time, the Many-Body state is a Slater Determinant  
$\left| \psi(t)  \right\rangle= \left| \phi \right\rangle$.  For a short time step
$\Delta t$, we have 
\begin{eqnarray}
\left| \psi(t+ \Delta t) \right\rangle &=& \exp \left( \frac{\Delta t}{i\hbar} \hat H \right)
\left| \phi (t) \right\rangle \simeq  \Big(1+ \frac{\Delta t}{i\hbar} \hat H + o(\Delta t) \Big) \left| \phi(t) \right\rangle.
\end{eqnarray} 
Due to the presence of a two-body interaction in $\hat H$, the state 
$\left| \psi(t+ \Delta t) \right\rangle$ differs from
a Slater Determinant. 
Let us show how it can be written as a sum of 
Slater Determinants. For simplicity, we 
first assume that $\hat H = \hat H_1 + \hat O^2$, where $\hat H_1$ and $\hat O$ are both one-body operators. Therefore $\hat O^2$ corresponds to a two-body interaction.
We introduce the notation $G(x)$ 
for a normalized gaussian probability of the variable $x$ with mean zero and variance $1$
\begin{eqnarray}
\overline{x} = \int_{-\infty }^{+\infty }\stb \d x \stf x G(x)   = 0, ~~~~~~~~ 
\overline{x^2} = \int_{-\infty }^{+\infty } \stb \d x \stf x^2 G(x)  = 1.
\end{eqnarray}
We define the complex number $\Delta \omega \equiv \sqrt{\frac{2\Delta t}{i\hbar}}$ 
and the one-body operator $ \hat S(\Delta t , x)$ written as 
\begin{eqnarray}
\hat S(\Delta t , x) \equiv \frac{\Delta t}{i\hbar} \hat H_1 + x \, \Delta \omega \, \hat O .
\end{eqnarray}
Considering the average value of $\hat S(\Delta t , x)$ and keeping 
only terms up to $\Delta t$, we obtain
\oeq
\int_{-\infty}^{+\infty}\stb \d x \stf e^{\hat S(\Delta t , x)} G(x) = 1 + \frac{\Delta t}{i\hbar} \hat H_1 
+ \overline{x} ~\Delta \omega \hat O + \frac{1}{2} \overline{x^2} ~(\Delta \omega)^2 \hat O^2  + o(\Delta t) 
= 1 + \frac{\Delta t}{i\hbar} \hat H + o(\Delta t) .
\ceq
By averaging over the different realizations of $x$, we recover the exact propagator over a short time step. Note 
that more general relations could be found using the Hubbard-Stratonovish transformation 
(see for instance \cite{{koo97}}).
Using the above relation, we see that
\begin{eqnarray}
\exp \left( \frac{\Delta t}{i\hbar} \hat H \right) \left| \phi (t) \right\rangle = 
\int_{-\infty}^{+\infty} \stb \d x \stf G(x) e^{\hat S(\Delta t , x)}\left| \phi(t) \right\rangle \equiv 
\int_{-\infty}^{+\infty} \stb \d x \stf G(x) \left| \phi_x (t+ \Delta t) \right\rangle .
\label{eq:evol_exact}
\end{eqnarray} 
Due to the one-body nature of $\hat S$, each $\left| \phi_x (t+ \Delta t) \right\rangle$ is a Slater determinant. Indeed, according to the Thouless theorem \cite{tho61} (see appendix 
\ref{annexe:Thouless}, an exponential of a one body operator transforms a Slater determinant 
into another Slater determinant.
We have therefore demonstrated that the evolution of the exact state could be replaced by an ensemble of Slater determinants. 
The method could then be iterated for each $\left| \phi_x (t+ \Delta t) \right\rangle$ 
to obtain the long-time evolution as a superposition of independent particle states.

Several comments are in order:
\begin{itemize}
\item[$\bullet$] Since $\hat S(\Delta t, x)$ is not a priori Hermitian, the dynamics does not preserves the orthogonality 
of the single-particle wave-function. Such a non-orthogonality should properly be treated during the 
time evolution \cite{jui02,lac05}.
\item[$\bullet$] Starting from a Many-Body density written as $D(t) = \left| \phi \right\rangle \left\langle \phi \right|$, at an 
intermediate time, the average density writes 
\begin{eqnarray}
D(t) = \overline{\left| \phi_1 (t) \right\rangle \left\langle \phi_2 (t) \right|} 
\end{eqnarray}  
with two states, where $\left| \phi_1 \right\rangle$ evolves according to Eq. (\ref{eq:SSEl}) while $\left\langle \phi_2 \right|$
evolves according to
\begin{eqnarray}
\left\langle \phi_2(t+ \Delta t) \right| &=& 
\left\langle \phi_2(t) \right| \exp\left\{ -\frac{\Delta t}{i\hbar} \hat H_1 + y \Delta \omega^* \hat O \right\}
\label{eq:SSEl} 
\end{eqnarray} 
where $ y $ is a noise independent of $x$, with mean zero and $\overline{y y} = 1$. Since the 
evolution is exact, any one-, two- or k-body observable $\hat A$ estimated through $\langle  \hat A \rangle \equiv 
\Tr(D(t) A)$ will follow the exact dynamics.
\end{itemize}

\subsubsection{General two-body Hamiltonian}

In previous section, we have illustrated how the exact dynamics of a system could be 
replaced by a set of stochastic evolutions of several Slater 
determinants using a schematic two-body 
Hamiltonian. We now start from the general expression of the Hamiltonian  
(\ref{eq:hphi})  to introduce the exact {\it Stochastic Mean-field} (SMF) technique.
In this expression, the Hamiltonian is naturally splitted into  a mean-field part
$\hat H_{MF}$ and a two-body residual interaction $\hat V_{res}[\rho]$. It turns out that this interaction can 
always be decomposed as a sum of square of one-body operators \cite{koo97}
\begin{equation}
\left\langle \bar{\alpha} \bar{\beta} 
\left| \hat{\bar v}_{12} \right| \alpha \beta \right\rangle = 
\sum_{\Lambda 
} c_\Lambda \langle \bar{\alpha}%
| \oO_{\Lambda } | \alpha \rangle \langle \bar{\beta}
| \oO_{\Lambda } | \beta  \rangle,
\label{eq:os1}
\end{equation}
where $\oO_\Lambda$ is a one-body operator 
while $c_\Lambda$  denotes a set of constants  (eventually  complex). 
The discussion in the previous section can then easily be generalized. The residual interaction $\hat V_{res}$ factorizes as
\begin{eqnarray}
\hat V_{res} = \frac{1}{4} \sum_\Lambda c_\Lambda \hat O^2_\Lambda   
\end{eqnarray}
where $\hat O_\Lambda \equiv \sum_{\alpha \bar \alpha} 
\langle \bar{\alpha} | \oO_{\Lambda } | \alpha \rangle \oa^\dagger_{\bar{\alpha}} \oa_\alpha$. 
Therefore, for realistic interactions one should introduce as many stochastic Gaussian independent 
variables as the number of operators entering in the sum. In practice, this number defines the 
numerical effort which in general is very large. For this reason, only few applications to the dynamics 
of rather simple systems exist so far.   
Finally, the extension of above stochastic theories to HFB states can be found in Ref.~\cite{lac06b} while 
an explicit link with observables evolutions is studied in Refs.~\cite{lac07,lac05}.

\subsection{Summary}

In this section, we have summarized some of the possible ways to extend TDHF, some of them are able 
to incorporate pairing correlations (like TDHFB or TDDM) while others concentrate on direct nucleon-nucleon
collisions (ETDHF). Though promising, these theories to the nuclear 
many-body problem remain very challenging to apply.

A second difficulty which has been largely hidden in the present lecture is that all applications 
of dynamical quantum transport theories, like TDHF,  to nuclear reactions was possible only because of the 
introduction of effective interactions (essentially Skyrme like). 
These interactions have led to the more 
general concept of Energy Density Functional (EDF) and are expected, in a similar way as the Density Functional 
Theory (DFT) in condensed matter, to incorporate most of the correlations already at the mean-field level. 
Then, the notion of ''beyond mean-field'' calculations becomes ill defined. For instance, all theories 
(extended, stochastic, functional...) start from a Many-Body Hamiltonian. In the EDF context, such an 
Hamiltonian, although it exists, is not simply connected to the EDF itself. As a consequence, the Hamiltonian 
derivation could serve as a guideline but a proper formulation in the EDF framework is mandatory. Large debates
exist nowadays on the validity and foundation of the nuclear EDF applied to static properties of nuclei. Similar 
discussion should be made on what we should definitively call Time-Dependent EDF (TDEDF) and not ''TDHF''.


\vspace{1cm}
\centerline{\bf Acknowledgments}
\addcontentsline{toc}{section}{Acknowledgments}
\vspace{0.5cm}

These lecture notes are dedicated to the memory of Paul Bonche, 
one of the most famous pioneer
in the application of TDHF to nuclear physics 
and author of the 3D TDHF code that has been 
used in most of the applications presented here.

We also thank our collaborators working in nuclear structure and using similar techniques
based on mean-field theory for various discussions. Last, we thank Andrea Vitturi for
his contribution on the fusion part.

\newpage


\addcontentsline{toc}{section}{APPENDIX}

\appendix

\section{Basics of quantum mechanics}
\label{annexe:rappelsMQ}

Here, we summarize some aspects of quantum mechanics 
and more specifically {\it second quantization}. 
Only concepts useful for the the present lecture notes are introduced
(see \cite{mes59,bla86} for more details).

\subsection{Single-particle creation/annihilation operators}
\label{sec:prop_2ndeQ}

The single-particle wave-function 
$|i\>$, with spin and isospin components denoted respively by  $s$ 
and $\tau$ can be written as 
\oeq
\az_i^{s\tau}(\vr)=\<\vr s \tau |i\>,
\label{eq:fosp}
\ceq
The associated creation operator $\oad_i$ acting on the vacuum $\kvac$ verifies
\oeq 
\oad_i \kvac = |i\>.
\label{eq:ad}
\ceq
Its Hermitian conjugated is the annihilation operator 
$
\oa_i |i\> = \kvac.
$
For all annihilation operator, we have
\oeq 
\oa_i \kvac = 0 \stf \stf \stf \stf \stf \forall i.
\label{eq:annihilation_vide}
\ceq
This property defines the particle vacuum associated to the set of operators $\oad$ and $\oa$.

We consider here fermions. The Pauli principle imposes that 
two identical fermions could not be created in the same state 
 $\oad_i |i\>= 0$.
Creation/annihilation operators verifies anti-commutation rules
\oeqn 
\{ \oad_i, \oad_j\} = \{ \oa_i, \oa_j\} &=& 0 \label{eq:anticom1}\\
\{ \oad_i, \oa_j\} &=& \del_{ij} \label{eq:anticom2}
\ceqn
where $\{a,b\} = ab + ba$.

It could be useful to go from one single-particle basis to another,
and to express creation/annihilation of the first basis in terms of  
creation/annihilation of the second basis $\{|i\>\}$.
To do so, the two basis should obviously represent the same Hilbert 
space (and be complete with respect to this space).
Fort instance, the relation of any single-particle basis to the 
coordinate representation is 
\oeqn
\oa(\vr s\tau) &=& \sum_i \az_i^{s\tau}(\vr) \sdf  \oa_i \label{eq:ar} \\
\oad(\vr s\tau) &=& \sum_i {\az_i^{s\tau}}^*(\vr) \sdf \oad_i . \label{eq:adr}
\ceqn
Using the closure relation
$\sum_{s\tau}\int \sdb \d \vr \stf |\vr s \tau\> \< \vr s \tau | = 1$, 
previous relations can be inverted as
\oeqn
\oa_i &=& \sum_{s,\tau} \int \sdb  \d \vr \stf {\az_i^{s\tau}}^*(\vr) \sdf  \oa(\vr s \tau) \label{eq:anu}\\
\oad_i &=& \sum_{s,\tau} \int \sdb  \d \vr \stf {\az_i^{s\tau}(\vr)} \sdf  \oad(\vr s \tau) \label{eq:adnu} .
\ceqn

\subsection{$N$ identical particles  state}

We denote by  $\kpsi$ a $N$-body state which is {\it a priori} correlated.
Occupation numbers of a correlated state, in opposite to an uncorrelated state, 
have values which might differ from zero and one.
A $N$ independent particles  state corresponds to 
a specific case where $N$ occupation numbers 
are exactly one while the other are zero. 
We not such a state  $\kfi$. It can be written as an antisymmetric product
of single-particle states (called {\it Slater determinant} due to its specific form, see Eq. 
(\ref{eq:fonction}))
\oeq 
\kfi \equiv  | \phi _{\nu_1 \cdots \nu_N} \>  = \sqrt{N!} \sdf \hat{\mA} \, | 1:\nu_1, 2:\nu_2,...,N:\nu_N\>,
 \label{eq:etat_antisym}
\ceq
where $|1:\al,2:\be,\cdots \>$ means that the particle 1 is in the state $\kal$, particle 2 occupies state 
 $|\be\>$,..  The $\sqrt{N!}$ coefficient insures the proper normalization of the wave 
while the anti-symmetrization operator writes 
\oeq
\hat{\mA} = \frac{1}{N!} \sum_{\mbox{\small permutation }P} \mbox{sign}(P) \sdf P.
\ceq
Here $\mbox{sign}(P)=1$ (resp. -1) for odd (resp. even) 
permutations of particles. For instance, its action on a two-particle wave-function reads  
\oeq
\hat{\mA} \sdf |1:\al,2:\be\> = \frac{1}{2} \(|1:\al,2:\be\> - |1:\be,2:\al\>\).
\label{eq:A_2part}
\ceq
In second quantization, an independent particle state reads 
\oeq
 | \phi _{\nu_1 \cdots \nu_N} \>  = \left( \prod_{i=1}^N \sdf \oad_{\nu_i} \right) \sdf  \kvac.
\ceq

\subsection{Wick's Theorem}
\label{subannexe:wick}

The Wick's theorem enables to simply express expectation values of observables 
for specific states (that could be considered as vacua). Let us 
 first define the term {\it contraction}. 
Consider two creation and/or annihilation operators, denoted by $\oA$ and $\oB$ (these operators could also be 
linear combinations of creation/annihilation operators) and $|0\>$ the associated vacuum.
In practice, this vacuum could be a single-particle vacuum, a HF state denoted by $|\phi\>$ or a quasi-particle (HFB) vacuum
(see section \ref{subsec:TDHFB}).
The contraction $\overline{\oA\oB}$ is defined as the expectation value of $\oA\oB$
on the vacuum
\oeq
\overline{\oA\oB} = \< 0 | \oA \oB | 0 \>.
\ceq

Then, the Wick's theorem states:

{\it The expectation value of a product of creation/annihilation operators 
on their vacuum $|0\>$ is equal to the sum of all possible products 
of contractions of pairs of operators, each 
product of contractions being multiplied by + or - depending on the
parity  of the permutation needed to put together the contracted operators.}

Let us illustrate this on the overlap between two anti-symmetric two independent 
particles states 
\oeq
\< \phi_{\al \be} | \phi_{\mu \nu} \> = \<-| \oa_\be \oa_\al \oad_\mu \oad_\nu |-\> .
\ceq
Contractions are given by
$
\overline{\oa_i \oa_j} = \overline{\oad_i \oad_j} = 0 
$ and $
\overline{\oa_i \oad_j} = \<i|j\>.
$
We then deduce
\oeq
\< \phi_{\al \be} | \phi_{\mu \nu} \> = \overline{\oa_\be \oad_\nu} \sdf \overline{\oa_\al \oad_\mu} - \overline{\oa_\be \oad_\mu} \sdf \overline{\oa_\al \oad_\nu} 
= \<\be | \nu\> \, \<\al | \mu\> - \<\be | \mu\> \,  \<\al | \nu\>.
\ceq

This result can be generalized to the overlap between two $N$-particles Slater determinants constructed 
from two different basis $\{\knu \}$ and $\{ \kal\}$
\oeq
 \< \phi _{\nu_1 \cdots \nu_N}   | \phi _{\al_1 \cdots \al_N} \>  = \bvac \oa_{\nu_N} \cdots \oa_{\nu_1} \sdf  \oad_{\al_1} \cdots \oad_{\al_N} \kvac.
\ceq
We can then use the Wick's theorem to express this overlap in terms of a determinant 
made of overlaps between single-particle states  $\<\nu_j|\al_i\>$
\oeq
 \< \phi _{\nu_1 \cdots \nu_N}   | \phi _{\al_1 \cdots \al_N} \> = \sdf  \left|
 \begin{array}{ccc}
\overline{\oa_{\nu_1} \oad_{\al_1}} & \cdots & \overline{\oa_{\nu_N} \oad_{\al_1}} \\
\vdots & &\vdots \\
\overline{\oa_{\nu_1} \oad_{\al_N}}& \cdots & \overline{\oa_{\nu_N} \oad_{\al_N}} \\
\end{array}
\right|
=\sdf  \left|
 \begin{array}{ccc}
\<\nu_1|\al_1\> & \cdots &\<\nu_N|\al_1\>\\
\vdots & &\vdots \\
\<\nu_1|\al_N\>& \cdots & \<\nu_N|\al_N\> \\
\end{array}
\right|.
\label{eq:recouv}
\ceq

\subsection{Basis of $N$ particles states }

Starting from a single-particle complete basis, we can always construct 
a basis of $N$-body Slater determinants.     
For instance, we calculate the overlap between 
two Slaters formed from {\it a priori} different
single-particle states of the same basis.  
According to Eq. (\ref{eq:recouv}), we have  
\oeqn
 \< \phi _{\nu_1 \cdots \nu_N}   | \phi _{\nu'_1 \cdots \nu'_N} \> &=&\sdf  \left|
 \begin{array}{ccc}
\del_{\nu_1 \nu'_1} & \cdots & \del_{\nu_N \nu'_1} \\
\vdots & &\vdots \\
\del_{\nu_1 \nu'_N}& \cdots & \del_{\nu_N \nu'_N} \\
\end{array}
\right|.
\ceqn
This overlap is  $\pm 1$ if $ | \phi _{\nu_1 \cdots \nu_N} \>$ and $ | \phi _{\nu'_1 \cdots \nu'_N} \>$ contain exactly 
the same occupied states and $0$ if at least one of the single-particle states is different. 
Therefore, two different Slater determinants are orthogonal. It could be shown finally that the set of Slaters 
constructed by picking up all possible combination of $N$ single-particle states forms a complete basis 
of the $N$-body Hilbert space.
In other words, any  many-body state, correlated or not, can be written in the form 
\oeq
\kpsi = \sum_{\nu_1 \cdots \nu_N} \sdf C_{\nu_1 \cdots \nu_N} \sdf | \phi _{\nu_1 \cdots \nu_N} \> .
\label{eq:etat_correle}
\ceq

\subsection{Independent particles wave-functions}
\label{subsubsec:fonction}

The Wick's theorem is also helpful to express the wave function of 
a $N$ particles state. 
We want to write a wave function in position-spin-isospin representation. 
Associated single-particle states are 
 $|\xi_n\> \equiv |\vr_n s_n \tau_n\>$, where $\vr$ is the position, $s$ 
and $\tau$ the spin and isospin projections respectively. 
Using Eq.~(\ref{eq:etat_antisym}), a Slater 
built using this basis reads
\oeq
|\xi_1\cdots \xi_N\> = \sqrt{N!} \sdf \hat{\mA} \, | 1:\xi_1\cdots N:\nu_N\>.
\label{eq:base_Slater}
\ceq
The $N$ particles wave function is then given by
\oeq
\psi (\xi_1 \cdots \xi_N) = \frac{1}{\sqrt{N!}} \sdf \<\xi_1\cdots \xi_N\sdf  |\psi\> .
\label{eq:def_fo} 
\ceq
For an independent particle system, the wave function  reduces to  
\oeq
\phi_{\nu_1 \cdots\nu_N} (\xi_1 \cdots \xi_N) = 
\frac{1}{\sqrt{N!}} \sdf  \bvac \oa(\xi_N) \cdots \oa(\xi_1) \sdf \oad_{\nu_1} \cdots\oad_{\nu_N} \kvac.
\ceq
From Eq.~(\ref{eq:recouv}), we see that we simply obtain the standard Slater determinant 
formula 
\oeq
\phi_{\nu_1\cdots\nu_N} (\xi_1\cdots \xi_N) =   \frac{1}{\sqrt{N!}} \sdf  \left|
 \begin{array}{ccc}
\az_{\nu_1}(\xi_1) & \cdots & \az_{\nu_1}(\xi_N) \\
\vdots & &\vdots \\
\az_{\nu_N}(\xi_1) & \cdots & \az_{\nu_N}(\xi_N) \\
\end{array}
\right|.
\label{eq:fonction}
\ceq
Let us check that this wave-function is properly normalized. 
Introducing the notation $ \int \sdb \d \xi \equiv \sum_{s\tau}\int \sdb \d \vr$,
the nomalization reads
\oeqn
\mbox{Norm}[\phi]&=& \int \stb \d \xi_1\cdots \d \xi_N \stf \phi_{\nu_1...\nu_N}^*(\xi_1\cdots \xi_N) 
\sdf \phi_{\nu_1...\nu_N}(\xi_1\cdots \xi_N) \label{eq:def_norm}\\
&=& \frac{1}{N!} \sdf \int \stb \d \xi_1\cdots \d \xi_N \stf 
\left|
 \begin{array}{ccc}
{\az_{\nu_1}}^*(\xi_1) & \cdots & {\az_{\nu_1}}^*(\xi_N) \\
\vdots & &\vdots \\
{\az_{\nu_N}}^*(\xi_1) & \cdots & {\az_{\nu_N}}^*(\xi_N) \\
\end{array}
\right| \times 
\left|
 \begin{array}{ccc}
{\az_{\nu_1}}(\xi_1) & \cdots & {\az_{\nu_1}}(\xi_N) \\
\vdots & &\vdots \\
{\az_{\nu_N}}(\xi_1) & \cdots & {\az_{\nu_N}}(\xi_N) \\
\end{array}
\right|.
\ceqn
Developing  the determinants and using 
$\int \sdb \d \xi \sdf \az_i^*(\xi) \, \az_j(\xi)=\del_{ij}$, we finally get
\oeqn
\mbox{Norm}[\phi]
&=& \frac{1}{N!} \sdf \int \stb \d \xi_1\cdots \d \xi_N \stf  \sum_{\mbox{\small Permutation }P} \sdf P\{ \az_{\nu_1}^*(\xi_1) \az_{\nu_1}(\xi_1) \cdots \az_{\nu_N}^*(\xi_N) \az_{\nu_N}(\xi_N)\} \nonumber \\
&=&1.
\ceqn
Finally, it is worth to mention, that, using Eq. (\ref{eq:def_fo}) and (\ref{eq:def_norm}), 
the closure relation for the $N$ particles states reads 
\oeq 
\hat{1}_N = \frac{1}{N!} \sdf \int \stb \d \xi_1\cdots \d \xi_N \stf | \xi_1 \cdots \xi_N \>  \< \xi_1 \cdots \xi_N |.
\label{eq:fermeture}
\ceq

\section{One-body observables}
\label{annexe:1corps}

One-body observables are generally written as a sum of operators acting on each 
single-particle components $i$ independently, $i. e.$ 
\oeq
\oF = \sum_{i=1}^N \of{(i)}.
\label{eq:un_corps}
\ceq
In second quantization, one-body operators reads 
\oeqn
\oF = \sum_{ij} \sdf & \underbrace{\<i|\, \of \, |j\>} &\sdf \oad_i \oa_j. \label{eq:un_corpsb}
 \\
& f_{ij} & \nonumber
\label{eq:un_corps_2ndQ}
\ceqn
Both expressions lead to the same action on a $N$-body wave function. For instance, considering the independent 
particle case, using Eqs. (\ref{eq:un_corps}),
(\ref{eq:etat_antisym}), (\ref{eq:A_2part}) and the notation $\of |j\> = \sum_i f_{i j} |i\>$, we have  
\oeqn
\oF | \phi_{\nu_1 \cdots \nu_N} \> &=& \sum_{i=1}^{N} \sdf \of{(i)} \sdf\sqrt{N!} \sdf\hat{\mA}\sdf | 1: \nu_1, \cdots , N: \nu_N \> \nonumber \\
&=& \sum_{i=1}^{N} \sdf \sqrt{N!} \sdf\sum_\mu \sdf f_{\mu\nu_i}\sdf  \hat{\mA} \sdf| 1: \nu_1, \cdots,i:\mu,\cdots , N: \nu_N \> \nonumber \\
&=&  \sum_{i=1}^{N} \sdf \sum_\mu \sdf f_{\mu\nu_i}\sdf \oad_{\nu_1} \cdots \oad_{\nu_{i-1}} \sdf \oad_{\mu} \sdf \oad_{\nu_{i+1}} \cdots \oad_{\nu_N} \sdf \kvac.
\label{eq:methode1}
\ceqn
Eq.~(\ref{eq:un_corpsb}) leads to
\oeq
\oF | \phi_{\nu_1 \cdots \nu_N} \> = \sum_{\mu\nu}\sdf f_{\mu\nu} \oad_\mu \sdf 
\oa_\nu\sdf \oad_{\nu_1}  \cdots \oad_{\nu_N} \kvac .
\ceq
Only occupied  $\nu$ contribute due to Eq.~\ref{eq:annihilation_vide} and we get
\oeqn
\oF | \phi_{\nu_1 \cdots \nu_N} \> &=& \sum_{i=1}^{N}\sdf \sum_{\mu}\sdf f_{\mu\nu_i} \sdf\oad_\mu \sdf 
\sdf \oad_{\nu_1}  \cdots \oad_{\nu_{i-1}} \sdf (-1)^{i+1}\sdf  \oa_{\nu_i} \oad_{\nu_i} \sdf \oad_{\nu_{i+1}} 
\cdots \oad_{\nu_N}\sdf\kvac \nonumber\\
&=&  \sum_{i=1}^{N}\sdf \sum_{\mu}\sdf f_{\mu\nu_i}  \sdf 
\sdf \oad_{\nu_1}  \cdots \oad_{\nu_{i-1}} \sdf\oad_\mu \sdf \oad_{\nu_{i+1}} 
\cdots \oad_{\nu_N}\sdf\kvac
\ceqn
where we have also used Eqs.~(\ref{eq:anticom1}) and
(\ref{eq:anticom2}). This result is nothing but Eq.~(\ref{eq:methode1}).
The table \ref{tab:op_1corps} gives few examples of one-body operators commonly used 
in the nuclear reaction context.
\begin{table}[h]
\begin{center}
\begin{tabular}{|c|c|c|}
\hline
Observable & Standard form & Second quantization \\
\hline
&&\\
Center of mass position $\oR$ & $\frac{1}{N} \sum_{i=1}^{N} \ovr{(i)} $&$ \frac{1}{N} \sum_{s\tau}\int \sdb \d \vr \stf \vr \sdf \oad(\vr s \tau) \sdf \oa(\vr s \tau)$ \\
&&\\
\hline
&&\\
Center of mass momentum$\oP$ & $\sum_{i=1}^{N} \ovp{(i)}$ & $\sum_{s\tau}\int \sdb \d \vp \stf \vp \sdf \oad(\vp s \tau)\sdf \oa(\vp s \tau)$ \\
&&\\
\hline
&&\\
Particle number $\oN$ & $\sum_i \hat{1}(i)$  & $\sum_i \oad_i \oa_i $ \\
&&\\
\hline
&&\\
Monopole operator $\oQ_0$ &  $ \frac{1}{\sqrt{4\pi}} \sum_{i=1}^N \ovr(i)^2 $ & $ \frac{1}{\sqrt{4\pi}} \sum_{s\tau } \int \sdb \d \vr \stf r^2 \sdf \oad(\vr s \tau) \sdf \oa(\vr s \tau)$ \\
&&\\
\hline
&&\\
Quadrupole operator $\oQ_{20}$ &  $ \sqrt{\frac{5}{16\pi}} \sum_{i=1}^N \(2\oz(i) \!-\! \ox(i)\! -\!\oy(i)\) $ & $ \sqrt{\frac{5}{16\pi}}  \sum_{s\tau } \! \int \sdb \d \vr \, \( 2z\!-\!x\!-\!y\) \, \oad(\vr s \tau) \, \oa(\vr s \tau)$ \\
&&\\
\hline
\end{tabular}
\end{center}
\caption{Typical example of one-body operators.}
\label{tab:op_1corps}
\end{table}

\section{Density matrices}
\label{annexe:densite}

The $N$-body density matrix of a $N$ particle system (described by $\kpsi$) 
contains all the information on the system. It could be written as an operator 
acting on the $N$-body Hilbert space as 
$\oD = \kpsi \bpsi$ (see for example \cite{sur95,abe96,lac04}). 
If we are only interested in the $M$-body observables ($M\le N$),
we do not need all the information carried by $\hat{D}$. A natural way to reduce 
the information is to introduce the $M$-body density 
matrix of the state $\kpsi$. It could be defined from $\oD$ as 
\oeq
\oro^{(M)} = \frac{N!}{(N-M)!} \Tr_{M+1...N} \oD.
\ceq
Equivalently, the $M$-body density matrix components are defined by the 
relation 
\oeq
\ro^{(M)}_{\nu_1...\nu_M,\mu_1...\mu_M} = 
\<\nu_1...\nu_M|\oro^{(M)}|\mu_1...\mu_M\> 
= \bpsi \oad_{\mu_M}...\oad_{\mu_1}\sdf \oa_{\nu_1}...\oa_{\nu_M}\kpsi.
\label{eq:defroM}
\ceq
In coordinate space ($\xi\equiv\{\vr s \tau\}$), it writes 
\oeq
\ro^{(M)} (\xi_1...\xi_M, \xi'_1...\xi'_M)= \frac{N!}{(N-M)!} \sdf  \int \stb \d \xi_{M+1} ... \d \xi_N \stf 
\psi^*(\xi'_1...\xi'_M \xi_{M+1}  ... \xi_N) \sdf  \psi(\xi_1... \xi_N).
\ceq
Other notations, like $\ro^{(M)}\equiv \ro_{1...M} \equiv \ro(1...M)$, are also sometimes used.
The main advantage of the $M$-body density is that it contains all the information 
on the system as far as $M$-body or lower observables are concerned.
For instance, the expectation value of a $M$-body observable $\oO^{(M)}$ reads 
\oeqn
\<\oO^{(M)}\>_\psi &=&    \int \stb \d \xi_1... \d \xi_M \d \xi'_1...\d \xi'_M  \stf
 \ro^{(M)}(\xi'_1...\xi'_M,\xi_1...\xi_M)  \sdf O^{(M)}(\xi_1...\xi_M,\xi'_1...\xi'_M) \nonumber \\
&=& \Tr_{1...M} [\ro^{(M)} O^{(M)}].
\label{eq:TrroM}
\ceqn

\section{Two-body correlations}
\label{annexe:correl_2corps}

When we want to get the expectation value of a two-body observable, as it is the case for the 
Hamiltonian $\oH$, we do {\it a priori}
need the two-body density matrix $\ro^{(2)}$  of the system (see appendix~\ref{annexe:densite}). The two-body density matrix can always be decomposed into 
an uncorrelated part (anti-symmetric product of one-body densities)
and a correlation part, denoted by $C^{(2)}$ 
\oeq
{\ro_{{ijkl}}^{(2)}} = \< \oad_l \, \oad_k \, \oa_i\, \oa_j\>_\psi  
= {\ro_{{jl}}^{(1)}} \sdf {\ro_{{ik}}^{(1)}} - {\ro_{{il}}^{(1)}} \sdf {\ro_{{jk}}^{(1)}} + {C_{{ijkl}}^{(2)}}
\label{eq:ropsi}
\ceq
The correlation operator $C^{(2)}$ can be seen as the part of the two-body density 
which cannot be written as a product of single-particle operators. 
It is also possible to use the notation
\oeq
\ro_{12} = \ro_1 \ro_2 \(1-P_{12}\) + C_{12}
\label{eq:ro12}
\ceq
where $P_{12}$ corresponds to the permutation operator between particles 1 and 2.
We recall that notations $O^{(2)} \equiv O(1,2) \equiv O_{12}$ are equivalent. 
Note that higher $M$-body correlations  matrices (with $M \ge 3$) could be defined 
in a similar way \cite{lac04}.

\section{Hartree-Fock and quasi-particle vacua}
\label{annexe:videHF}
A Slater determinant is sometimes called HF vacuum.
The state $|\phi \>$ is indeed a vacuum for the creation operators $\obd_\mu $ and 
annihilation operators $\ob_\mu $ written, in the basis that has served to construct the many-body 
state, as
\oeqn
\obd_\mu &=& (1-n_\mu)\sdf  \oad_\mu +n_\mu \sdf\oa_\mu \label{eq:obd_mu}\\
\ob_\mu &=& (1-n_\mu) \sdf\oa_\mu +n_\mu \sdf\oad_\mu \label{eq:ob_mu}
\ceqn
where $n_\mu=1$ for occupied states (also called hole states) and 0 for unoccupied 
states (particle states). It could easily be checked that these new operators verify 
 Eqs. (\ref{eq:anticom1}) and  
(\ref{eq:anticom2}) and  that $|\phi\>$ is a vacuum for them, $i.e.$
\oeq
\ob_\mu |\phi\> = 0 \stf \stf \stf \stf \stf \forall \mu.
\label{eq:bmuvac}
\ceq 

Slater determinants are a specific case of quasi-particle vacua.
More generally,  quasi-particle creation/annihilation operators are defined 
through a linear combination of single-particle creation/annihilation operators 
$(\oad_i, \oa_i)$
\begin{eqnarray}
\left\{
\begin{array} {cc}
\obe_\alpha   = & \sum_{i} U^*_{i \alpha} \oa_i + V^*_{i\alpha} \oad_i \\
\obed_\alpha = & \sum_{i} U_{i\alpha } \oad_i + V_{i\alpha } \oa_i 
\end{array}
\right.
\label{transfoBogo}
\end{eqnarray}
where matrices $U$ et $V$ are chosen in such a way that the quasiparticle operators 
verify the fermionic anti-commutation rules (Eqs.  (\ref{eq:anticom1}-\ref{eq:anticom2})). 
The vacua associated to these operators can be written as 
\begin{equation}
\ket{\psi_{bogo}}={\cal C} \prod_{k}\hba_k \ket{-}
\end{equation}
where ${\cal C}$ is a normalization constant. We see that the latter expression insures  
$\obe_i \ket{\psi_{bogo}} = 0$.

\section{Two-body density for independent particle systems }
\label{annexe:correl_part_indep}
We show here, using the Wick's theorem, that it is equivalent to have 
the property $C^{(2)} = 0$ and to consider a system of independent particles. 
Starting from a Slater determinant  $|\phi\> $,
using the fact that this state is a vacuum for the $\ob_\mu$
(see appendix \ref{annexe:videHF}), 
and the inverse of Eqs. (\ref{eq:obd_mu}-\ref{eq:ob_mu}) in a given basis denoted by $|i\>$
\oeqn
\oad_i &=& \sum_\mu \sdf \bmu i \> \sdf \[n_\mu \sdf \ob_\mu + (1-n_\mu) \sdf\obd_\mu \] \label{eq:oadi}\\
\oa_i &=& \sum_\mu \sdf \<i \kmu  \sdf \[n_\mu \sdf \obd_\mu + (1-n_\mu) \sdf\ob_\mu \] \label{eq:oai}
\ceqn
with $n_\mu=0$ or $1$, we can use Wick's theorem to express two-body matrix elements :
\oeq
{\ro^{(2)}_{ijkl}} = \< \oad_l \, \oad_k \, \oa_i\, \oa_j\>_{\phi} = \overline{\oad_l \oad_k}\sdf  \overline{\oa_i^{ } \oa_j} + \overline{\oad_l \oa_j}\sdf  \overline{\oad_k \oa_i}
- \overline{\oad_l \oa_i}\sdf  \overline{\oad_k \oa_j}.
\label{eq:ro2}
\ceq
Here contractions are made with the HF state  $|\phi\>$. Using $\overline{\oad_i \oa_j} =\<\oad_i \oa_j \>_{\phi} =  \ro_{{ji}}$
and $\overline{\oad_i \oad_j} = 0$, we finally 
 obtain that the two-body matrix elements read 
\oeq
\ro_{{ijkl}}^{(2)} = \ro_{{jl}}^{(1)} \sdf \ro_{{ik}}^{(1)} - \ro_{{il}}^{(1)} \sdf \ro_{{jk}}^{(1)}.
\label{eq:ro_slater}
\ceq
Here, the two-body correlation $C^{(2)}$ is strictly zero.
In the notes, we often used contractions in the specific basis of occupied states, which simply read
\begin{eqnarray}
\overline{\oad_\alpha  \oa_\beta } =\<\oad_\alpha  \oa_\beta  \>_{\phi} =  n_\alpha \delta_{\alpha \beta },~~~~ \overline{\oad_\alpha  
\oad_\beta } = 0.
\label{eq:etats_occ}
\end{eqnarray}  

\section{Mean-field dynamics from the Thouless Theorem} 
\label{annexe:Thouless}

Starting from Eq. (\ref{eq:hphi}) and an initial Slater determinant, we neglect 
the residual interaction, i.e. we only keep $\hat H_{MF}[\rho]$ in the dynamical evolution. 
Then, over an infinitesimal time step $dt$, the new state is approximated by  
\begin{eqnarray}
\left| \psi (t_0+dt) \right\rangle \simeq \exp \left(\frac{dt}{i\hbar} 
\hat H_{MF}[\rho] \right) 
\left| \phi \right\rangle.
\end{eqnarray} 
where $\hat H_{MF}[\rho] = E[\ro]+ \sum_{\bar \beta \alpha} \sdf h[\rho]_{\bar{\beta} \al} \sdf
\hat a_{\bar{\beta}}^{\dagger} \hat a_{\alpha}$, 
and $E[\ro] = \Tr \[\rho \( t+\frac{1}{2}U[\rho] \)\]$. 
We recall that ${\al}$ denotes an occupied state and $\bar{\al}$ an unoccupied one. 
As we will discuss below, to recover the single-particle evolutions, 
it is convenient to rewrite the Mean-Field Hamiltonian in a slightly different
manner (using $\hat a_{\beta}^{\dagger} \hat a_{\alpha} \left| \phi \right\rangle = \delta_{\alpha \beta } \left| \phi \right\rangle$ )
\begin{eqnarray}
\hat H_{MF}[\rho]\left| \phi \right\rangle 
&=& \Big( -\frac{1}{2} \Tr(\rho U[\rho] ) + \sum_{\beta \alpha} \left\langle \beta | h[\rho] | \alpha \right\rangle
\hat a_{\beta }^{\dagger} \hat a_{\alpha}
+ \sum_{\bar \beta \alpha} \sdf \left\langle \bar{\beta} | h[\rho] | \alpha \right\rangle
\hat a_{\bar{\beta}}^{\dagger} \hat a_{\alpha} \Big) \left| \phi \right\rangle \nonumber \\
&=& \Big( -\frac{1}{2} \Tr(\rho U[\rho] ) + \sum_{i \alpha} \left\langle i | h[\rho] | \alpha \right\rangle
\hat a_{i}^{\dagger} \hat a_{\alpha} \Big) \left| \phi \right\rangle
\label{eq:thoulesstowave}
\end{eqnarray} 
where "i" denotes a complete single-particle basis. The term $-\frac{1}{2} \Tr(\rho U[\rho] )$ does not influence  
the single particle evolution (we omit its contribution in the following).   
The propagator expressed as an exponential of one-body operators and therefore, 
according to the Thouless Theorem 
\cite{tho61}, 
transforms a Slater Determinant into another Slater determinant. 
Indeed, the fact that $e^{-\frac{dt}{i\hbar} 
\hat H_{MF}} e^{\frac{dt}{i\hbar} 
\hat H_{MF}} = 1$  and $e^{\frac{dt}{i\hbar} 
\hat H_{MF}}  \left| - \right\rangle \propto \left| - \right\rangle$ allow us to write
\begin{eqnarray}
e^{\frac{dt}{i\hbar} 
\hat H_{MF}} \left| \phi \right\rangle &=& 
e^{\frac{dt}{i\hbar} 
\hat H_{MF}} \Pi_\alpha \hat a^{\dagger}_\alpha \left| - \right\rangle \nonumber \\
&=& e^{\frac{dt}{i\hbar} 
\hat H_{MF}} a^{\dagger}_{\alpha_1} 
e^{-\frac{dt}{i\hbar} 
\hat H_{MF}} e^{\frac{dt}{i\hbar} 
\hat H_{MF}} a^{\dagger}_{\alpha_2} e^{-\frac{dt}{i\hbar} 
\hat H_{MF}} \cdots e^{+\frac{dt}{i\hbar} 
\hat H_{MF}} a^{\dagger}_{\alpha_N} e^{-\frac{dt}{i\hbar} 
\hat H_{MF}} e^{+\frac{dt}{i\hbar} 
\hat H_{MF}}\left| - \right\rangle . \nonumber  
\end{eqnarray}
Considering the transformation of each creation operator separately, we have 
\begin{eqnarray}
 e^{\frac{dt}{i\hbar} 
\hat H_{MF}}  a^{\dagger}_{\alpha}  e^{-\frac{dt}{i\hbar} 
\hat H_{MF}} &=& a^{\dagger}_{\alpha} + \frac{dt}{i\hbar} [\hat H_{MF}, a^{\dagger}_{\alpha}] + o(dt) \nonumber \\
&=& a^{\dagger}_{\alpha} + \frac{dt}{i\hbar} \sum_i  a^\dagger_i \left\langle i \left| h[\rho] \right| \alpha 
\right\rangle + o(dt) \equiv a^{\dagger}_{\alpha + d \alpha } + o(dt) ,
\end{eqnarray}
where the expression of the mean-field operator defined in Eq.~(\ref{eq:hphi}) has been used. 
From the above identity, we see that the propagated many-body 
state writes $\left| \Psi(t+dt) \right\rangle \propto \Pi_{\alpha} a^{\dagger}_{\alpha + d \alpha } 
\left| - \right\rangle$  where, using $\sum_i \left| i \right\rangle\left\langle i \right|=1$, the single-particle 
states evolve according to 
\begin{eqnarray}
i\hbar \frac{d\left| \alpha \right\rangle}{dt} = h[\rho] \left| \alpha \right\rangle
\label{eq:standardtdhf}
\end{eqnarray}
which is nothing but the standard mean-field evolution. Therefore, we have shown in this appendix that 
the mean-field evolution is recovered by neglecting the two-body residual interaction. 

The fact that we exactly recover the standard single-particle wave-function evolution is due to the specific 
way we have written the Mean-Field Hamiltonian in Eq. (\ref{eq:thoulesstowave}). If, instead, we take the original 
expression and use the fact that $E[\rho]$ induces only a phase,
then, using the same technique, we would get
\begin{eqnarray}
i\hb \frac{d\left| \alpha \right\rangle}{dt}=  (1-\oro) \sdf \oh[\rho]\sdf  \left| \alpha \right\rangle.
\end{eqnarray}
These single-particle equation differs from the standard Eq. (\ref{eq:standardtdhf}) but 
contains the same information. Indeed, we have 
\oeq
i\hb \partial_t \oro =
i \hb \sdf \sum_\al \sdf \[ (\partial_t |\al \> )\sdf  \<\al | + |\al \> \sdf (\partial_t \< \al |) \] 
= (1-\oro) \sdf \oh[\ro] \sdf \oro - \oro \sdf \oh[\ro] \sdf(1-\oro) = \[\oh[\ro],\oro\]
\ceq
where we have recovered the correct one-body density evolution.
\newpage


\end{document}